\documentclass[
aps,
pra,
showpacs
amsmath,
amssymb,
reprint,
floatfix,
longbibliography,
letterpaper,
lengthcheck,
superscriptaddress,
nofootinbib,
]{revtex4}

\usepackage{graphicx}
\usepackage{rotating}
\usepackage{amsmath}
\usepackage{bbm}
\usepackage{subfigure}
\usepackage{color}
\usepackage{dsfont}
\usepackage{ulem}

\graphicspath{
  {figures/pdf/}
  {figures/eps/}
}

\def\diff{\mathrm{d}}
\def\imagi{\mathrm{i}}

\def\p{\,|\,}

\newcommand{\ext}{{\rm ext}}

\newcommand{\ket}[1]{\left| \, #1 \, \right\rangle}
\newcommand{\braket}[2]{\left\langle \, #1 \,|\, #2 \, \right\rangle}

\def\im{{\mbox{Im}}}

\def\bea{\begin{eqnarray}} 
\def\eea{\end{eqnarray}} 
\def\ben{\begin{equation}} 
\def\een{\end{equation}} 
\def\benu{\begin{enumerate}} 
\def\enu{\end{enumerate}}

\def\s{_{\sss S}}

\def\p{\,|\,}

\definecolor{DarkGreen}{rgb}{0.000000,0.392157,0.000000}

\def\J{\hat{J}}

\def\A{\hat{A}}

\def\s{\hat{\sigma}}

\def\a{\hat{a}}
\def\ad{\hat{a}^{\dagger}}

\def\dspin{\hat{\bar{\psi}}}
\def\spin{\hat{\psi}}

\def\E{\hat{E}}

\def\p{\hat{\phi}}
\def\c{\hat{\chi}}

\def\ext{\mathrm{ext}}

\def\braket#1#2{\langle #1 \vert #2 \rangle}

\begin{document}

\title{Quantum Electrodynamical Density-Functional Theory: Bridging Quantum Optics and Electronic-Structure Theory}

\author{Michael Ruggenthaler}
  \email[Electronic address:\;]{michael.ruggenthaler@uibk.ac.at}
  \affiliation{Institut f\"ur Theoretische Physik, Universit\"at Innsbruck, Technikerstra{\ss}e 25, A-6020 Innsbruck, Austria}
\author{Johannes Flick}
   \affiliation{Fritz-Haber-Institut der Max-Planck-Gesellschaft, Faradayweg 4-6, D-14195 Berlin-Dahlem, Germany}
\author{Camilla Pellegrini}
  \affiliation{Nano-Bio Spectroscopy Group and ETSF Scientific Development Centre, Departamento de F\'isica de Materiales, Centro de F\'isica de Materiales CSIC-UPV/EHU-MPC and DIPC, Universidad del Pa\'is Vasco UPV/EHU, E-20018 San Sebasti\'an, Spain}
\author{Heiko Appel}
   \affiliation{Fritz-Haber-Institut der Max-Planck-Gesellschaft, Faradayweg 4-6, D-14195 Berlin-Dahlem, Germany}
\author{Ilya V. Tokatly}
   \affiliation{Nano-Bio Spectroscopy Group and ETSF Scientific Development Centre, Departamento de F\'isica de Materiales, Centro de F\'isica de Materiales CSIC-UPV/EHU-MPC and DIPC, Universidad del Pa\'is Vasco UPV/EHU, E-20018 San Sebasti\'an, Spain}
   \affiliation{IKERBASQUE, Basque Foundation for Science, 48011 Bilbao, Spain}
\author{Angel Rubio}
    \email{angel.rubio@ehu.es}
    \affiliation{Nano-Bio Spectroscopy Group and ETSF Scientific Development Centre, Departamento de F\'isica de Materiales, Centro de F\'isica de Materiales CSIC-UPV/EHU-MPC and DIPC, Universidad del Pa\'is Vasco UPV/EHU, E-20018 San Sebasti\'an, Spain}
    \affiliation{Fritz-Haber-Institut der Max-Planck-Gesellschaft,
    Faradayweg 4-6, D-14195 Berlin-Dahlem, Germany}

\date{\today}

\begin{abstract}
In this work we give a comprehensive derivation of an exact and numerically feasible method to perform ab-initio calculations of quantum particles interacting with a quantized electromagnetic field. We present a hierachy of density-functional-type theories that describe the interaction of charged particles with photons and introduce the appropriate Kohn-Sham schemes. We show how the evolution of a system described by quantum electrodynamics in Coulomb gauge is uniquely determined by its initial state and two reduced quantities. These two fundamental observables, the polarization of the Dirac field and the vector potential of the photon field, can be calculated by solving two coupled, non-linear evolution equations without the need to explicitly determine the (numerically infeasible) many-body wave function of the coupled quantum system. To find reliable approximations to the implicit functionals we present the according Kohn-Sham construction. In the non-relativistic limit this density-functional-type theory 
of quantum 
electrodynamics reduces to the density-functional reformulation of the Pauli-Fierz Hamiltonian, which is based on the current density of the electrons and the vector potential of the photon field. By making further approximations, e.g. restricting the allowed modes of the photon field, we derive further density-functional-type theories of coupled matter-photon systems for the corresponding approximate Hamiltonians. In the limit of only two sites and one mode we deduce the according effective theory for the two-site Hubbard model coupled to one photonic mode. This model system is used to illustrate the basic ideas of a density-functional reformulation in great detail and we present the exact Kohn-Sham potentials for our coupled matter-photon model system.
\end{abstract}


\date{\today}

\maketitle

\section{Introduction}
\label{sec:Intro}
\noindent

The behavior of elementary charged particles, like electrons and positrons is governed by quantum electrodynamics (QED). In this theory the quantum particles interact via the exchange of the quanta of light, i.e. the photons \cite{ryder-QFT, greiner-FQ, greiner-QED}. Thus in principle we have to consider the quantum nature of the charged particles as well as of the light field. However, in several important cases we can focus almost exclusively on either the charged particles or the photons, while employing crude approximations for the other degrees of freedom.

In condensed matter physics and quantum chemistry the quantum nature of light can usually be ignored and the interaction between the charged quantum particles is approximated by the instantaneous Coulomb interaction. However, even then the resulting quantum mechanical equations (usually the many-body Schr\"odinger equation), where the electromagnetic fields are treated classically through the solution of the Maxwell equations, are solvable only for very simple systems. This lies ultimately in our incapability of handling the huge number of degrees of freedom of many-particle systems and consequently in our inability to determine the many-body states. This so-called many-body problem spawned a lot of interest into the question whether one can devise a closed set of equations for reduced quantities which do not involve the explicit solution of the full quantum mechanical equations and in which the many-body correlations can be approximated efficiently. Pursuits in this direction have led to various approaches 
such as, among others, many-body Green's function theories \cite{fetter-MBT, stefanucci-MBT}, (reduced) density-matrix theories \cite{bonitz-QKT} and density-functional theories \cite{dreizler-DFT, engel-DFT, ullrich-TDDFT, marques-TDDFT}. These approaches differ in the complexity of the reduced quantity, which is used to calculate the various observables of interest. Especially density-functional theories, which are based on the simplest of those (functional) variables, the one-particle density (current), have proven to be exceptionally successful \cite{burke-2012}. Their success can be attributed to the unprecedented balance between accuracy and numerical feasibility \cite{bleiziffer-2013}, which allows at present to treat several thousands of atoms \cite{andrade-2012}. Although the different flavors of density-functional theories cover most of the traditional problems of physics and chemistry (including approaches that combine classical Maxwell dynamics with the quantum particles \cite{gupta-2001, iwasa-2009, chen-2010, fratalocchi-2011, yabana-2012}), by construction these theories cannot treat problems involving the quantum nature of light. 

In quantum optics, on the other hand, the focus is on the photons, while usually simple approximations for the charged particles are employed, e.g. a few-level approximation. However, even in this situation the solution of the resulting equations \cite{scully-QO,gardiner-QO} is only possible in simple cases (again due to the large number of degrees of freedom) and usually simplified model Hamiltonians, e.g. the Dicke model realized in a cavity \cite{dicke-1954, chen-2008, braak-2013}, are employed to describe these physical situations. Already the validity of these effective Hamiltonians and their properties can be a matter of debate \cite{rzazewski-1975, rzazewski-1991, vukics-2012} and often further simplifications are adopted such as the Jaynes-Cummings model in the rotating-wave approximation. The rapid progress in quantum-optical experiments on the other hand, especially in the field of cavity QED \cite{raimond-2001, walther-2006, mekhov-2012, ritsch-2013} and circuit QED \cite{blais-2004,wallraff-2004}
, allows to study and control multi-particle systems ultra-strongly coupled to photons \cite{todorov-2010, you-2011, schwartz-2011, imorral-2012}, where such a simple approximative treatment is no longer valid \cite{ciuti-2006}. This new regime of light-matter interaction is widely unexplored for, e.g., molecular physics and material sciences \cite{hutchison-2012}. Possibilities like altering and strongly influencing the chemical reactions of a molecule in the presence of a cavity mode or setting the matter into new non-equilibrium states with novel properties, e.g. light-induced superconductivity \cite{fausti-2011}, arise. Specifically in such situations an oversimplified treatment of the charged particles may no longer be allowed and an approach that considers both, the quantum nature of the light field as well as of the charged particles is needed.

In this work we give a comprehensive derivation of an exact and numerically feasible method that generalizes ideas of time-dependent density-functional theory (TDDFT). This method bridges the gap between the above two extreme cases and provides a scheme to perform ab-initio calculations of quantum particles coupled to photons. The electron-photon generalization of TDDFT in describing non-relativistic many-electron systems coupled to photon modes of mesoscopic cavities was introduced in Ref.~\cite{tokatly-2013}.  Here we provide a general framework describing fully coupled electron-photon systems in most possible regimes/systems ranging from effective model Hamiltonians to strongly relativistic cases, which has been introduced in Refs.~\cite{rajagopal-1994, ruggenthaler-2011b}. For clarity we divide the following presentation in two parts: We first demonstrate the basic ideas in a simple model system and then show how these concepts can be used in the case of general coupled matter-photon problems. A summary 
of all findings of the present work for the time-dependent density-functional description of QED at different levels of approximations, namely the basic variables, initial conditions and fundamental Kohn-Sham multicomponent equations is given in appendix~\ref{app:overview}.

We start considering a simple model system for charged matter coupled to photons: the two-site Hubbard model interacting with one photonic mode. By employing density-functional ideas we show how one can solve this quantum-mechanical problem without the need to explicitly calculate the complex many-body wave function. Instead, we derive equations of motion for a pair of reduced quantities from which all physical observables can by determined. We demonstrate that these equations have unique solutions and can be used to calculate the basic reduced quantities (here the basic pair of reduced quantities is the charge density of the particle and the potential induced by the photons) of the coupled problem. Therefore we here reformulate the coupled matter-photon problem in terms of an effective theory, that we call in the following a model of quantum electrodynamical density-functional theory (QEDFT). Since an explicit calculation of the coupled wave function is not needed, this approach allows to determine 
properties of the matter-photon system in a numerically feasible way. We introduce an new Kohn-Sham scheme to approximate the unknown functionals in the basic equations of motion and present results for a simple approximation. We compare these results to the exact Kohn-Sham functionals and identify shortcomings and indicate improvements.

Based on the ideas developed in the first part of this work we repeat the steps illustrated in our example but now we construct a density-functional reformulation for the full theory of QED \cite{rajagopal-1994,ruggenthaler-2011b}. We show that a straightforward approach based on the current and the potential leads to problems and that a consistent density-functional reformulation of QED has to be based on the polarization and the potential which is generated by the photons. This approach to the fully coupled QED problem we denote as relativistic QEDFT, and we present the corresponding Kohn-Sham construction and give the simplest approximation to the unknown functionals. In the following we then demonstrate how relativistic QEDFT reduces in the non-relativistic limit to its non-relativistic version of the corresponding non-relativistic Hamiltonian. By employing further approximations on the matter system or on the photon field a family of different approximate QEDFTs is introduced, which are consistent with 
their respective approximate Hamiltonians. At this level we recover the theory of Ref.~\cite{tokatly-2013}. In lowest order we rederive the model QEDFT of the first part of this work. Therefore, we demonstrate how all different flavours of QEDFT are just approximations to relativistic QEDFT in the same manner as different physical Hamiltonians are merely approximations to the QED Hamiltonian. Furthermore, by ignoring all photonic degrees of freedom, we find the standard formulations of TDDFT which are extensively used in the electronic-structure community \cite{marques-TDDFT, ullrich-TDDFT}.

\textbf{Outline} In Sec.~\ref{sec:Model} we investigate the QEDFT reformulation of a simple model of one particle coupled to one mode in great detail. The developed ideas are then employed in Sec.~\ref{sec:RelQEDFT} to derive a QEDFT reformulation of QED. In Sec.~\ref{sec:NonRelQEDFT} we show how all different QEDFT reformulations are approximations to relativistic QEDFT. We conclude and give an outlook in Sec.~\ref{sec:Conclusion}.


\section{Model of QEDFT}
\label{sec:Model}

In this section, we introduce the basic formulation and underlying ideas of QEDFT. By employing a model Hamiltonian, we can almost exclusively focus on the density-functional ideas that allow a reformulation of the wave-function problem in terms of simple effective quantities. We first identify the pair of external and internal variables and then show that both are connected via a bijective mapping. As a consequence, all expectation values become functionals of the initial state and the internal pair. This allows for a reformulation of the problem in terms of two coupled equations for the internal pair. Then we introduce the Kohn-Sham construction as a way to find approximations to the unknown functionals, and show first numerical results.

To describe the dynamics of particles coupled to photons we solve an evolution equation of the form
\begin{align}
\label{Schr�dinger}
 \imagi \hbar c \partial_0 \ket{\Psi(t)} = \hat{H}(t) \ket{\Psi(t)}                                                            
\end{align}
for a given initial state $\ket{\Psi_0}$. Here $\partial_0 = \partial/\partial x^0$ with $x^{0} = ct$ and the standard relativistic (covariant) notation $x \equiv (ct, \vec{r})$ (see also appendix \ref{app:con} for notational conventions). The corresponding hermitean Hamiltonian has the general form
\begin{align}
\label{BasicHam}
 \hat{H}(t)=& \hat{H}_{\mathrm{M}} + \hat{H}_{\mathrm{EM}} + \frac{1}{c}\int \diff^3 r \; \J_{\mu}(x) \A^{\mu}(x) 
\\
&+ \frac{1}{c} \int \diff^3 r \left( \J_{\mu}(x)a_{\ext}^{\mu}(x) + \A_{\mu}(x)j_{\ext}^{\mu}(x) \right), \nonumber
\end{align}
where the depedence of the total Hamiltonian on $t$ indicates an explicit time-dependence. Here the (time-independent) Hamiltonian $\hat{H}_{\mathrm{M}}$ describes the kinetic energy of the particles, i.e. how they would evolve without any perturbation, and $\hat{H}_{\mathrm{EM}}$ is the energy of the photon-field. The third term describes the coupling between the (charged) particles and the photons by the charge current $\J_{\mu}$ and the Maxwell-field operators $\A_{\mu}$ (where the Einstein sum convention with the Minkowski metric $g_{\mu \nu}\equiv(1,-1,-1,-1)$ is implied and greek letters refer to four vectors, e.g. $\mu \in \{0,1,2,3 \}$, while roman letters are restricted to spatial vectors only, e.g. $k \in\{1,2,3\}$). This term is frequently called the minimal-coupling term and arises due to the requirement of a gauge-invariant coupling between the particles and the photon field. The specific form of the operators $\J_{\mu}$ and $\A_{\mu}$ depends on the details of the physical situation. Finally, 
the last term describes how the particles interact with a  (in general time-dependent) classical external vector potential $a_{\ext}^{\mu}$ and how the photons couple to a (in general time-dependent) classical external current $j_{\ext}^{\mu}$. 

While we usually have no control over how the particles and photons evolve freely or interact, i.e. the first three terms of the Hamiltonian (\ref{BasicHam}), we have control over the preparation of the initial state $\ket{\Psi_0}$ and the external fields $(a_{\ext}^{\mu},j_{\ext}^{\mu})$. Therefore, all physical wave functions, i.e. found by solving Eq.~(\ref{Schr�dinger}), can be labeled by their initial state and \textit{external pair} $(a_{\ext}^{\mu},j_{\ext}^{\mu})$, 
\begin{align*}
 \ket{\Psi([\Psi_0,a_{\ext}^{\mu},j_{\ext}^{\mu}];t)}.
\end{align*}
However, for any but the simplest systems the (numerically exact) solution of Eq.~(\ref{Schr�dinger}) is not feasible. Even if we decouple the matter part from the photons by employing the Coulomb-approximation (i.e. describing the exchange of photons by the respective lowest-order propagator) the resulting problem is far from trivial.


\subsection{Two-level system coupled to one mode} 

\label{subsec:Model}

In this subsection, we introduce a simple model of charged particles coupled to photons. We discuss the basic concepts of a density-functional-type reformulation, identify the pair of conjugate variables and then deduce the fundamental equations of motion on which we base our QEDFT reformulation.

In order to demonstrate the basic ideas of a QEDFT we employ the simplest yet non-trivial realization of one charged particle coupled to photons: a two-site Hubbard model coupled to one photonic mode. The resulting Hamiltonian (see appendix \ref{app:Hubbard} for a detailed derivation) reads as
\begin{align}
\label{Hubbard}
 \hat{H}(t) = \hat{H}_{\mathrm{M}} + \hat{H}_{\mathrm{EM}}  -  \frac{\lambda}{c} \J \A -   \frac{1}{c} \left[ \J a_{\ext}(t) + \A j_{\ext}(t)  \right],
\end{align}
where the kinetic energy of the charged particle is given by 
\begin{align*}
 \hat{H}_{\mathrm{M}} = -t_{\mathrm{kin}} \s_x,
\end{align*}
and the energy of the photon mode reads
\begin{align*}
 \hat{H}_{\mathrm{EM}} = \hbar \omega \ad \a.
\end{align*}
Here $t_{\mathrm{kin}}$ is the hopping parameter between the two sites, $\omega$ is the frequency of the photonic mode and $(\s_x, \s_y,\s_z)$ are the Pauli matrices that obey the usual fermionic anti-commutation relations. The photon creation and annihilation operators ($\a^\dagger$ and $\a$, respectively) obey the usual bosonic commutation relations. The current operator \footnote{To be precise, $\J$ is proportional to the dipole-moment operator, i.e. it is connected to the zero component $\J_{0}$ of the general four-current operator $\J_{\mu}$. To highlight the analogy in structure to the general case discussed in the later sections, we give it the units of a current and denote it by $\J$} is defined by 
\begin{align*}
 \J = e \omega l \s_z,
\end{align*}
where $l$ is a characteristic length-scale of the matter-part and $\lambda$ is a dimensionless coupling constant.
\\
The operator for the conjugate potential \footnote{To be precise, $\A$ is actually proportional to the electric field as can be seen from the derivations in appendix \ref{app:Hubbard}. This is, because in the course of approximations one employs the length-gauge and thus transforms from the potential to the electric field. However, to highlight the analogy in structure to the general case discussed in the later sections, we give it the units of the potential and denote it by $\A$} is given by
\begin{align*}
 \A = \left(\frac{\hbar c^2}{\epsilon_0 L^3}  \right)^{1/2} \frac{(\a + \ad)}{\sqrt{2 \omega}},
\end{align*}
where $L$ is the length of the cubic cavity.
Further, the current operator couples to the external potential $a_{\ext}(t)$ and the potential operator to the external current $j_{\ext}(t)$. These are the two (classical) external fields that we can use to control the dynamics. 

If we then fix an initial state $\ket{\Psi_0}$ and choose an external pair $(a_{\ext},j_{\ext})$, we usually want to solve Eq.~(\ref{Schr�dinger}) with the Hamiltonian given by Eq.~(\ref{Hubbard}). The resulting wave function, given in a site basis $\ket{x}$ for the charged particle and a Fock number-state basis $\ket{n}$ for the photons
\begin{align*}
 \ket{\Psi([\Psi_0,a_{\ext},j_{\ext}];t)} = \sum_{x=1}^{2} \sum_{n=0}^{\infty} c_{xn}(t) \ket{x}\otimes \ket{n},
\end{align*}
depends on the initial state and the external pair $(a_{\ext},j_{\ext})$. Thus, by varying over all possible combinations of pairs $(a_{\ext},j_{\ext})$, we scan through all physically allowed wave functions starting from a given initial state. Hence, we parametrize the relevant, i.e. physical, time-dependent wave functions by $\ket{\Psi_0}$ and $(a_{\ext},j_{\ext})$. Since the wave functions have these dependencies, also all derived expressions, e.g. the expectation values for general operators $\hat{O}$
\begin{align*}
O([\Psi_0,a_{\ext},j_{\ext}],t) = \braket{\Psi(t)|\hat{O}}{\Psi(t)},
\end{align*}
are determined by the initial state and the external pair $(a_{\ext},j_{\ext})$. 

The idea of an \textit{exact} effective theory like QEDFT is now, that we identify a different set of fundamental variables, which also allow us to label the physical wave functions (and their respective observables), and that we have a closed set of equations for these new (functional) variables, which do not involve the full wave functions explicitly. Such a functional-variable change is similar to a coordinate transformation, say from Cartesian coordinates to spherical coordinates. This can only be done if every point in one coordinate system is mapped uniquely to a point in the other coordinate system. For a functional-variable change we thus need to have a one-to-one correspondence, i.e. bijective mapping, between the set of (allowed) pairs $(a_{\ext},j_{\ext})$ and some other set of functions (while we keep the initial state fixed). To identify the simplest new functional variables one usually employs arguments based on the Legendre transformation \cite{vanleeuwen-2001}. That is why these new 
functional variables are often called conjugate variables. We will consider this approach in the next sections where we investigate general QEDFT, and also show how one can determine the conjugate variables of this model system from more general formulations of QEDFT. For this simple model we simply state that a possible pair of conjugate variables is $(J,A)$. 
In the next subsection we show that this functional variable-transformation is indeed allowed, i.e.
\begin{align*}
 \ket{\Psi([\Psi_0,J,A];t)}.
\end{align*}
The main consequence of this result is, that from only knowing these three basic quantities we can (in principle) uniquely determine the full wave function. Accordingly, every expectation value becomes a unique functional of $\ket{\Psi_0}$ and $(J,A)$. Thus, instead of trying to calculate the (numerically expensive) wave function, it is enough to determine the \textit{internal pair} $(J,A)$ for a given initial state. An obvious route to then also find a closed set of equations for these new variables is via their respective equations of motion. These equations will at the same time be used to prove the existence of the above change of variables, i.e. that the wave function is a unique functional of the initial state and the internal pair $(J,A)$.

To find appropriate equations we first apply the Heisenberg equation of motion once and find
\begin{align*}
 \imagi \partial_0 \J &= -\imagi \frac{2 t_{\mathrm{kin}}e \omega l}{ \hbar c} \s_y,
\\
 \imagi \partial_0 \A& = -\imagi  \hat{E},
\end{align*}
where $\hat{E}= i \sqrt{\frac{\hbar\omega}{2\epsilon_0 L^3}} \left( a - a^\dag\right)$. Yet these two equations are not sufficient for our purposes: we need equations that explicitly connect $(a_{ext},j_{\ext})$ and $(J,A)$. Therefore we have to go to the second order in time,
\begin{align}
\label{DiscerteForce}
 \left(\imagi \partial_0\right)^2 \J &=  \frac{4 t^2_{\mathrm{kin}}}{\hbar^2 c^2} \J - \lambda \hat{n} \A - \hat{n}a_\mathrm{ext}(t),
\\
\label{DiscerteMaxwell}
\left( \imagi \partial_0\right)^2 \A &=  k^2 \A - \frac{\mu_0 c}{L^3}\left( \lambda \J + j_\mathrm{ext}(t)\right).
\end{align}
where 
\begin{align}
 \hat{n} = \frac{4 t_{\mathrm{kin}}(e \omega l)^2}{ \hbar^2 c^3} \s_x,
\end{align}
$k=\frac{\omega}{c}$ and $\epsilon = \frac{1}{\mu_0c^2}$. Here, Eq.~(\ref{DiscerteForce}) is the discretized version of $\partial_t^2 n$ of standard TDDFT \cite{leeuwen-1999, farzanehpour-2012}, and Eq.~(\ref{DiscerteMaxwell}) is the inhomogeneous Maxwell equation for one photon mode \cite{tokatly-2013}. 


\subsection{Foundations of the model QEDFT} 
\label{subsec:FoundModel}
 
In the previous subsection we have stated that $(J,A)$ and $(a_{\ext},j_{\ext})$ are the possible conjugate pair of the model Hamiltonian (\ref{Hubbard}). In this subsection we want to demonstrate that indeed this holds true and that we can perform a variable transformation from the external pair $(a_{\ext},j_{\ext})$ to the internal pair $(J,A)$. What we need to show is, that for a fixed initial state $\ket{\Psi_0}$ the mapping   
\begin{align}
\label{PairMapping}
(a_{\ext},j_{\ext})  \; {\buildrel\rm 1:1 \over \leftrightarrow }  \; (J,A)
\end{align}
is bijective, i.e. if $(a_{\ext},j_{\ext}) \neq (\tilde{a}_{\ext},\tilde{j}_{\ext})$ then necessarily for the according expectation values $(J,A) \neq (\tilde{J}, \tilde{A})$. To do so we first note, that in the above equations of motion every expectation value is by construction a functional of $(a_{\ext},j_{\ext})$ for a fixed initial state, 
\begin{align}
\label{DiscerteForceExp}
\partial_0^2 J&([a_{\ext},j_{\ext}];t) =  - \frac{4 t^2_{\mathrm{kin}}}{
\hbar^2 c^2} J([a_{\ext},j_{\ext}];t) 
\\
&+ \lambda \langle \hat{n} \A \rangle ([a_{\ext},j_{\ext}];t) + n ([a_{\ext},j_{\ext}];t) a_{\ext}(t), \nonumber
\\
\label{DiscerteMaxwellExp}
\partial_0^2 A&([a_{\ext},j_{\ext}];t) =  - k^2 A([a_{\ext},j_{\ext}];t) \nonumber
\\
& + \frac{\mu_0 c}{L^3}\left( \lambda J([a_{\ext},j_{\ext}];t) + j_{\ext}(t)\right),
\end{align}
i.e. they are generated by a time-propagation of $\ket{\Psi_0}$ with a given external pair $(a_{\ext},j_{\ext})$. Suppose now, that we fix the expectation values of the internal variables $(J, A)$, i.e. we do not regard them as functionals but rather as functional variables. Then the above Eqs.~(\ref{DiscerteForceExp}) and (\ref{DiscerteMaxwellExp}) become equations for the pair $(a_{\ext},j_{\ext})$ that produce the given internal pair $(J,A)$ via propagation of the initial state $\ket{\Psi_0}$, i.e.
\begin{align}
\label{DiscerteForceFun}
 \partial_0^2  J(t) =&  -\frac{4 t^2_{\mathrm{kin}}}{  \hbar^2 c^2} J(t) + \lambda \langle \hat{n} \A \rangle ([a_{\ext},j_{\ext}];t) \nonumber
\\
 &+ n ([a_{\ext},j_{\ext}];t) a_{\ext}(t),
\\
\label{DiscerteMaxwellFun}
\partial_0^2  A(t) =&  -k^2 A(t)  + \frac{\mu_0 c}{L^3}\left( \lambda J(t) + j_{\ext}(t)\right).
\end{align}
Obviously these equations can only have a solution, if the given internal variables are consistent with the initial state, i.e.
\begin{align}
\label{InitialCurrent}
J^{(0)} &= \braket{\Psi_0| \J }{\Psi_0}, \; J^{(1)} = - \frac{2 t_{\mathrm{kin}}e \omega l}{\hbar c} \braket{\Psi_0|\s_y}{\Psi_0},
\\
\label{InitialPotential}
A^{(0)} &=  \braket{\Psi_0| \A}{\Psi_0}, \; A^{(1)} =  -  \braket{\Psi_0| \hat{E}}{\Psi_0}.
\end{align}
Here we have used the definition
\begin{align}
 A^{(\alpha)} = \left.\partial_{0}^{\alpha} A(t)\right|_{t=0},
\end{align}
and every internal pair $(J,A)$ that we consider is subject to these boundary conditions. Thus, the mapping (\ref{PairMapping}) is bijective, if the corresponding Eqs.~(\ref{DiscerteForceFun}) and (\ref{DiscerteMaxwellFun}), which connect the internal pair $(J,A)$ with the external pair $(a_{\ext},j_{\ext})$, allow for one and only one solution pair.

Let us first note that for a given pair $(J,A)$, Eq.~(\ref{DiscerteMaxwellFun}) uniquely determines the external current $j$ by
\begin{align}
j_{\ext}(t) =   \frac{L^3}{\mu_0 c} \left( \partial_0^2 + k^2 \right) A(t)  - \lambda J(t).
\end{align}
Thus, the original problem reduces to the question whether Eq.~(\ref{DiscerteForceFun}) determines $a_{\ext}(t)$ uniquely. The most general approach to answer this question is via a fixed-point procedure similar to Ref. \cite{ruggenthaler-2011}. In the case of a discretized Schr\"odinger equation like Eq.~(\ref{Hubbard}) it should also be possible to apply a rigorous approach based on the well established theory of nonlinear ordinary differential equations \cite{farzanehpour-2012}. However, for simplicity we follow Ref.~\cite{tokatly-2013} and employ the standard strategy of \cite{runge-1984} which restricts the allowed external potentials $a_{\ext}$ to being Taylor-expandable in time, i.e.
\begin{align}
 a_{\ext}(t) = \sum_{\alpha=0}^{\infty} \frac{a_{\ext}^{(\alpha)}}{\alpha !}(ct)^{\alpha}.
\end{align}
From Eq.~(\ref{DiscerteForce}) we can find the Taylor-coefficients of $J$ (if they exist) by
\begin{align}
 J^{(\alpha+2)} = -\frac{4 t^2_{\mathrm{kin}}}{c^2 \hbar^2} J^{(\alpha)} + \lambda \langle \hat{n} \A \rangle ^{(\alpha)} + \sum_{\beta=0}^{\alpha} {\alpha \choose \beta} n^{(\alpha-\beta)}a_{\ext}^{(\beta)},
\end{align}
where the terms $\langle \hat{n} \A \rangle ^{(\alpha)}$ and $n^{(\alpha)}$ are given by their respective Heisenberg equations at $t=0$ and only contain Taylor coefficients of $a_{\ext}^{(\beta)}$ for $\beta < \alpha$.

Now, assume that we have two different external potentials $a_{\ext}(t) \neq \tilde{a}_{\ext}(t)$. This implies, since we assumed Taylor-expandability of $a_{ext}$ and $\tilde{a}_{\ext}$, that there is a lowest order $\alpha$ for which
\begin{align}
 a_{\ext}^{(\alpha)} \neq \tilde{a}_{\ext}^{(\alpha)}.
\end{align}
For all orders $\beta < \alpha$ (even though the individual $J^{(\beta)}$ and $\tilde{J}^{(\beta)}$ might not exist) it necessarily holds that
\begin{align}
 J^{(\beta+2)}- \tilde{J}^{(\beta+2)} = 0.
\end{align}
But for $\alpha$ we accordingly find that
\begin{align}
 J^{(\alpha+2)}- \tilde{J}^{(\alpha+2)} = n^{(0)}\left( a_{ext}^{(\alpha)} - \tilde{a}_{\ext}^{(\alpha)} \right) \neq 0,
\end{align}
provided we choose the initial state such that ($n^{(0)} \neq 0$). Consequently, $J(t) \neq \tilde{J}(t)$ infinitesimally later for two different external potentials $a_{\ext}(t) \neq \tilde{a}_{\ext}(t)$. Therefore, Eq.~(\ref{DiscerteForceFun}) allows only one solution and the mapping $(a_{\ext},j_{\ext}) \rightarrow (A,J)$ is bijective.

As a consequence, since every expectation value of the quantum system becomes a functional of the internal pair $(J,A)$, in the above Eqs.~(\ref{DiscerteForceFun}) and (\ref{DiscerteMaxwellFun}) we can perform a change of variables and find
\begin{align}
\label{DiscerteForceIntFun}
 \partial_0 ^2  J(t) &=  -\frac{4 t^2_{\mathrm{kin}}}{ \hbar^2 c^2} J(t) + \lambda \langle \hat{n} \A \rangle ([J,A];t) + n ([J,A];t) a_{\ext}(t),
\\
\label{DiscerteMaxwellIntFun}
\partial_0 ^2  A(t) &=  -k^2 A(t)  + \frac{\mu_0 c}{L^3}\left( \lambda J(t) + j_{\ext}(t)\right).
\end{align}
These coupled evolution equations have unique solutions $(J, A)$ for the above initial conditions (\ref{InitialCurrent}) and (\ref{InitialPotential}). Therefore we can, instead of solving for the many-body wave function, solve these non-linear coupled evolution equations for a given initial state and external pair $(a_{\ext},j_{\ext})$, and determine the current and the potential of the combined matter-photon system from which all observables could be computed. This is an exact reformulation of the model in terms of the current and the potential of the combined system only.


\subsection{Kohn-Sham approach to the model QEDFT} 
\label{subsec:KSModel}

In the previous subsection we have derived a QEDFT reformulation in terms of the current and the potential. While the equation that determines the potential $A$ is merely the classical Maxwell equation, and every term is known explicitly, the equation for the current contains implict terms. Therefore, to solve these coupled equations in practice, we need to give appropriate explicit approximations for the implict terms. Approximations based on $(J,A)$ directly would correspond to a Thomas-Fermi-type approach to the model. As known from standard density-functional theory, such approximations are in general very crude and hard to improve upon. A more practical scheme is based on the Kohn-Sham construction, where an auxiliary quantum system is used to prescribe explicit approximations. However, the numerical costs of a Kohn-Sham approach compared to a Thomas-Fermi-type approach are increased.

The details of the Kohn-Sham construction depend on the actual auxiliary quantum system one wants to employ. The only restriction of the auxiliary system is that one can control the current and the potential by some external variables. Thus, one could even add further (unphysical) external fields to make approximations of the coupled quantum system easier. However, here we only present the simplest and most natural Kohn-Sham scheme, which is to describe the coupled quantum system by an uncoupled quantum system. To this end, we assume that we can find a factorized initial state
\begin{align*}
 \ket{\Phi_0} = \ket{\mathrm{M}_0} \otimes \ket{\mathrm{EM}_0}
\end{align*}
that obeys the same initial conditions as the coupled problem (\ref{InitialCurrent}) and (\ref{InitialPotential}).
Especially, if the initial state of the coupled system is the same as in the uncoupled problem, then this condition is trivially fulfilled. In a next step we note, that for the uncoupled system subject to the external pair $(a_{\mathrm{eff}}, j_{\mathrm{eff}})$ the equations of motion become (since $\lambda = 0$)
\begin{align}
\label{DiscerteForceUn}
 \partial_0^2 J([a_{\mathrm{eff}},j_{\mathrm{eff}}];t) &=  - \frac{4 t^2_{\mathrm{kin}}}{ \hbar^2 c^2} J([a_{\mathrm{eff}},j_{\mathrm{eff}}];t) 
\\& + n ([a_{\mathrm{eff}},j_{\mathrm{eff}}]; t) a_{\mathrm{eff}}(t), \nonumber
\\
\label{DiscerteMaxwellUn}
\partial_0^2 A([a_{\mathrm{eff}},j_{\mathrm{eff}}];t) &=  - k^2 A([a_{\mathrm{eff}},j_{\mathrm{eff}}];t) + \frac{\mu_0 c}{L^3} j_{\mathrm{eff}}(t),
\end{align}
Now, obviously if one would choose $(a_{\mathrm{eff}}, j_{\mathrm{eff}}) = (a_{\ext}, j_{\ext})$, i.e. the external pair of the coupled problem, the uncoupled quantum system will in general lead to a different internal pair. However, can we find an effective pair that \textit{reproduces} the internal pair $(J,A)$ of the coupled system? The existence of such an effective pair can be based on equations for the uncoupled system similar to Eqs.~(\ref{DiscerteForceFun}) and (\ref{DiscerteMaxwellFun}). Note that before we were considering the question of uniqueness, i.e. can one have two external pairs leading to the same $(J,A)$. Thus any internal pair $(J,A)$ was apriori associated with an external pair $(a_{\ext},j_{\ext})$. If on the other hand we are given some internal pair $(J,A)$, say from a different (coupled) quantum system, we do not apriori know that this internal pair can be represented by propagation of an initial state with some $(a_{\mathrm{eff}}, j_{\mathrm{eff}})$. Thus this problem is equivalent 
to the existence 
of a solution to 
\begin{align}
\label{DiscerteForceFunUn}
 \partial_0^2  J(t) &=  -\frac{4 t^2_{\mathrm{kin}}}{\hbar^2 c^2} J(t) + n ([a_{\mathrm{eff}},j_{\mathrm{eff}}];t) a_{\mathrm{eff}}(t),
\\
\label{DiscerteMaxwellFunUn}
\partial_0^2  A(t) &=  -k^2 A(t)  + \frac{\mu_0 c}{L^3} j_{\mathrm{eff}}(t).
\end{align}
for a given pair $(J,A)$ and $\ket{\Phi_0}$. As before, $j_{\mathrm{eff}}$ is uniquely determined by simply rearranging Eq.~(\ref{DiscerteMaxwellFunUn}) as
\begin{align*}
j_{\mathrm{eff}}(t) =   \frac{L^3}{\mu_0 c} \left( \partial_0^2 + k^2 \right) A(t),
\end{align*} 
while the existence of an $a_{\mathrm{eff}}$ that reproduces $(J,A)$ is less clear. Again, the most general approach to answer this question can rely on a fixed-point scheme similar to \cite{ruggenthaler-2011}, or on mapping the problem to a special nonlinear Schr\"odinger equation \cite{Tokatly2011ChemPhys,tokatly-2011,farzanehpour-2012}. Importantly, in the discretized case certain subtleties arise \cite{li-2008, baer-2008, tokatly-2011,kurth-2011} that have to be treated with care \cite{tokatly-2011, farzanehpour-2012}. Disregarding these more subtle points, we follow a simpler approach based on the assumption of Taylor-expandability in time of $J$. Then one can successively construct the Taylor coefficients of the effective potential from  
\begin{align*}
a^{(\alpha)}_{\mathrm{eff}}  = \frac{1}{n^{(0)}}\left(\frac{4 t^2_{\mathrm{kin}}}{\hbar^2 c^2} J^{(\alpha)} + J^{(\alpha+2)} - \sum_{\beta=0}^{\alpha-1} {\alpha \choose \beta} n^{(\alpha-\beta)}a^{(\beta)}_{\mathrm{eff}} \right),
\end{align*}
assuming that for the initial state $\ket{\Phi_0}$ the expectation value $n{(0)} \neq 0$. Further assuming that this Taylor-series converges \cite{vignale-2008, leeuwen-1999}, we have constructed a pair
\begin{align*}
 (a_{\mathrm{eff}}[\Phi_0, J,A],j_{\mathrm{eff}}[A]),
\end{align*}
that reproduces $(J,A)$ via propagation of $\ket{\Phi_0}$. 

The above defined pair $(a_{\mathrm{eff}}[\Phi_0, J,A],j_{\mathrm{eff}}[A])$ actually describes the mapping
\begin{align*}
 \left(J,A\right) \; {\buildrel\rm \ket{\Phi_0} \over \mapsto }  \; \left(a_{\mathrm{eff}},j_{\mathrm{eff}}\right).
\end{align*}
Now, in order to actually \textit{predict} the physical pair $(J,A)$ via the Kohn-Sham system (and thus solve Eqs.~(\ref{DiscerteForceIntFun}) and (\ref{DiscerteMaxwellIntFun})) we have to connect the coupled and the auxiliary system. To do so we make the composite mapping
\begin{align*}
 (a_{\ext}, j_{\ext}) \; {\buildrel\rm \ket{\Psi_0} \over \mapsto }  \; (J,A)  \; {\buildrel\rm \ket{\Phi_0} \over \mapsto }  \; (a_{\mathrm{eff}},j_{\mathrm{eff}}),
\end{align*}
i.e. we employ the fact that $(J,A)$ are unique functionals of the initial state $\ket{\Psi_0}$ and $(a_{\ext}, j_{\ext})$. The definition of the resulting Kohn-Sham potentials and currents are then found by equalizing the functional Eqs.~(\ref{DiscerteForceIntFun}) and (\ref{DiscerteMaxwellIntFun}) with the according equations of the uncoupled auxiliary system. This leads to (now also indicating the appropriate dependence on the initial states) \cite{ruggenthaler-2011b, tokatly-2013}
\begin{align}
\label{KSpotential}
n([\Phi_0,J,A];t) a_{\mathrm{KS}}(t) =& \lambda \langle \hat{n} \A \rangle ([\Psi_0,J,A];t)
\\
 &+ n ([\Psi_0,J,A];t) a_{\ext}(t) \nonumber
\\
\label{KScurrent}
 j_{\mathrm{KS}}(t) =&  j_{\ext}(t) + J(t).
\end{align}
Therefore, they are functionals of the two initial states, $(J,A)$ and $(a_{\ext}, j_{\ext})$, i.e.
\begin{align*}
 \left( a_{\mathrm{KS}}[\Psi_0,\Phi_0,J,A,a_{\ext}], j_{\mathrm{KS}}[J, j_{\ext}] \right).
\end{align*} 
With these definitions the coupled problem, starting from $\ket{\Psi_0}$ and subject to the external pair $(a_{\ext}, j_{\ext})$, can be formally solved by the solution of an uncoupled, yet non-linear problem with initial state $\ket{\Phi_0}$ and the Kohn-Sham pair $(a_{\mathrm{KS}}, j_{\mathrm{KS}})$. The resulting equations are
\begin{align}
& \imagi \hbar c \partial_0 \ket{\mathrm{M}(t)} = \left[-t_{\mathrm{kin}} \s_x - \frac{1}{c} \J a_{\mathrm{KS}}(t)  \right]\ket{\mathrm{M}(t)},
\label{ModelKSCurrent}
\\
& \left( \partial_0^2 + k^2\right) A(t) =   \frac{\mu_0 c}{L^3}\left( \lambda J(t) + j(t)\right).
\label{ModelKSField}
\end{align}
The self-consistent solutions of the Kohn-Sham Eqs.~(\ref{ModelKSCurrent}) and (\ref{ModelKSField}) by construction obey Eqs.~(\ref{KSpotential}) and (\ref{KScurrent}), as well as equations of motion similar to Eqs.~(\ref{DiscerteForceFunUn}) and (\ref{DiscerteMaxwellFunUn}). By combining these equations we see that the solutions to the Kohn-Sham equations generate the solutions to the coupled Eqs.~(\ref{DiscerteForceIntFun}) and (\ref{DiscerteMaxwellIntFun}).

We point out, that in the equation for the photonic mode we do not need any approximate functional. We merely need to solve a classical Maxwell equation. However, in practice it might be useful, especially when calculating non-trivial photonic expectation values, that one solves an actual (uncoupled) photon problem to have a first approximation to the photonic wave function. 


\begin{figure}[t]
  \begin{center}
    \includegraphics[width=0.5\textwidth]{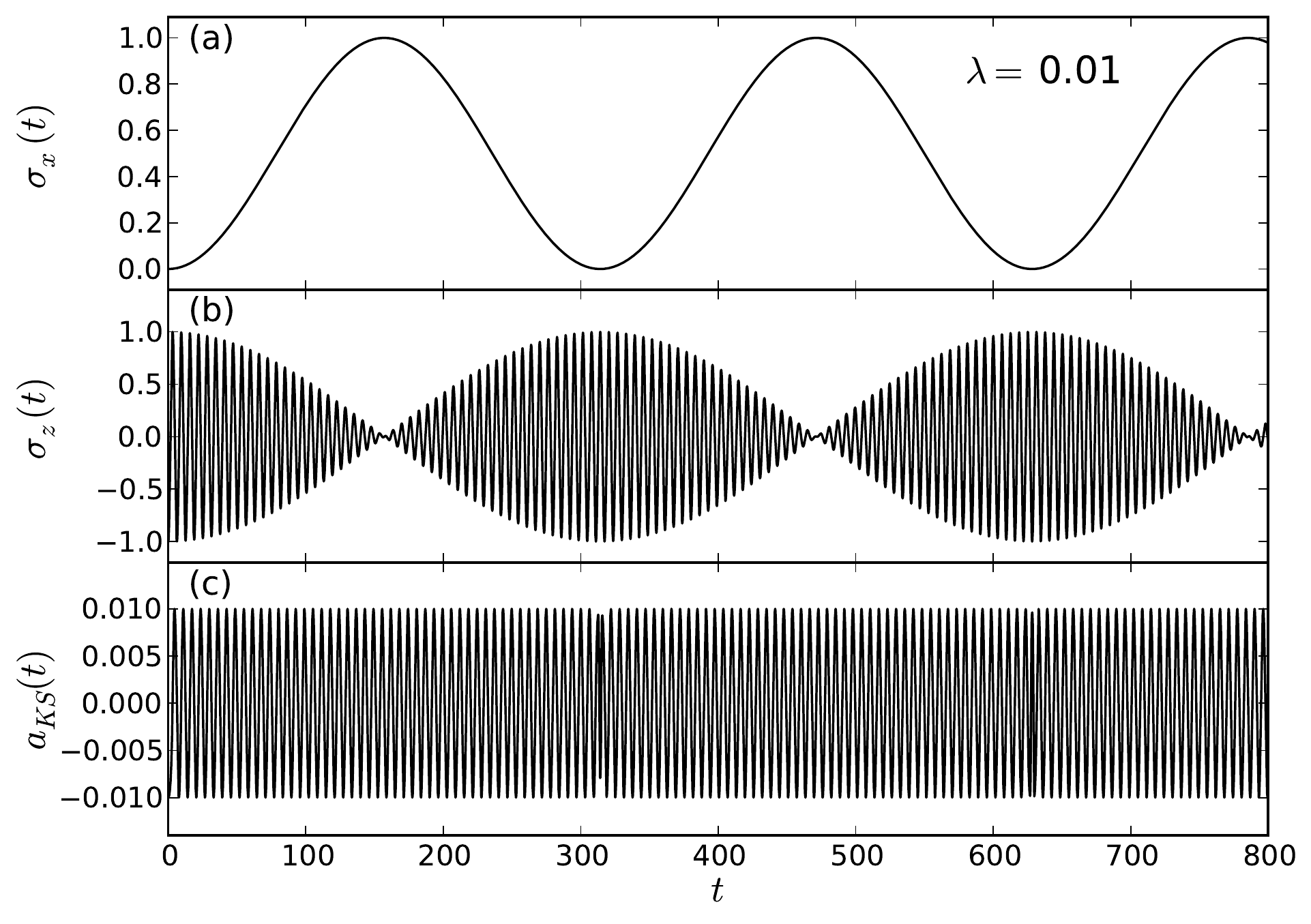}
  \end{center}
\caption{Exact results for the Rabi-Hamiltonian of Eq.~(\ref{rabi-hamiltonian}) in the weak-coupling limit:
(a) Inversion $\sigma_x(t)$, (b) density $\sigma_z(t)$ and (c) exact Kohn-Sham potential $a_{\mathrm{KS}}(t)$ in the case of regular Rabi oscillations.}
  \label{fig:rabi_exact}
\end{figure}

\begin{figure*}[t]
  \begin{center}
    \includegraphics[width=\textwidth]{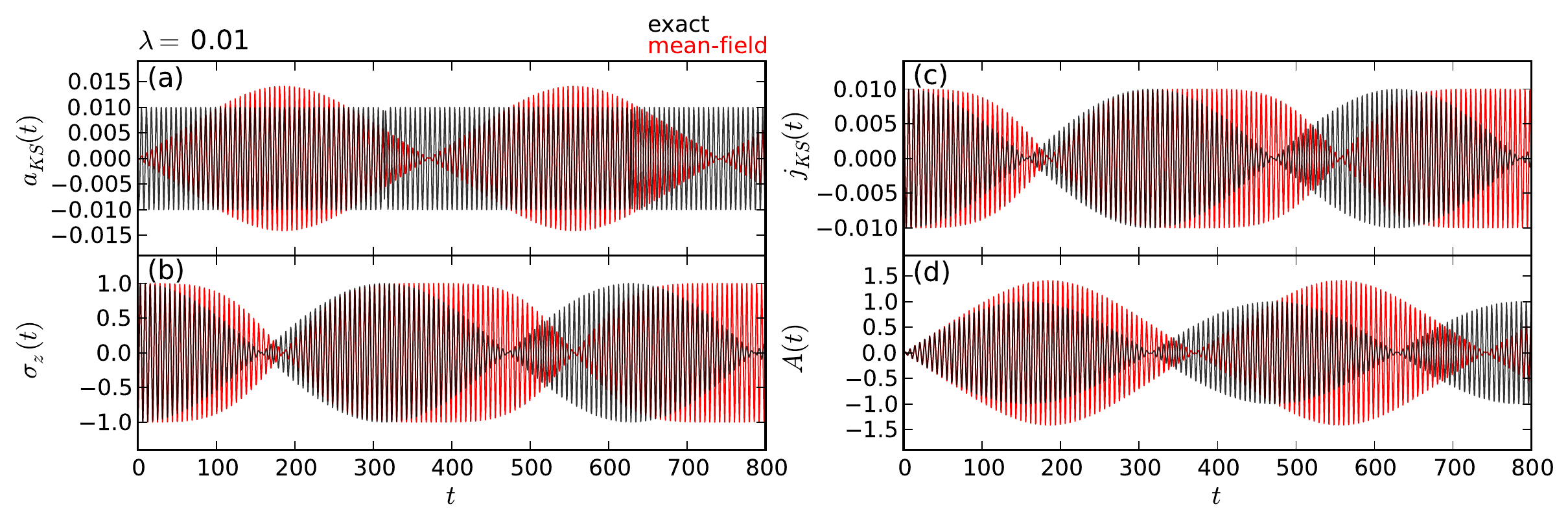}
  \end{center}
\caption{Exact potentials and densities (in black) compared to mean-field potentials and densities (in red) in the case of regular Rabi oscillations in the weak-coupling limit: Left: (a) Kohn-Sham potential $a_{\mathrm{KS}}(t)$ and (b) density $\sigma_z(t)$. Right: (c) Kohn-Sham potential $j_{\mathrm{KS}}(t)$ and (d) density $A(t)$.}
  \label{fig:rabi_weak}
\end{figure*}

\begin{figure*}[t]
  \begin{center}
    \includegraphics[width=\textwidth]{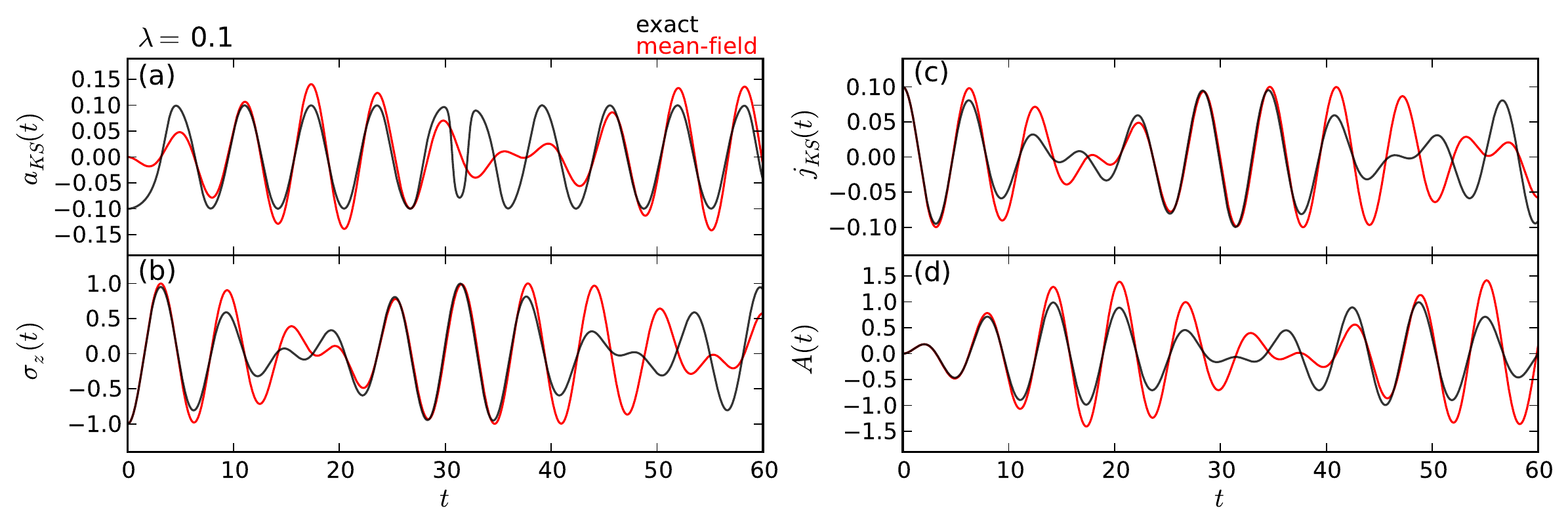}
  \end{center}
\caption{Exact potentials and densities (in black) compared to mean-field potentials and densities (in red) in the case of regular Rabi oscillations in the strong-coupling limit: Left: (a) Kohn-Sham potential $a_{\mathrm{KS}}(t)$ and (b) density $\sigma_z(t)$. Right: (c) Kohn-Sham potential $j_{\mathrm{KS}}(t)$ and (d) density $A(t)$.}
  \label{fig:rabi_strong}
\end{figure*}

\subsection{Numerical example for the model QEDFT}

In this section, we show numerical examples for our model system. We use the density-functional framework introduced in the previous sections
and we explicitly construct the corresponding exact Kohn-Sham potentials.
To illustrate our QEDFT approach, we focus mainly on two different examples: The first example treats a setup in resonance, where regular Rabi oscillations occur. We show results in a weak-coupling limit and in a strong-coupling limit. The second example includes the photon field initially in a coherent state. For this case, we study collapses and revivals of the Rabi oscillations.

The Hamiltonian in Eq.~(\ref{Hubbard}) is directly connected to the famous Jaynes-Cummings-Hamiltonian and the Rabi Hamiltonian \cite{shore1990theory,Shore1993,gerry2005introductory,Braak2011}, which is heavily investigated in quantum optics. It has been studied in the context of Rabi oscillations, field fluctuations, oscillation collapses, revivals, coherences and entanglement (see Ref.\cite{Shore1993} and references therein). 


To directly see the connection between the two-site Hubbard model coupled to one photon mode and the Rabi Hamiltonian, we transform the Hamiltonian in Eq.~(\ref{Hubbard}) by dividing with $I = n \left( \frac{e\omega l}{c} \right) \left(\frac{\hbar c^2}{2 \epsilon_0 L^3 \omega} \right)^\frac{1}{2}$, where $n$ is an arbitary (dimensionless) scaling factor. Thus we make the Hamiltonian and the according Schr\"odinger equation dimensionless. The Hamiltonian of Eq.~(\ref{Hubbard}) can then be rewritten in a similar form as usually found in the literature
\begin{align}
\label{rabi-hamiltonian}
\hat{H}(t) = &-\frac{t_{\text{kin}}}{I} \hat{\sigma}_x + \frac{\hbar\omega}{I} \hat{a}^\dagger\hat{a} - \lambda \left(\hat{a} + \hat{a}^\dagger \right)\hat{\sigma}_z \\
&- j_{\ext}(t)\left(\hat{a} + \hat{a}^\dagger \right) - a_{\ext}(t) \hat{\sigma}_z, \nonumber
\end{align}
where we transformed to the dimensionless external potential $\frac{1}{n} \left(\frac{\hbar c^2}{2\epsilon_0 L^3\omega}\right)^{-\frac{1}{2}}{a}_{\ext} \rightarrow a_{\ext}$ and the dimensionless external current $\frac{1}{n} \left(\frac{1}{e\omega l} \right) j_{\ext} \rightarrow j_{\ext}$. Further, we also transform to a dimensionless time variable $\frac{I}{\hbar} t \rightarrow t$. To actually perform numerical calculations, we have to choose values for the free parameters. Here, we choose typically used values from the literature: $t_{\text{kin}}/I=0.5$, $\hbar\omega/I=1$, $\lambda=(0.01,0.1)$ and external fields which are set to zero $j_{\ext}(t) = a_{\ext}(t) =0$. This set of parameters allow for a resonance situation, with no detuning between the transition energy of the atomic levels and the frequency of the field mode.

As discussed above the basic variables (densities) are the current operator $\hat{J}$ and the operator for the field potential $\hat{A}$. In this two-level example $\hat{J}$ reduces to $\hat{\sigma}_z$ and $\hat{A}$ reduces to $\left(\hat{a} + \hat{a}^\dagger\right)$.

If the rotating-wave approximation is applied to the Rabi Hamiltonian in Eq.~(\ref{rabi-hamiltonian}), one recovers the Jaynes-Cummings Hamiltonian. This Hamiltonian is then analytically solvable. The rotating-wave approximation is only valid in the weak-coupling limit ($\lambda \approx 0.01$). In the strong-coupling limit ($\lambda \geq 0.1$), however, the rotating-wave approximation breaks down. Only recently, analytic results without the rotating-wave approxmation have been published \cite{Braak2011}. Here we emphasize that the QEDFT approach presented in this paper is exact and does not rely on the rotating-wave approximation and hence also allows to treat strong-coupling situations.

In our first example we choose as initial state for both, the coupled many-body system and the uncoupled Kohn-Sham problem
\begin{align*}
 \ket{\Psi_0} = \ket{\Phi_0} = \ket{1} \otimes \ket{0},
\end{align*}
meaning the electron initially populates site one and the field is in the vacuum state. Therefore, no photon is present in the field initially. In Fig.~\ref{fig:rabi_exact}, we show the inversion $\sigma_x(t)$, the density $\sigma_z (t)$ and the corresponding exact Kohn-Sham potential $a_{\mathrm{KS}}(t)$ for the weak-coupling case. The atomic inversion $\sigma_x(t)$ shows regular Rabi oscillations. Rabi oscillations are also visible in ${\sigma}_z(t)$, where we observe the typical neck-like features \cite{Fuks2013} at $t \approx 150$ and at later points in time.

To determine the exact Kohn-Sham potential for this case, we follow a fixed-point construction similar to \cite{nielsen-2013}. As input for the fixed-point construction, we use the exact many-body densities. In addition, we also compare to an analytic formula for the Kohn-Sham potential for a one-electron two-site Hubbard model given in \cite{li-2008,farzanehpour-2012}. This expression gives an explicit formula for the dependence of the Kohn-Sham potential on the density. Such an explicit formula is only known in a few cases, while the fixed-point construction is generally valid. However, both methods yield in the present case the same results. A detailed discussion of the fixed-point construction for multicomponent systems of electrons and photons will be presented in a forthcoming work \cite{flick-2014}.

We emphasize that a propagation of the uncoupled Kohn-Sham system with the exact Kohn-Sham potential $a_{\mathrm{KS}}(t)$ obtained in Fig.~\ref{fig:rabi_exact} reproduces by construction the exact many-body density ($\sigma_z (t)$ in the present case). However, as illustrated in Sec. IIC, if a Kohn-Sham propagation is used, the numerical expenses can be drastically reduced, since the Kohn-Sham construction effectively decouples the quantum system.

In practical calculations the exact Kohn-Sham potentials are normally not available and one has to rely on approximations. In the present case, the simplest approximation for $v_{\mathrm{KS}}[\Psi_0,\Phi_0,J,A,a_{\ext}]$ is straightforward if we assume $n[\Phi_0,J,A] \approx n[\Psi_0,J,A]$ and $\langle \hat{n} \A \rangle \approx \langle \hat{n}\rangle\langle \A \rangle = n A$. Then, from Eq.~(\ref{KSpotential}) we find the mean-field approximation to the Kohn-Sham potential
\begin{align}
a_{\mathrm{MF}}([A,a_{\ext}];t) = \lambda A(t) + a_{\ext}(t).
\label{Eq:MeanFieldPotential}
\end{align}
The mean-field approximation is actually identical to the Maxwell-Sch\"odinger approach, i.e. we treat the electromagnetic field as being essentially classical. Further, for $\lambda \rightarrow 0$ and for $\lambda \rightarrow \infty$ the mean-field approximation becomes asymptotically exact. In Fig. \ref{fig:rabi_weak} and Fig.~\ref{fig:rabi_strong}, we compare exact densities and exact Kohn-Sham potentials to densities and potentials, which were obtained by a self-consistend mean-field propagation. Already in the weak-coupling limit, Fig. \ref{fig:rabi_weak}, quite sizable differences between exact results and mean-field results become visible: Already at $t=0$ the exact Kohn-Sham potential deviates from the mean-field potential. In the case of the densities, this leads to a frequency shift, where the mean-field density oscillates slower than the exact density. In the strong-coupling limit shown in Fig.~\ref{fig:rabi_strong}, effects beyond the rotating-wave approximation are visible. In the exact 
Kohn-Sham potential, we see a non-regular feature at $t=30$, which is also not coverd by the mean-field approximation. However, the mean-field approximation already covers at least some dynamical features of the propagation.

\begin{figure}[t]
  \begin{center}
    \includegraphics[width=0.5\textwidth]{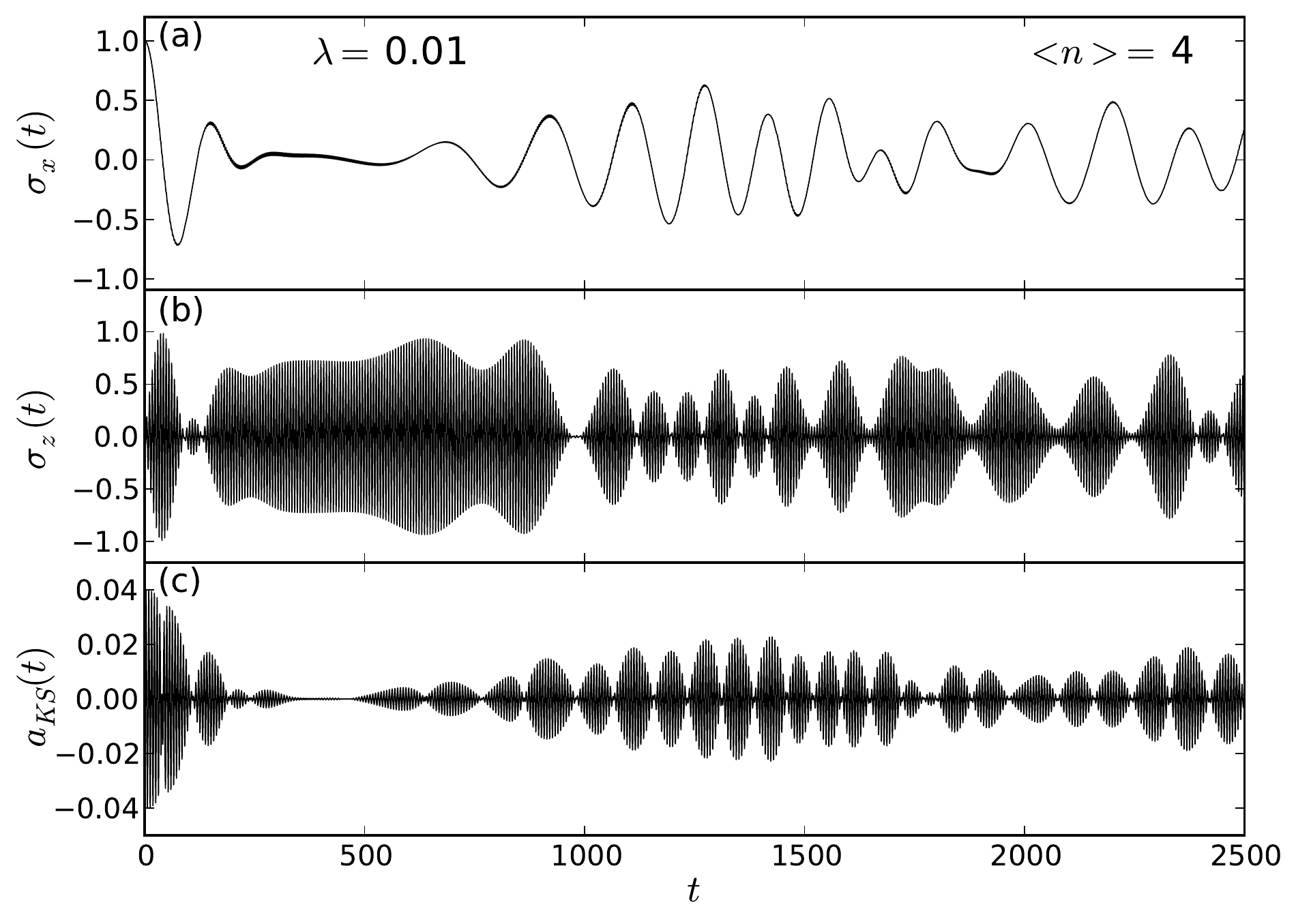}
  \end{center}
\caption{Exact results for the Rabi-Hamiltonian in the weak-coupling limit:
(a) Inversion $\sigma_x(t)$, (b) density $\sigma_z(t)$ and (c) exact Kohn-Sham potential $a_{\mathrm{KS}}(t)$ in the case of coherent states (in spirit of panel 3 in Fig. 4 in Ref. \cite{Shore1993})}
  \label{fig:cs_exact}
\end{figure}

\begin{figure*}[t]
  \begin{center}
    \includegraphics[width=\textwidth]{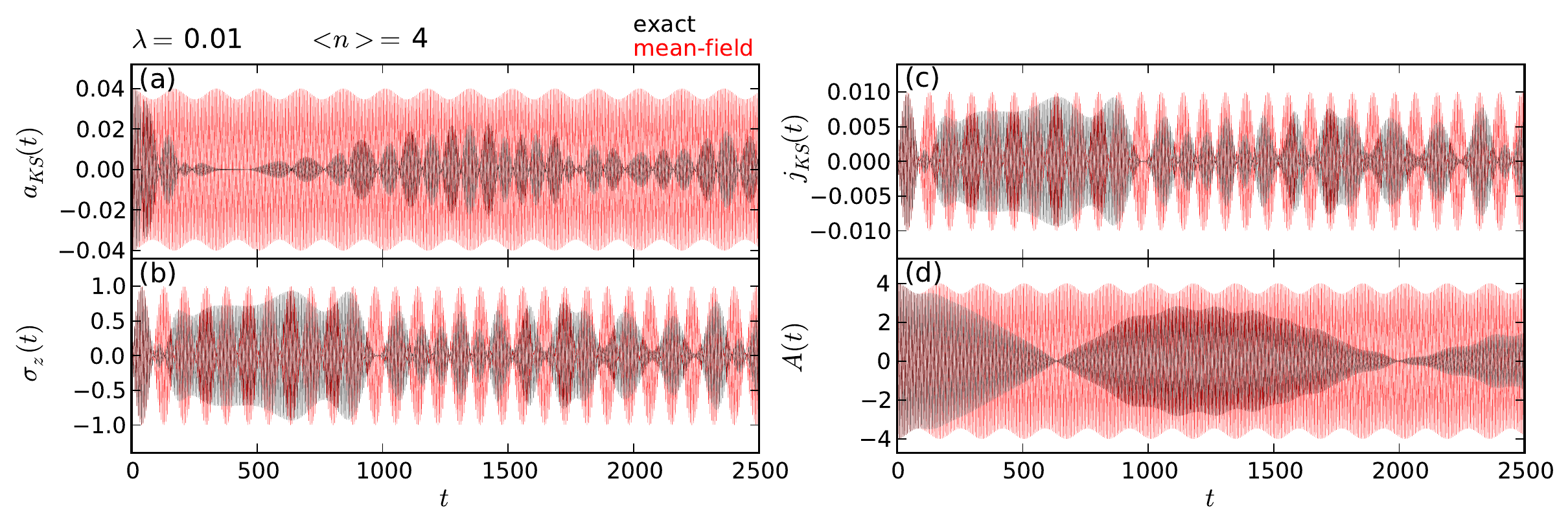}
  \end{center}
\caption{Exact densities and potentials (in black) compared to mean-field densities and potentials (in red) in the case of regular Rabi oscillations in the case of coherent states: Left: (a) Kohn-Sham potential $a_{\mathrm{KS}}(t)$ and (b) density $\sigma_z(t)$. Right: (c) Kohn-Sham potential $j_{\mathrm{KS}}(t)$ and (d) density $A(t)$.}
  \label{fig:cs_weak}
\end{figure*}

For the second example in this section, we start with the field initially in a coherent state. For a single field mode, coherent states~\cite{Glauber1963,Glauber1963a} can be written as follows:
\begin{align*}
\ket{a} = \sum_{n=0}^\infty f_n(\alpha) \ket{n}, \; \; \; \text{with} \; \; \; f_n(\alpha) = \frac{\alpha^n}{\sqrt{n!}}\exp\left(-\frac{1}{2} {|\alpha|}^2\right)
\end{align*}
In this example, we use as initial state for the many-body propagation and the Kohn-Sham propagation
\begin{align*}
 \ket{\Psi_0} = \ket{\Phi_0} = \ket{g} \otimes \ket{\alpha}.
\end{align*}
Here, the atomic state $\ket{g}$ is the ground state of the electronic Hamiltonian ($\ket{g} = \frac{1}{\sqrt{2}}\left(\ket{1} + \ket{2} \right)$. For the field state we choose $|\alpha|^2 = \langle\hat{a}^\dagger\hat{a}\rangle=4$. This example is in the spirit of the calculation in panel 3 in Ref.~\cite{Shore1993}. Hence, as shown in Fig.~\ref{fig:cs_exact}, we obtain a similar time-evolution of the inversion $\sigma_x(t)$ as in Ref.~\cite{Shore1993}. We see the Cummings collapse of Rabi oscillations at $t=250$ followed by a quiescence up to $t=500$ occuring. After $t=500$, we see a revival of the Rabi oscillations. We also observe, as shown in \cite{Narozhny1981}, that the atomic dipole operator (here the density $\sigma_z(t)$) continues to change during the interval of quiescence after the inversion collapse. As before, we show in the lowest panel the corresponding exact Kohn-Sham potential obtained via fixed-point iterations.

In Fig.~\ref{fig:cs_weak}, we show a comparison of the exact Kohn-Sham potentials and densities to the mean-field propagation. Here, we see that the mean-field approximation performs rather poorly. For this case the simple ansatz in Eq.~(\ref{Eq:MeanFieldPotential}) is not sufficient and more sophisticated approximations to the exact Kohn-Sham potential are necessary to reach a better agreement \cite{maitra-2001,maitra-2002b}.

In summary, we have shown in this section the exact Kohn-Sham potentials which reproduce the dynamics of the exact many-body densities. In particular the coherent state example shows that there is a clear need for better approximations to the exact Kohn-Sham potential \cite{tokatly-2013} that go beyond the mean-field level and that include correlation contributions. One possibility along these lines is provided by an approach based on the optimized effective potential (OEP) method \cite{kuemmelRMP, ullrich-TDDFT, marques-TDDFT}. We have already implemented such an OEP approach for the present model system and the corresponding results improve quite considerably over the mean-field approximation. The details of this general OEP approach to QEDFT are beyond the scope of the present paper and will be presented in a separate publication \cite{pellegrini-2014}.


\section{Relativistic QEDFT}
\label{sec:RelQEDFT}

After having presented the basic concepts of a QEDFT reformulation of a coupled matter-photon problem in a model system, we apply the very same ideas to the full theory of QED. While no new density-functional-type ideas have to be introduced, the intricacies of QED make the actual details more involved. A first subtlety is the gauge freedom of the photon field. In this work we choose Coulomb gauge to fix the superfluous degrees of freedom. This gauge has two distinct advantages over the other gauges: it reduces the independent components of the photon field to the two transversal (physical) polarizations, and it singles out the classical Coulomb interaction between the charged particles. Since we want to connect QEDFT to derived theories like cavity QED, where usually Coulomb-gauged photons are employed, and condensed-matter theory, where Coulomb interactions play a dominant role, the Coulomb gauge is for the present purpose the natural gauge to work in. However, we emphasize that also other gauges can be 
used as well \cite{rajagopal-1994, ruggenthaler-2011, engel-DFT}.

We first present the standard approach to identify possible conjugate variables and introduce the basic equations of motions. While in the usual non-relativistic setting this route works just fine, in the fully relativistic situation the internal structure of the "Dirac particles", i.e. the electronic and positronic degrees of freedom, give rise to certain subtleties when performing a density-functionalization. Therefore, instead of using the current, we employ the polarization as a basic fundamental variable in relativistic QEDFT.


\subsection{Equations of Quantum Electrodynamics}

\label{subsec:EOMQED}

In the following we define the basic quantities of QED in Coulomb gauge and derive the equations of motion for the fundamental (functional) variables of the theory. We employ SI units throughout, since in the next section we perform the non-relativistic limit which is most easily done if we keep the physical constants explicit. A detailed discussion of quantizing QED in Coulomb gauge is given in appendix \ref{app:Coulomb}.

The full QED Hamiltonian in Coulomb gauge (indicating explicit time-dependence of the Hamiltonians by $t$) is given by
\begin{align}
\label{QED}
 \hat{H}(t) = \hat{H}_{\mathrm{M}} + \hat{H}_{\mathrm{E}} +  \hat{H}_{\mathrm{C}}(t) + \hat{H}_{\mathrm{ext}}(t) + \hat{H}_{\mathrm{int}}.
\end{align}
Here
\begin{align}
\label{DiracHam}
 \hat{H}_{\mathrm{M}} = \int \diff^{3} r : \dspin(\vec{r}) \left( -\imagi \hbar c  \; \vec{\gamma}\cdot \vec{\nabla} + m c^2 \right) \spin(\vec{r}):
\end{align}
is the normal ordered (::) free Dirac Hamiltonian in the Schr\"odinger picture, where $\spin$ and $\dspin$ denote the Dirac-field operators and $\gamma^{k}$ the Dirac matrices (see appendix~\ref{app:Coulomb} for definitions). The energy of the free photon field is given by 
\begin{align}
\label{MaxwellHam}
\hat{H}_{\mathrm{E}}= \frac{\epsilon_0}{2}  \int  \diff^{3} r \, :  \left( \hat{\vec{E}}^{2}(\vec{r}) + c^2 \hat{\vec{B}}^{2}(\vec{r})\right):,
\end{align}
where $\hat{\vec{E}}$ and $\hat{\vec{B}}$ are the (vector-valued) electric and magnetic field operators defined as in appendix \ref{app:Coulomb} in terms of the Maxwell-field operators $\A^{k}$. We note that due to the Coulomb-gauge condition $\vec{\nabla}\cdot \vec{A} = 0$ only the spatial components of the Maxwell field are quantized. The time component $A^0$ is given by the classical Coulomb field of the total charge density, which is the sum of the charge density of the Dirac field and the classical external current, and gives rise to the Coulomb term 
\begin{align}
\label{CoulombHam}
 \hat{H}_{\mathrm{C}}(t)\!  = \! \frac{1}{2 c^2} \!\! \int \!\! \frac{\diff^{3}r \, \diff^{3} r'}{4 \pi \epsilon_0 |\vec{r} -\vec{r}'|} \left( 2 j^0_{\mathrm{ext}}(x') \J_0(\vec{r})   + :\J^0(\vec{r}) \J_0(\vec{r}'): \right).
\end{align}
Here $\J^0$ is the zero component of the Dirac current
\begin{align}
\label{4current}
  \J^{\mu}(\vec{r}) = e c : \dspin(\vec{r}) \gamma^{\mu} \spin(\vec{r}):, 
\end{align}
and $j^0_{\mathrm{ext}}$ is the zero component of a given external current $j^\mu_{\mathrm{ext}}$. In the Coulomb term the energy due to the Coulomb interaction of the external current with itself is elided. Since this term is purely multiplicative, i.e. it is equivalent to the identity operator times some real number, it does not influence the dynamics of the system and can be discarded. The rest of the coupling to the external fields is given by
\begin{align}
\label{ExternalHam}
 \hat{H}_{\mathrm{ext}}(t) = \frac{1}{c} \int \diff^{3}r \; \left( \J_{\mu}(\vec{r}) a^{\mu}_{\mathrm{ext}}(x) - \vec{j}_{\mathrm{ext}}(x) \cdot \hat{\vec{A}}(\vec{r}) \right).
\end{align}
Finally, the coupling between the quantized fields in Coulomb gauge reads as
\begin{align}
\label{InteractionHam}
 \hat{H}_{\mathrm{int}} = - \frac{1}{c} \int \diff^{3} r \; \hat{\vec{J}}(\vec{r}) \cdot \hat{\vec{A}}(\vec{r}). 
\end{align}
Comparing to the Lorentz-gauge QED Hamiltonian \cite{ruggenthaler-2011b} the main difference lies in the Coulomb term, that treats the zero component of the photon field explicitly.

Without further refinements the above QED Hamiltonain is not well-defined, since it gives rise to infinities \cite{greiner-FQ, ryder-QFT, greiner-QED}. These infinities can be attributed, with the help of perturbation theory, to three divergent types of Feynman diagrams: the self-energy of the Fermions, the self-energy of the photons (also called vacuum polarization) and the vertex corrections. These divergences vanish if we regularize the theory, e.g., by introducing frequency cut-offs in the plane-wave expansions of the fermionic as well as the bosonic field operators or by dimensional regularization \cite{ryder-QFT}. Such procedures make the above Hamiltonian self-adjoint \cite{takaesu-2009}, but we have introduced a dependence on parameters that change the theory at smallest and largest length scales. Perturbatively one can remove these dependencies by renormalizing the theory, i.e. we first identify and then subtract the part of each of these three terms that 
diverges due to these parameters. The resulting 
three divergent counter-terms \footnote{Note that these counter-terms are defined by the vacuum expectation value \cite{engel-DFT, ryder-QFT}. This allows to compare Hamiltonians with different external potentials and currents.} can be recast as a renormalization of the mass and the field-operators of the Fermions (due to the self-energy), as a renormalization of the photonic field-operators (due to vacuum polarization) and a renormalization of the charge (due to the vertex corrections). We can do this to any order in perturbation theory due to the Ward-Takahashi identities \cite{ryder-QFT}. Thus, the above QED Hamiltonian is at least perturbatively renormalizable. For simplicity, we assume in the following that one can fully renormalize the QED Hamiltonian (as has been shown for certain limits \cite{hainzl-2003}) and interpret it as a bare Hamiltonian, i.e. we use the renormalized quantities \footnote{Note that an exhaustive discussion of renormalization is beyond the scope of the present work. Nevertheless,
 to comprehensibly connect the different formulations of matter-photon systems, a general field-theoretical approach is advantageous. If we want to avoid the difficulties due to renormalization, we have to keep the cut-offs. Since we are interested exclusively in condensed-matter systems, a physical (highest) cut-off would be at energies that allow for pair-creation.}. That a full renormalization is possible has been shown, e.g., for the Nelson model of QED \cite{nelson-1964, bachmann-2012}, where the divergent self-energy term shifts the spectrum of the Hamiltonian to infinity. Thus, subtracting this infinite shift, i.e. introducing a counter-term, makes the Hamiltonian well-defined (when removing the cut-offs), provided the energy of the system is below the pair-creation limit. The same condition, i.e. a stable vacuum, we need to impose also on our QED considerations as discussed in \cite{engel-DFT, rajagopal-1994, ruggenthaler-2011b}.   

In a next step we identify the \textit{possible} conjugate (functional) variables of the above QED Hamiltonian. Here the physical, time-dependent wave function $\ket{\Psi(t)}$ depends on the initial state and the external pair $(a_{\mu}^{\mathrm{ext}},j_{\mu}^{\mathrm{ext}})$, which is indicated by
\begin{align*}
 \ket{\Psi([\Psi_0,a_{\mu}^{\mathrm{ext}},j_{\mu}^{\mathrm{ext}}];t)}.
\end{align*}

Thus, with $\int \!\! \equiv\!\! \int_{0}^{T} \!\! \diff t \! \int  \!\! \diff ^3 r $, the (negative) QED action \cite{rajagopal-1994, ruggenthaler-2011b} 
\begin{align}
\label{QEDaction}
\tilde{\mathcal{A}}[\Psi_0,a_{\mu}^{\mathrm{ext}},j_{\mu}^{\mathrm{ext}}] \!=\! -\!\! \int\!\! \mathcal{L}_{\mathrm{QED}}  = \!- \mathcal{B} 
\!+\! \frac{1}{c} \! \int \! \left( j^{\mu}_{\mathrm{ext}} A_{\mu} + J_{\mu} a^{\mu}_{\mathrm{ext}} \right) 
\end{align}
becomes a functional of these variables ($T$ corresponds to an arbitrary time). Here we employed the definition of the QED Lagrangian of Eq.~(\ref{QEDLagrangian}) and defined the internal QED action with help of Eq.~(\ref{QEDHam'}) by
\begin{align*}
\mathcal{B} \!= \! \int_{0}^{T} &\!\!\! \diff t \braket{\Psi(t)| \imagi \hbar c \partial_0 \! - \! \hat{H}_{\mathrm{M}} \! - \! \hat{H}_{\mathrm{E}} \! - \! \hat{H'}_{\mathrm{int}}(t) }{\Psi(t)}.
\end{align*}
Eq.~(\ref{QEDaction}) looks like a Legendre transformation between $J_{\mu} \leftrightarrow a_{\ext}^{\mu}$ and $A_{\mu} \leftrightarrow j_{\ext}^{\mu}$. Since a Legendre transformation amounts to a change of variables, this indicates (for a fixed initial state) the possibility of transforming from $(a_{\ext}^{\mu},j_{\ext}^{\mu})$ to the conjugate variables $(J_{\mu}, A_{\mu})$ \footnote{One should not confuse these conjugate variables with the conjugate momenta that are used in field theory to quantize the system. In the case of Coulomb-gauge QED the pair of conjugate momenta are $(\vec{A},\psi)$ and $(\epsilon_0 \vec{E}^{\perp}, \imagi \hbar c \psi^{\dagger})$ \cite{greiner-FQ}}. If these variables would indeed be connected via a standard Legendre transformation the functional derivative with respect to $a^{\mu}_{\ext}$ and $j^{\mu}_{\ext}$ should give the respective conjugate variables. However, following derivations similar to \cite{vignale-2008} we find the appearence of extra terms, i.e.
\begin{align}
\label{Conjugatea}
 &\frac{\delta \tilde{\mathcal{A}}}{\delta a^{\mu}_{\mathrm{ext}}(x)} + \imagi \hbar c \braket{\Psi(T)}{\frac{\delta \Psi(T) }{\delta a^{\mu}_{\mathrm{ext}}(x)}} =  \frac{1}{c}J_{\mu}(x),
\\
\label{Conjugatej}
 &\frac{\delta \tilde{\mathcal{A}}}{\delta j^{\mu}_{\mathrm{ext}}(x)} +  \imagi \hbar c \braket{\Psi(T)}{\frac{\delta \Psi(T) }{\delta j^{\mu}_{\mathrm{ext}}(x)}} = \frac{1}{c} A_{\mu}(x).
\end{align}
These non-trivial boundary terms are due to the fact, that variations of the external fields give rise to non-zero variations of the wave function at the (arbitrary) upper boundary $T$ (in contrast to direct variations of the wave function that are supposed to obey $\ket{\delta \Psi(T)} = 0$) \cite{vanleeuwen-2001}. These boundary terms are necessary to guarantee the causality of $J_{\mu}$ and $A_{\mu}$ \cite{vignale-2008}. Thus, Eqs.~(\ref{Conjugatea}) and (\ref{Conjugatej}) show that a straightforward approach to demonstrate a one-to-one correspondence between $(a_{\ext}^{\mu},j_{\ext}^{\mu})$ and $(J_{\mu}, A_{\mu})$ based on a Legendre transformation becomes difficult \cite{ruggenthaler-2011b}. Nevertheless, usually this Legendre-transformation arguments work well to identify the possible conjugate variables. 

However, in the relativistic situation a further problem arises: the current has an internal structure due to the electronic and positronic degrees of freedom. The current $J_{\mu}$ describes the net-charge flow of the negatively charged electrons and the positively charged positrons \cite{greiner-FQ}. Therefore, the current expectation value can not differ between the situation of, e.g., the movement of two electrons and one positron or three electrons and two positrons. This fact, which is absent in the non-relativistic situation, will lead to problems when employing the ideas developed in subsection \ref{subsec:FoundModel}.

For the moment, however, we follow the above identification scheme and  derive the basic equations of motion for $\J_{\mu}$ and $\A_{\mu}$. Since $\int \diff ^3 r' \; [\J_{\mu}(\vec{r}),\J_{0}(\vec{r}')] f(\vec{r}') = 0$, where $f(\vec{r}')$ is any testfunction,  the term $\hat{H}_{\mathrm{C}}$ commutes with $\J_{\mu}$ and the equation of motion for the four current is the same as in Lorentz gauge \cite{ruggenthaler-2011b}
\begin{align}
\label{HeisenbergCurrent}
&\imagi  \partial_0 \J^{k}(\vec{r}) = \frac{e}{\hbar} m c^2 \dspin(\vec{r})\left[\gamma^{k} \gamma^{0} - \gamma^{0} \gamma^{k}  \right]\spin(\vec{r})
\\
& + e c \dspin(\vec{r})\left[ \gamma^{k}\gamma^{0}\left(- \imagi  \vec{\gamma} \cdot \vec{\nabla}\right) + \left(- \imagi  \vec{\gamma} \cdot \overset{\smash{\raisebox{-1.5pt}{\tiny$\leftarrow$}}}{\nabla} \right) \gamma^0 \gamma^{k}\right] \spin(\vec{r}) \nonumber
\\
& + \frac{e^2}{\hbar}\dspin(\vec{r})\left[\gamma^{k} \gamma^{0} \gamma^{l}- \gamma^{l} \gamma^{0} \gamma^{k}  \right]\spin(\vec{r})\! \left( \A_{l}(\vec{r}) + a_{l}^{\mathrm{ext}}(x)  \right) \nonumber,
\end{align}
where the zero component is given by $\imagi \partial_0 \J^{0} = - \imagi \; \vec{\nabla} \cdot \hat{\vec{J}}$, i.e. the current obeys the conservation of charge. A different equation that determines the charge current $J_{\mu}$ is found by the Gordon-decomposition \cite{engel-DFT}, which is the evolution equation of the polarization 
\begin{align*}
 \hat{P}^{\mu}(\vec{r}) = e c :\spin^{\dagger}(\vec{r}) \gamma^{\mu} \spin(\vec{r}):,
\end{align*}
\begin{align}
\label{GordonDecomposition}
\imagi \partial_0 \hat{P}^{k}(\vec{r}) =&  \frac{2emc}{\hbar} \J^{k}(\vec{r}) + \imagi e c \dspin(\vec{r})\! \left(\partial^{k} -\overset{\smash{\raisebox{-1.5pt}{\tiny$\leftarrow$}}}{\partial}^{k}  \right) \spin(\vec{r}) \nonumber
\\
&- ec \epsilon^{klj} \partial_l \left(\dspin(\vec{r})\Sigma_{j}\spin(\vec{r}) \right) \nonumber
\\
&+ \frac{2e}{\hbar c} \hat{P}_{0}(\vec{r})\left(\hat{A}^{k}(\vec{r}) + a^{k}_{\mathrm{ext}}(x)  \right),
\end{align}
where $\epsilon^{klj}$ is the Levi-Cevita symbol and
\begin{align*}
\Sigma^{k} =
 \begin{pmatrix}
  \sigma^{k} & 0 \\
  0 & \sigma^{k}
 \end{pmatrix}.
\end{align*}
With the definition of bigger and smaller components of the Dirac-field operators $\spin^{\dagger}(\vec{r}) = \left(\hat{\phi}^{\dagger}(\vec{r}), \hat{\chi}^{\dagger}(\vec{r})  \right)$ we find that the current and the polarization are the real and imaginary part of the same operator
\begin{align*}
 \J^{k}(\vec{r})= 2 \Re\left\{ ec :\hat{\phi}^{\dagger}(\vec{r})\sigma^k \hat{\chi}(\vec{r}): \right\},
\\
 \hat{P}^{k}(\vec{r})= 2 \Im \left\{ ec :\hat{\phi}^{\dagger}(\vec{r})\sigma^k \hat{\chi}(\vec{r}): \right\}.
\end{align*}
The change of gauge only affects the equation for the photon-field operator which becomes 
\begin{align}
\label{FirstDerivPot}
 \partial_0 \hat{\vec{A}}(x) = -\hat{\vec{E}}(x),
\end{align}
and accordingly
\begin{align}
\label{SecondDerivPot}
 \left(\partial_0^2 + \partial_l \partial^l  \right) \A^{k}(\vec{r}) &- \partial^{k} \partial_0\left(\frac{1}{c} \int \diff^3 r' \frac{j^{0}_{\mathrm{ext}}(x') + \J^{0}(\vec{r}')}{4 \pi \epsilon_0 |\vec{r} - \vec{r}'|}  \right) \nonumber
\\
&=  \mu_0 c \left(j^{k}_{\ext}(x) + \J^{k}(\vec{r}) \right).
\end{align}
This is indeed the quantized Maxwell equation in Coulomb gauge. 


\subsection{Foundations of relativistic QEDFT}

\label{subsec:FoundRelQEDFT}

In this subsection we first reexamine the previous approach to relativistic QEDFT \cite{rajagopal-1994,ruggenthaler-2011b} and identify its shortcomings. We then show why physically the polarization is better suited as fundamental variable of the matter part and reformulate QED in terms of $(P_{\mu},A_{\mu})$. Already here we point out that both, a relativistic QEDFT based on the current or on the polarization, lead to the same density-functional-type theory in the non-relativistic limit.

A first restriction we impose is to fix a specific gauge for the external fields $a^{\mu}_{\ext}$. Since by construction external fields that only differ by a gauge transformation, i.e. $\tilde{a}_{\ext}^{\mu} = a_{\ext}^{\mu} + \partial^{\mu} \Lambda $, lead to the same current density (and polarization) \footnote{This is most easily seen by considering the commutator $[\J^{\mu}; \int \J_{\nu} \partial^{\nu} \Lambda]$ which determines the effect of a gauge on the equation of $\J_{\mu}$, i.e. Eq.~(\ref{HeisenbergCurrent}). By partial integration, application of the continuity equation and the fact that $[\J^{\mu}; \J^{0}] \equiv 0$ this term becomes zero and therefore has no effect on the current. The same reasoning shows that also $\hat{P}_{\mu}$ is gauge independent.}, the desired one-to-one correspondence can only hold modulo these transformations. Thus in principle we consider a bijective mapping between equivalence classes, and by fixing a gauge we take a unique respresentative of each class. For 
simplicity we impose a gauge condition similar to \cite{vignale-2004}
\begin{align}
\label{GaugeCondition}
 a^{0}_{\mathrm{ext}}(x) = 0.
\end{align}
In the following, any other gauge that keeps the initial state unchanged, i.e. the gauge function has to obey $\Lambda(0,\vec{r}) = 0$, is also allowed \cite{vignale-2004}. This condition is necessary for  our further investigations, since we will employ that the initial state is fixed (and thus the expectation values at $t=0$), in accordance to the derivations of subsection \ref{subsec:FoundModel}. 

Further, we assume that the external current obeys the continuity equation $\partial_0 j^{0}_{\ext} = - \vec{\nabla}\cdot \vec{j}_{\ext}$. This leaves the choice of the initial charge configuration $j_{\ext}^{0}(0,\vec{r})$. Since the photons only couple to moving charges this choice will not influence the dynamics of the system. Therefore, also in the case of the external currents we have an equivalence class (of possible zero components), and will therefore restrict to only prescribing $\vec{j}_{\ext}$.

In \cite{rajagopal-1994, ruggenthaler-2011b} the one-to-one correspondence was based on 
the corresponding Ehrenfest equations
\begin{align}
\label{IntFourCurrent}
\partial_0 J^{k}(x) &=q_{\mathrm{kin}}^{k}(x) + q_{\mathrm{int}}^{k}(x) + n^{k l}(x) a_{l}^{\mathrm{ext}}(x),
\\
\label{IntPotential}
\Box A^{k}(x) &- \partial^{k} \partial_0\left(\frac{1}{c} \int \diff^3 r' \frac{j^{0}_{\mathrm{ext}}(x') + J^{0}(x')}{4 \pi \epsilon_0 |\vec{r} - \vec{r}'|}  \right) \nonumber
\\
&=  \mu_0 c \left(j^{k}_{\ext}(x) + J^{k}(x) \right),
\end{align}
where
\begin{align*}
\hat{q}_{\mathrm{kin}}^{k}(\vec{r}) = & - ec \dspin(\vec{r})\!\!\left[ \gamma^{k}\gamma^{0}\!\!\left(\vec{\gamma} \cdot \vec{\nabla}\right) \!\!+\!\! \left( \vec{\gamma} \cdot \overset{\smash{\raisebox{-1.5pt}{\tiny$\leftarrow$}}}{\nabla} \right)\!\! \gamma^0 \gamma^{k}\right] \!\!\spin(\vec{r}) \nonumber
\\
& + \imagi \; \frac{e}{\hbar} m c^2 \dspin(\vec{r})\left[ \gamma^{0} \gamma^{k} - \gamma^{k} \gamma^{0}   \right]\spin(\vec{r}),
\end{align*}
\begin{align*}
\hat{n}^{k l}(\vec{r}) =  \frac{ \imagi e^2}{\hbar} \dspin(\vec{r})\left[ \gamma^{l} \gamma^{0} \gamma^{k} - \gamma^{k} \gamma^{0} \gamma^{l} \right]\spin(\vec{r}),
\end{align*}
\begin{align*}
\hat{q}_{\mathrm{int}}^{k}(\vec{r}) =  \hat{n}^{k l}(\vec{r}) \A_{l}(\vec{r}),
\end{align*} 
and the D'Alembert operator reads as $\Box = \partial_0^2 + \partial_k \partial^k$. We can then reexpress
\begin{align*}
 \hat{n}^{kl}(\vec{r}) = - \frac{2 e^2}{\hbar} \epsilon^{klj} \spin^{\dagger}(\vec{r}) \Sigma_{j} \spin(\vec{r}),
\end{align*}
and therefore
\begin{align*}
 \hat{n}^{k l}(\vec{r}) a_{l}^{\mathrm{ext}}(x) \rightarrow \frac{2 e^2}{\hbar} \left( \spin^{\dagger}(\vec{r}) \vec{\Sigma} \spin(\vec{r}) \right) \times \vec{a}_{\mathrm{ext}}(x).
\end{align*}
If we then want to show a possible one-to-one correspondence we can follow the reasoning of Sec.~\ref{subsec:FoundModel} and consider the uniqueness of solutions of the functional equations   
\begin{align}
\label{IntFourCurrentFun}
\partial_0 J^{k}(x) &=q_{\mathrm{kin}}^{k}([a_{\ext}^{m},j_{\ext}^{m}];x) + q_{\mathrm{int}}^{k}([a_{\ext}^{m},j_{\ext}^{m}];x) \nonumber
\\
&+ n^{k l}([a_{\ext}^{m},j_{\ext}^{m}];x) a_{l}^{\mathrm{ext}}(x),
\\
\label{IntPotentialFun}
\Box A^{k}(x) &+ \partial^{k} \left(\frac{1}{c} \int \diff^3 r' \frac{\vec{\nabla}'\cdot\vec{j}_{\mathrm{ext}}(x') + \vec{\nabla}'\cdot\vec{J}(x')}{4 \pi \epsilon_0 |\vec{r} - \vec{r}'|}  \right) \nonumber
\\
&=  \mu_0 c \left(j^{k}_{\ext}(x) + J^{k}(x) \right),
\end{align}
for given $J_{k}$ and $A_{k}$ \footnote{Note that in correspondence to the freedom of the external variable $a^{k}_{\ext}$ the freedom of the internal variable $J_{k}$ is also restricted, since $J_{0}$ is fixed by the initial state and the continuity equation for all times. Similarly the freedom of the external current $j^{k}_{\ext}$ is in correspondence to the freedom of the internal field $A_{k}$.}. As before in Sec.~\ref{subsec:FoundModel} we can construct the external current uniquely. By defining the vector field
\begin{align*}
 \zeta^{k}(x)\!= \!\Box A^{k}(x) \!+\! \partial^{k}\! \left(\!\frac{1}{c}\! \int\! \diff^3 r' \frac{\vec{\nabla}'\cdot\vec{J}(x')}{4 \pi \epsilon_0 |\vec{r} - \vec{r}'|}\!  \right) \!-\!  \mu_0 c J^{k}(x) ,
\end{align*}
we find from the Helmholtz decomposition of $\vec{\zeta} = - \vec{\nabla} \xi + \vec{\nabla} \times \vec{\Xi}$ and $\vec{j}_{\ext} = - \vec{\nabla} \upsilon + \vec{\nabla} \times \vec{\Upsilon}$ that
\begin{align}
\label{CurrentConstruction}
 \upsilon(x) = \frac{1}{2 \mu_0 c} \xi(x), \quad \vec{\nabla}\times\vec{\Upsilon}(x) = \frac{1}{\mu_0 c} \vec{\nabla}\times\vec{\Xi}(x).
\end{align}
Thus, we need to show that for given $(J_{k}, A_{k})$ there can only be a unique $a^{k}_{\ext}$ \footnote{We note at this point, that one can also fix the zero components $j^{0}_{ext}$ and $A_{0}$, respectively. Since $A_0$ is given by Eq.~(\ref{CoulombPotential}) its first order time-derivative is, due to the continuity equation, given in terms of $\vec{j}_{\ext}$ and $\vec{J}$. Thus, to determine $A_0$ we only have to choose an initial condition. However, due to Eq.~(\ref{CoulombPotential})
this choice also fixes automatically the initial condition for $j_{ext}^{0}$ for the continuity equation $\partial_0 j_{\ext}^{0} = - \vec{\nabla} \cdot \vec{j}_{\ext}$.}. To show this we first define
\begin{align*}
 J^{(\alpha)}_{\mu}(\vec{r}) = \left. \partial_{0}^{\alpha}J_{\mu}(x)\right|_{t=0},
\end{align*} 
formally construct the respective Taylor coefficients 
\begin{align}
 \label{CurrentTaylor}
J^{(\alpha + 1)}_{k}(\vec{r}) &=  \left[q_{\mathrm{kin},k}^{(\alpha)}(\vec{r}) + q_{\mathrm{int},k}^{(\alpha)}(\vec{r})\right] \\
& +\sum_{\beta=0}^{\alpha} {\alpha \choose \beta} \left( a^{l \, (\beta)}_{\mathrm{ext}}(\vec{r})  \right) \left(n_{k l}^{(\alpha-\beta)}(\vec{r}) \right),\nonumber 
\end{align}
and consider two external potentials $a^{k}_{\ext} \neq \tilde{a}^{k}_{\ext}$ that differ at lowest order $\alpha$. Accordingly we find in this order
\begin{align}
\label{PossibleContradiction}
 \vec{J}^{(\alpha + 1)}(\vec{r}) & -  \vec{\tilde{J}}^{(\alpha + 1)}(\vec{r}) 
\\
 &= \vec{n}^{(0)}(\vec{r})\times \left( \vec{a}^{(\alpha)}_{\ext}(\vec{r}) - \vec{\tilde{a}}^{(\alpha)}_{\ext}  (\vec{r}) \right)  \nonumber,
\end{align}
where
\begin{align*}
 \vec{n}^{(0)}(\vec{r}) = \frac{2 e^2}{\hbar} \braket{\Psi_0 | \spin^{\dagger}(\vec{r}) \vec{\Sigma} \spin(\vec{r}) }{\Psi_0}.
\end{align*}
While before we could conclude that the difference between the currents is necessarily non-zero provided $\vec{n}^{(0)} \neq 0$, here we find that this is not sufficient. Actually, we need to restrict the allowed potentials $\vec{a}_{\ext}$ to those that are perpendicular to $\vec{n}^{(0)}$. If we do this, then Eq.~(\ref{PossibleContradiction}) makes the currents necessarily different and we can conclude that we have a one-to-one correspondence. This aspect was not taken into account in previous work \cite{rajagopal-1994, ruggenthaler-2011b}, which is restricting effectively the one-to-one mapping to a smaller set of potentials and currents in these proofs. Still, it seems possible to find a different way to show the bijectivity of the complete mapping $(J_{k}, A_{k}) \leftrightarrow (a^{k}_{\ext}, j^{k}_{\ext})$. However, the true drawback of a relativistic QEDFT based on the current is found if we try to reproduce a given pair $(J_{k}, A_{k})$. If 
we choose a current that obeys
\begin{align*}
 \vec{J}^{(1)}(\vec{r}) = \vec{n}^{(0)}(\vec{r}) + \vec{q}_{\mathrm{kin}}^{(0)}(\vec{r}) + \vec{q}_{\mathrm{int}}^{(0)}(\vec{r}),
\end{align*}
then the resulting equation that defines the Taylor-coefficient of the external potential reads by employing Eq.~(\ref{CurrentTaylor}) and following the same strategy as in subsection \ref{subsec:KSModel}
\begin{align*}
 \vec{n}^{(0)}(\vec{r}) = \vec{n}^{(0)}(\vec{r}) \times \vec{a}^{(0)}_{\ext}(\vec{r}).
\end{align*}
This equation does not have a solution and therefore any current that obeys the above form cannot be reproduced by the respective quantum system. This does also call into doubt the possibility of exactly predicting the current of a coupled system by an uncoupled one, i.e. the Kohn-Sham construction of \cite{rajagopal-1994, ruggenthaler-2011b}. Of course we can remedy this problem by adding terms to the QED Hamiltonian that break the minimal-coupling prescription of the Lagrangian. Such procedures could then be alternatively used to provide a Kohn-Sham scheme to describe the fully coupled QED problem. The advantage of such an approach is, that still the equation for the vector potential is known explicitly in terms of the internal pair $(J_{k}, A_{k})$. This is not the case, when we use a different basic variable for the matter part of the QED system, as we will do in the following.


To avoid the problems with the relativistic current, we will in the following base our considerations on the polarization $P_{k}$. While the current describes the flow of the charge of the system (which is conserved), the polarization depends on the actual number of particles and anti-particles (which is not conserved). Therefore, the polarization can actually differ between a local current produced by, e.g., two electrons and one positron or three electrons and two positrons, in contrast to the current. To show now that for a fixed initial state $\ket{\Psi_0}$ we actually have
\begin{align}
\label{RelPairMapping}
(a^{k}_{\ext},j^{k}_{\ext})\; {\buildrel\rm 1:1 \over \leftrightarrow }  \; (P_{k}, A_{k}), 
\end{align} 
we demonstrate that for a given internal pair $(P_k, A_{k})$ the two coupled equations
\begin{align}
\label{IntPolFun}
\partial_0 &\vec{P}(x) = \vec{Q}_{\mathrm{kin}}([a_{\ext}^{k},j_{\ext}^{k}];x) +  \vec{Q}_{\mathrm{int}}([a_{\ext}^{k},j_{\ext}^{k}];x) 
\\
&+ \frac{2emc}{\imagi \hbar} \vec{J}([a_{\ext}^{k},j_{\ext}^{k}];x) + \frac{2e}{\imagi \hbar c} P_{0}([a_{\ext}^{k},j_{\ext}^{k}];x) \vec{a}_{\mathrm{ext}}(x) \nonumber 
\\
\label{IntPotFun}
\Box & \vec{A}(x) - \mu_0 c \left(\vec{j}_{\ext}(x) + \vec{J}([a_{\ext}^{k},j_{\ext}^{k}];x) \right) 
\\
&=  \vec{\nabla} \left(\frac{1}{c} \int \diff^3 r' \frac{\vec{\nabla}'\cdot\vec{j}_{\ext}(x') + \vec{\nabla}'\cdot\vec{J}([a_{\ext}^{k},j_{\ext}^{k}];x')}{4 \pi \epsilon_0 |\vec{r} - \vec{r}'|}  \right), \nonumber
\end{align}
allow only for a unique solution $(a_{\ext}^{k},j_{\ext}^{k})$. Here we used the definitions 
\begin{align*}
 \hat{Q}^{k}_{\mathrm{kin}}(\vec{r}) & = e c \dspin(\vec{r})\! \left(\partial^{k} -\overset{\smash{\raisebox{-1.5pt}{\tiny$\leftarrow$}}}{\partial}^{k}  \right) \spin(\vec{r}) \nonumber
\\
&+ \imagi ec \epsilon^{klj} \partial_l \left(\dspin(\vec{r})\Sigma_{j}\spin(\vec{r}) \right),
\end{align*}
\begin{align*}
 \hat{Q}^{k}_{\mathrm{int}}(\vec{r}) = \frac{2e}{\imagi \hbar c} \hat{P}_{0}(\vec{r})\hat{A}^{k}.
\end{align*}
These coupled equations can only have a solution if the pair $(P_k, A_{k})$ obeys the initial condition enforced by the fixed initial state $\ket{\Psi_0}$, i.e.
\begin{align}
\label{InitialPolarization}
P_k^{(0)}(\vec{r}) &= \braket{\Psi_0| \hat{P}_{k}(\vec{r}) }{\Psi_0}, 
\\
\label{InitialFourPotential}
A^{(0)}_{k}(\vec{r}) &=  \braket{\Psi_0| \A_{k}(\vec{r})}{\Psi_0}, \; A^{(1)}_{k}(\vec{r}) =  - \braket{\Psi_0| \hat{E}_{k}(\vec{r})}{\Psi_0}.
\end{align}
Since the current $J_{k}$ is now a functional of $(a_{\ext}^{k},j_{\ext}^{k})$ the previous explicit construction of $j^{k}_{\ext}$ is no longer valid. However, if we assume $(a_{\ext}^{k},j_{\ext}^{k})$ \textit{both} to be Taylor-expandable we find for the lowest order $\alpha$ on the one hand that
\begin{align}
\label{PolRelQEDFT}
 \vec{P}^{(\alpha+1)}(\vec{r})\! - \!\vec{\tilde{P}}^{(\alpha+1)}(\vec{r}) \!= \!\frac{2 e}{\imagi \hbar c} P^{(0)}_{0}(\vec{r})\! \left(\! \vec{a}^{(\alpha)}_{\ext}(\vec{r}) \!-\! \vec{\tilde{a}}^{(\alpha)}_{\ext}(\vec{r}) \!\right) \! \neq \! 0,
\end{align}
provided $P_{0}^{(0)}(\vec{r}) = \braket{\Psi_0|\hat{P}_0(\vec{r})}{\Psi_0} \neq 0$, which corresponds to the (local) total number of particles and anti-particles. On the other hand we have
\begin{align}
\label{PotRelQEDFT}
 &\vec{A}^{(\alpha+2)}(\vec{r})\! - \!\vec{\tilde{A}}^{(\alpha+2)}(\vec{r}) \!= \!- \mu_0 c \left(\! \vec{j}^{(\alpha)}_{\ext}(\vec{r}) \!-\! \vec{\tilde{j}}^{(\alpha)}_{\ext}(\vec{r}) \!\right)
\\
 & + \vec{\nabla} \left(\frac{1}{c} \int \diff^3 r' \frac{\vec{\nabla}'\cdot\vec{j}^{(\alpha)}_{\ext}(\vec{r}') - \vec{\nabla}'\cdot\vec{\tilde{j}}^{(\alpha)}_{\ext}(\vec{r}')}{4 \pi \epsilon_0 |\vec{r} - \vec{r}'|}  \right) \neq \! 0. \nonumber
\end{align}
Otherwise there would exist a current $\vec{j}_{\ext}(\vec{r}) \neq 0$ that fulfills
\begin{align}
\label{CurrentUniqueness}
 \mu_0 c \vec{j}_{\ext}(\vec{r}) - \vec{\nabla} \left(\frac{1}{c} \int \diff^3 r' \frac{\vec{\nabla}'\cdot\vec{j}_{\ext}(\vec{r}') }{4 \pi \epsilon_0 |\vec{r} - \vec{r}'|}  \right) = 0.
\end{align}
However, due to the definition of the Coulomb potential as the Green's function of the Laplacian (see Eq.~(\ref{CoulombPotential}) and (\ref{A0})) we find from the divergence of Eq.~(\ref{CurrentUniqueness}) that $\vec{\nabla} \cdot \vec{j}_{\ext} = 0$ and thus the only possible current that fulfills Eq.~(\ref{CurrentUniqueness}) is $\vec{j}_{\ext} = 0$. 
Thus, the mapping (\ref{RelPairMapping}) is bijective (at least for Taylor-expandable external pairs $(a^{k}_{\ext}, j^{k}_{\ext})$). Therefore we can, instead of solving the fully coupled QED problem for the (numerically infeasible) wave function $\ket{\Psi(t)}$, determine the exact internal pair $(P_k, A_{k})$ from the coupled non-linear equations
\begin{align}
\label{IntPolFunBasic}
\partial_0 &\vec{P}(x) = \vec{Q}_{\mathrm{kin}}([P_{k},A_{k}];x) +  \vec{Q}_{\mathrm{int}}([P_{k},A_{k}];x) 
\\
&+ \frac{2emc}{\imagi \hbar} \vec{J}([P_{k},A_{k}];x) + \frac{2e}{\imagi \hbar c} P_{0}([P_{k},A_{k}];x) \vec{a}_{\ext}(x) \nonumber 
\\
\label{IntPotFunBasic}
\Box &\vec{A}(x) - \vec{\nabla} \left(\frac{1}{c} \int \diff^3 r' \frac{\vec{\nabla}'\cdot\vec{j}_{\ext}(x') + \vec{\nabla}'\cdot\vec{J}([P_{k},A_{k}];x')}{4 \pi \epsilon_0 |\vec{r} - \vec{r}'|}  \right) \nonumber
\\
&=  \mu_0 c \left(\vec{j}_{\ext}(x) + \vec{J}([P_{k},A_{k}];x) \right),
\end{align}
for the initial conditions (\ref{InitialPolarization}) and (\ref{InitialFourPotential}). In order to solve these equations simultaneously we need to find approximations for the unknown functionals. The only drawback in this more general approach than the ones used in \cite{rajagopal-1994, ruggenthaler-2011b}, is that now we also have an unknown functional in the classical Maxwell equation, i.e. $\vec{J}[P_{k},A_{k}]$.


\subsection{Kohn-Sham approach to relativistic QEDFT}

In this subsection we provide the adopted Kohn-Sham construction based on the internal pair $(P_k,A_{k})$ and give the simplest approximation for the Kohn-Sham potential and current. As in Sec.~\ref{subsec:KSModel}, we choose our auxiliary Kohn-Sham system to be an uncoupled system. While different Kohn-Sham constructions are possible, this approach is the numerically least demanding.

In a first step, in accordance to Sec.~\ref{subsec:KSModel}, we first construct an uncoupled system that can \textit{reproduce} a given internal pair $(P_k, A_{k})$ of the fully coupled QED system. To do so, we first need an initial state $\ket{\Phi_0}$ that fulfills the initial condition (\ref{InitialPolarization}) and (\ref{InitialFourPotential}) of the full QED system. This allows that the coupled equations 
\begin{align}
\label{IntPolFunEff}
\partial_0 \vec{P}(x)& = \vec{Q}_{\mathrm{kin}}([a^{k}_{\mathrm{eff}},j^{k}_{\mathrm{eff}}];x) + \frac{2emc}{\imagi \hbar} \vec{J}([a^{k}_{\mathrm{eff}},j^{k}_{\mathrm{eff}}];x) \nonumber 
\\
& + \frac{2e}{\imagi \hbar c} P_{0}([a^{k}_{\mathrm{eff}},j^{k}_{\mathrm{eff}}];x) \vec{a}_{\mathrm{eff}}(x) 
\\
\label{IntPotFunEff}
\Box \vec{A}(x) &- \vec{\nabla} \left(\frac{1}{c} \int \diff^3 r' \frac{\vec{\nabla}'\cdot\vec{j}_{\mathrm{eff}}(x')}{4 \pi \epsilon_0 |\vec{r} - \vec{r}'|}  \right) \nonumber
\\
&=  \mu_0 c \vec{j}_{\mathrm{eff}}(x) ,
\end{align}
can only have a unique solution. Obviously, for the case of the uncoupled problem we can use a construction similar to Eqs.~(\ref{CurrentConstruction}) to determine the unique $j_{\mathrm{eff}}^{k}$. To show the existence of a solution to Eq.~(\ref{IntPolFunEff}) we perform the standard Taylor-expansion construction and assume that the series converges \cite{leeuwen-1999, vignale-2004, ruggenthaler-2011b}. A more general approach would be to follow a fixed-point procedure \cite{ruggenthaler-2011}. The respective Taylor-coefficients of the effective potential are given by
\begin{align*}
&P_{0}^{(0)}(\vec{r}) \vec{a}^{(\alpha)}_{\mathrm{eff}}(\vec{r}) = \frac{\imagi \hbar c}{2 e}\left( \vec{P}^{(\alpha + 1)}(\vec{r}) - \vec{Q}_{\mathrm{kin}}^{(\alpha)}(\vec{r})  \right. 
\\
& \left. - \frac{2 e m c}{\imagi \hbar} \vec{J}^{(\alpha)}(\vec{r}) \right) - \sum_{\beta=0}^{\alpha-1} {\alpha \choose \beta} \left( \vec{a}^{(\beta)}_{\mathrm{eff}}(\vec{r})  \right) \left(P^{(\alpha-\beta)}_{0}(\vec{r}) \right).\nonumber 
\end{align*}
This construction makes plausible that there exists an uncoupled system subject to the effective external fields $(a_{\mathrm{eff}}^k,j_{\mathrm{eff}}^{k})$ that reproduces a given pair of a fully coupled QED problem. The above construction actually resembles the mapping
\begin{align*}
 \left(P_k,A_{k} \right) \; {\buildrel\rm \ket{\Phi_0} \over \mapsto }  \; \left(a^{k}_{\mathrm{eff}},j^{k}_{\mathrm{eff}}\right).
\end{align*}
for a given pair $\left(P_k,A_{k} \right)$. Now, to \textit{predict} the internal pair $\left(P_k,A_{k} \right)$ of the full QED problem we again introduce a composite mapping
\begin{align*}
 \left(a^{k}_{\ext},j^{k}_{\ext}\right) \; {\buildrel\rm \ket{\Psi_0} \over \mapsto }  \;  \left(P_k,A_{k} \right) \; {\buildrel\rm \ket{\Phi_0} \over \mapsto }  \; \left(a^{k}_{\mathrm{eff}},j^{k}_{\mathrm{eff}}\right). 
\end{align*}
The resulting Kohn-Sham potential and Kohn-Sham current are then given by the functional equations
\begin{align}
\label{IntPolFunKS}
P_{0}([& \Phi_0, P_k ,A_{k}];x) \vec{a}_{\mathrm{KS}}(x)  = \frac{\imagi \hbar c} {2e}\left( \vec{Q}_{\mathrm{kin}}([\Psi_0, P_k,A_{k}];x) \right. \nonumber
\\
&\left. - \vec{Q}_{\mathrm{kin}}([\Phi_0, P_k,A_{k}];x) +  \vec{Q}_{\mathrm{int}}([\Psi_0, P_k,A_{k}];x) \right)  \nonumber 
\\
&+ mc^2 \left( \vec{J}([\Psi_0, P_k,A_{k}];x) - \vec{J}([\Phi_0, P_k,A_{k}];x) \right) \nonumber
\\
&+ P_{0}([\Psi_0, P_k,A_{k}];x) \vec{a}_{\mathrm{ext}}(x) 
\\
\label{IntPotFunKS}
\vec{j}_{\mathrm{KS}}&(x) =  \vec{j}_{\ext}(x) + \vec{J}([\Psi_0, P_k,A_{k}];x).
\end{align}
This allows to solve an uncoupled system instead of the fully coupled QED problem. However, as also pointed out in \cite{ruggenthaler-2011b}, we can only fully decouple the matter from the photon part if also the initial state is of product form, i.e. $\ket{\Phi_0} = \ket{\mathrm{M}_0} \otimes \ket{\mathrm{EM}_0}$. And if we further assume that $\ket{\mathrm{M}_0}$ is given in terms of a Slater-determinant we can actually map the whole problem to solving a Dirac equation with the above Kohn-Sham potential $a^{k}_{\mathrm{KS}}$ and simultaneously a classical Maxwell equation with $j_{\mathrm{KS}}^{k}$. The mean-field approximation recovers the approximation introduced in \cite{ruggenthaler-2011b} and reads as
\begin{align}
\label{RelMFApprox}
\vec{a}_{\mathrm{MF}}(x) &= \vec{a}_{\mathrm{ext}}(x) + \vec{A}(x),
\\
\vec{j}_{\mathrm{MF}}(x) &=  \vec{j}_{\ext}(x) + \vec{J}(x).
\end{align}
Since for simplicity we used a gauge where $a_{\ext}^{0} = 0$ while for the photon field we employed Coulomb gauge, we have to perform an according gauge transformation to have the mean field $a^{\mu}_{\mathrm{MF}}$ in either the one or the other gauge completely.
This approximation is similar to the Maxwell-Schr\"odinger approach, that assumes the photon field to behave essentially classically.


\section{Non-relativistic QEDFT}
\label{sec:NonRelQEDFT}

While for the sake of generality we have been considering the full QED problem in the previous section, we are actually mainly interested in the behaviour of condensed-matter systems or atoms and molecules that interact with photons. In such situations the external fields are usually small compared to the Schwinger-limit, i.e. we do not have pair-production is such situations. Further, we want to investigate systems, where the quantum nature of the photons becomes important. Most prominently this happens for the case of a cavity, where different boundary conditions for the Maxwell field have to be considered. These quantum-optical situations also naturally restrict the available photonic modes. Such physical situations are then well described by models of non-relativistic particles interacting with a quantized electromagnetic field, such as the Pauli-Fierz Hamiltonian (see e.g. \cite{hainzl-2002, hiroshima-2002}) or the Nelson model \cite{nelson-1964, bachmann-2012}. In the lowest order of approximations 
we find the situation of a two-level system interacting with one photonic mode, similar to the one presented in Sec.~\ref{sec:Model}. This simplest of models is 
the prime example of a quantum-optical problem.    

We realize at this point, that all the conditions we had to impose in order to make our starting QED Hamiltonian well-defined, are naturally met in the situations we aim at investigating. Actually, we even do not need to adopt a field-theoretical treatment for the particles in the first place and usually only need to take into account a few photonic modes. Such an approach would avoid a lot of unpleasent problems in connection with renormalization and regularization of these theories. However, one would then need to introduce a new QEDFT approach for every new type of model Hamiltonian. Therefore, in this section we want to demonstrate how naturally all lower lying QEDFT reformulations are just approximations to the fully relativistic QEDFT that we presented in the previous sections. In lowest order we then recover the two-site Hubbard model coupled to one mode of Sec.~\ref{sec:Model}.


\subsection{Equations of motion in the non-relativistic limit}

In this subsection we derive the non-relativistic limit of the basic equations of motion, on which the QEDFT reformulations are based. We show how approximations in the Hamiltonian correspond to approximations in the basic equations of the corresponding QEDFT approaches.

\begin{figure*}[t]
    \begin{center}
      \includegraphics[width=0.41\textwidth]{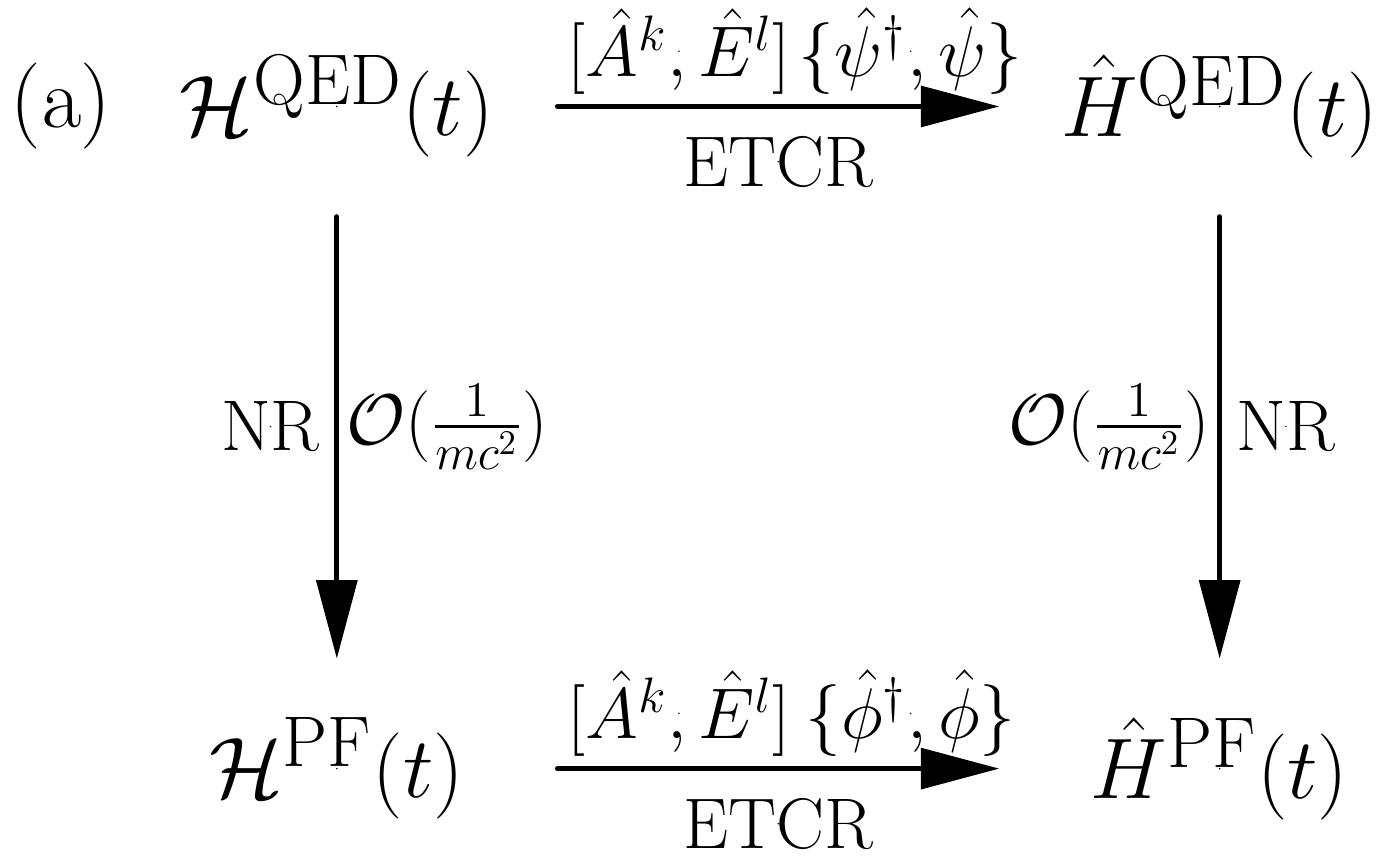}
        \hspace{20mm}
      \includegraphics[width=0.41\textwidth]{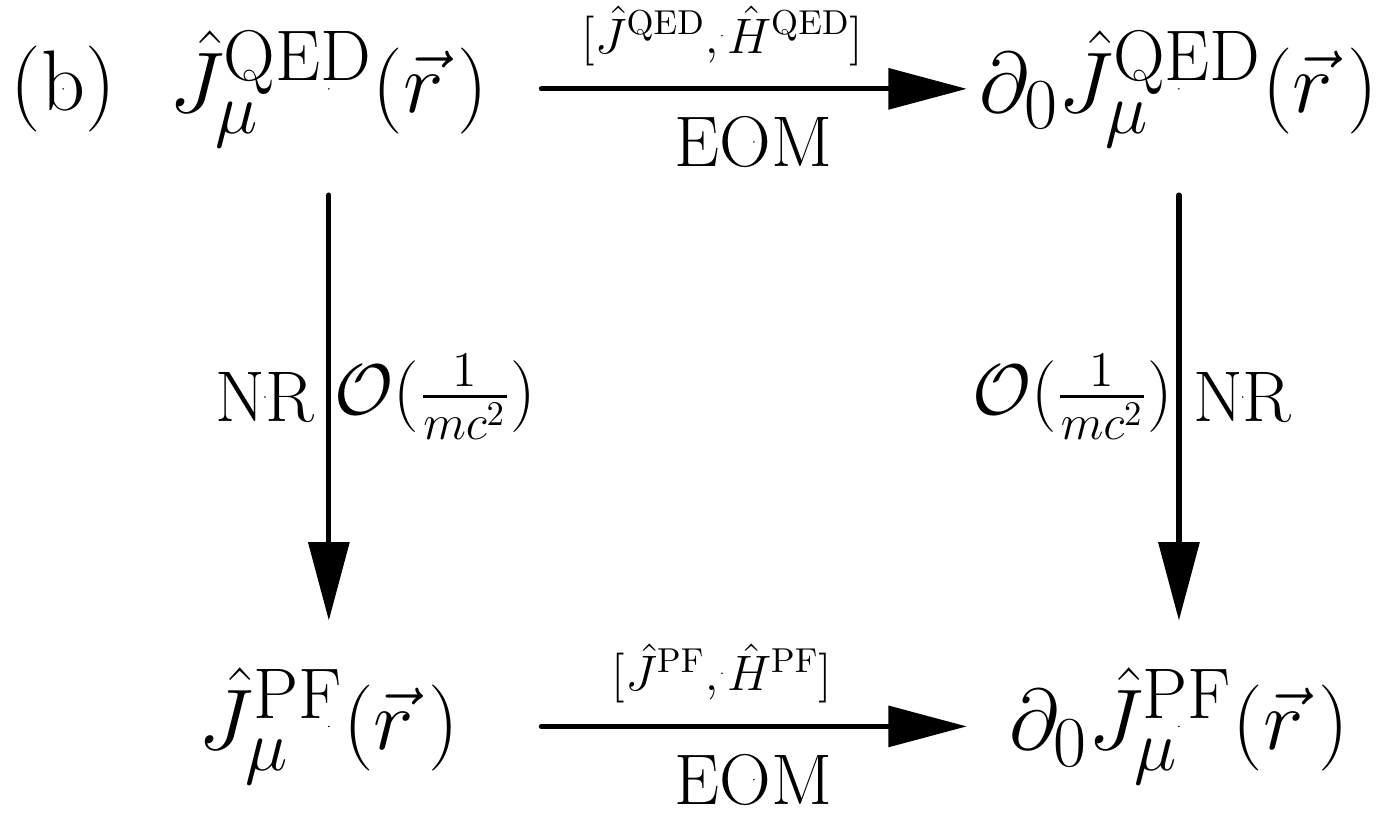}
    \end{center}
    \caption{The non-relativistic (NR) limits do not depend on the order of operations. First performing the limit and then quantizing with the equal-time (anti)commutation relations (ETCR) leads to the same (Pauli-Fierz) Hamiltonian as the opposite ordering (see (a)). Further, first performing the limit in the current and then calculating the equations of motion (EOM) leads to the same result as performing the limit directly on the relativistic EOM (see (b)).}
    \label{fig:qed_mapping}
\end{figure*}

Let us first start with the non-relativistic limit of the fully coupled QED Hamiltonian in Coulomb gauge. From the Heisenberg equation of motion, defining
\begin{align*}
 \A^k_{\mathrm{tot}}(x) &= \A^{k}(x) + a^{k}_{\ext}(x),
\\
 A^{0}_{\mathrm{tot}}(x) & = a^0_{\ext} (x) + \frac{1}{c} \int \diff^3 r' \frac{j^0_{\ext}(x')}{4 \pi \epsilon_0 |\vec{r}-\vec{r}'|},
\end{align*}
and $\alpha^k = \gamma^0 \gamma^k$, we find the quantized Dirac equation (in the Heisenberg picture) 
\begin{align}
\label{QuantizedDirac}
 \imagi &\hbar c \partial_0 \spin(x)  \nonumber
\\
&= \left[\alpha^k  \left( -\imagi \hbar c \partial_k + e \A^{\mathrm{tot}}_k(x)\right)  + \gamma^0 mc^2 + e A^{\mathrm{tot}}_{0}(x)\right] \spin(x) \nonumber
\\
& \qquad + e^2 \int \diff^3 r' \frac{:\spin^{\dagger}(x') \spin(x'):}{4 \pi \epsilon_0 |\vec{r}-\vec{r}'|} \spin(x) ,
\end{align}
and accordingly for $\spin^{\dagger}$. We see that the electronic components $\p$ of the four-spinor are mixed with the positronic components $\c$. Of course, for small energies only the electronic component of the four-spinor is important, and therefore we would like to find an equation based solely on $\p$. So naturally we would like to decouple the upper component $\p$ from the lower component $\c$. A possible way would be to find a unitary transformation of the Dirac Hamiltonian that does this, at least perturbatively. A possible expansion parameter for auch a perturbative transformation would be $(mc^2)^{-1}$, since we know that the energies involved in non-relativistic processes are small compared to the rest-mass energy. This energy also represents the spectral gap between the electronic and positronic degrees of freedom, which effectively decouples the dynamics of the particles and anti-particles for small enough energies. The resulting unitary transformations are known as the Foldy-Wouthuysen 
transformations \cite{strange-RQM} and are routinely used to generate the non-relativistic limits of the Dirac equation to any order desired. Here, we employ an equivalent but different procedure to decouple the electronic from the positronic degrees of freedom. To do so, we first rewrite Eq.~(\ref{QuantizedDirac}) componentwise 
\begin{align}
\left(\hat{D}(x) -  mc^2 \right) \p(x) = \!\vec{\sigma} \cdot \left(\! -\imagi \hbar c \vec{\nabla}\! -\! e \hat{\vec{A}}_{\mathrm{tot}}(x)\!\right) \c(x), \nonumber
\\
\label{ChiComponent}
\left(\hat{D}(x) +  mc^2 \right) \c(x) = \!\vec{\sigma} \cdot \left(\! -\imagi \hbar c \vec{\nabla}\! -\! e \hat{\vec{A}}_{\mathrm{tot}}(x)\!\right) \p(x), 
\end{align}
where we defined
\begin{widetext}
\begin{align*}
 \hat{D} (x) = \left(\imagi \hbar c \partial_0 - e A^{\mathrm{tot}}_0(x) -   e^2  \int \! \! \diff^3 r' \frac{:\p^{\dagger}(x')\p(x')  +  \c^{\dagger}(x')\c(x'):}{4 \pi \epsilon_0 |\vec{r}-\vec{r}'|} \right).
\end{align*}
\end{widetext}
And thus we (formally) find that
\begin{align*}
\c(x) =\! \left[\hat{D}(x) +  mc^2 \right]^{-1} \vec{\sigma} \cdot \left(\! -\imagi \hbar c \vec{\nabla}\! -\! e \hat{\vec{A}}_{\mathrm{tot}}(x)\!\right) \p(x).
\end{align*}
If we assume non-relativistic energies, the main contribution to the energy of the system stems from $mc^2$, i.e. $\imagi \hbar c \partial_0 \approx mc^2$. Accordingly, from the Neumann series of the resulting operator we find the inverse operator to lowest order as $\left[ \hat{D}(x) + \! mc^2  \! \right]^{-1} \approx 1/2mc^2$, and consequently
\begin{align}
\label{Decoupling}
\c(x) \approx \frac{\vec{\sigma}}{2mc^2} \cdot \left( -\imagi \hbar c \vec{\nabla} - e \hat{\vec{A}}_{\mathrm{tot}}(x) \right) \p(x). 
\end{align}
At this level of approximation to the full QED problem we find the Pauli-Fierz Hamiltonian (already transformed back to the Schr\"odinger picture),
\begin{align}
\label{NonRelHam}
 \hat{H}(t) &= \hat{H}_{\mathrm{M}} +\hat{H}_{\mathrm{EM}} + \hat{H}_{\mathrm{C}} -\frac{1}{c} \int \diff^3 r \; \hat{\vec{J}}(x)\cdot \hat{\vec{A}}(\vec{r})
\\
&+ \frac{1}{c}\int \diff^3 r \J_{0}(\vec{r})\left( A^{0}_{\mathrm{tot}}(x) - \frac{e}{2mc^2} \hat{\vec{A}}_{\mathrm{tot}}^2(\vec{r}) \right) \nonumber
\\
& - \frac{1}{c} \int \diff^3 r  \left(\hat{\vec{J}}(x)\cdot \vec{a}_{\ext}(x) + \hat{\vec{A}}(\vec{r})\cdot \vec{j}_{\ext}(x)  \right), \nonumber
\end{align}
where the non-relativistic kinetic energy reads as
\begin{align*}
 \hat{H}_{\mathrm{M}} = \int \diff^3 r \p^{\dagger}(\vec{r}) \left(- \frac{1}{2m} \vec{\nabla}^2  \right) \p(\vec{r}),
\end{align*}
the energy of the electromagnetic field is given as before, the Coulomb energy is given by
\begin{align*}
\hat{H}_{\mathrm{C}} = \frac{e^2}{2} \int \diff^3 r' \, \frac{\p^{\dagger}(\vec{r})\p^{\dagger}(\vec{r}') \p(\vec{r}') \p(\vec{r})}{4 \pi \epsilon_0 |\vec{r}-\vec{r}'|}, 
\end{align*}
and the non-relativistic current is defined by
\begin{align}
\label{NonRelCurrent}
 \J^{k}(x) &= 2 ec \Re\left\{ \p^{\dagger}(\vec{r}) \frac{\vec{\sigma}}{2 m c^2} \cdot \left(- \imagi \hbar c \vec{\nabla} - e \hat{\vec{A}}^{\mathrm{tot}}(x)  \right) \p (\vec{r})\right\} \nonumber
\\
& = \J^{k}_{\mathrm{p}}(\vec{r}) - \epsilon^{klj} \partial_l \hat{M}_{j}(\vec{r}) - \frac{e}{mc^2} \J_{0}(\vec{r})\A^{k}_{\mathrm{tot}}(x).
\end{align}
Here we used the definition of the paramagnetic current
\begin{align*}
 \J^{k}_{\mathrm{p}}(\vec{r}) = \frac{e \hbar}{2 m \imagi} \left[ \left(\partial^{k} \p^{\dagger}(\vec{r})\right) \p(\vec{r}) - \p^{\dagger}(\vec{r})\partial^{k} \p(\vec{r}) \right],
\end{align*}
the magnetization density
\begin{align*}
 \hat{M}^{k}(\vec{r}) = \frac{e \hbar}{2 m} \p^{\dagger}(\vec{r})\sigma^{k} \p(\vec{r}),
\end{align*}
and the zero component of the current 
\begin{align*}
 \J_{0}(\vec{r}) = e c \p^{\dagger}(\vec{r}) \p(\vec{r}).
\end{align*}
By construction the current obeys the continuity equation $\partial_0 \J_{0}(x) = - \vec{\nabla}\cdot \hat{\vec{J}}(x)$. We note here, that due to the non-relativistic limit the physical current defined in Eq.~(\ref{NonRelCurrent}) becomes explicitly time-dependent \cite{stefanucci-MBT}. Further, we point out that the result of the above (formal) derivations is the same as the result obtained by first performing the non-relativistic limit of the classical Hamiltonian $\mathcal{H}^{\mathrm{QED}}(t)$ (constructed from the classical Lagranian density of Eq.~(\ref{QEDLagrangian})) and then canonically quantizing the Schr\"odinger field, as shown in Fig.~\ref{fig:qed_mapping}~(a).

In the non-relativistic limit the resulting Hamiltonian commutes with the particle-number operator $\hat{N} = \int \diff^3 r \p^{\dagger}(\vec{r}) \p(\vec{r})$, as can be seen directly from the continuity equation. Accordingly we do not need to employ a field-theoretical description for the electrons and all matter-operators can be expressed in first-quantized notation (while still being a many-particle problem). Nevertheless, we can still encounter infinities due to the interaction between the non-relativistic particles and the quantized Maxwell field \cite{hainzl-2002, hiroshima-2002}. While we do no longer have vacuum polarization (no electron-positron pairs are possible) and vertex corrections, we still have an infinite self-energy \cite{hainzl-2002}. To first order in the coupling the ground-state energy (for $\vec{a}_{\ext} = \vec{j}_{\ext}=0$) diverges as
\begin{align*}
 E_0 \sim  \frac{2 e}{\pi} \left(\Lambda - \ln(1 + \Lambda)  \right),
\end{align*}
where $\Lambda$ is the ultra-violet cut-off for the photon modes. By subtracting the infinite self-energy of the ground-state, which is equivalent to introducing a renormalized mass, we can renormalize the Hamiltonian perturbatively. In the following we assume, that the Pauli-Fierz Hamiltonian can be fully renormalized. For instance, in the limit of only scalar photons (the Nelson model) we know that we can perform a full renormalization of the Hamiltonian by subtracting the self-energy (provided that the kinetic energy of the problem is smaller than $mc^2$) \cite{nelson-1964,bachmann-2012}. Therefore, we interpret the electron mass in the Hamiltonian as a bare mass, i.e. we subtract the infinite self-energy.

Now, the equation of motion for $\J_{k}$ can be either found by the Heisenberg equation with the Pauli-Fierz Hamiltonian or by the non-relativistic limit of Eq.~(\ref{HeisenbergCurrent}) (see appendix (\ref{app:NonRelEOM})). We have explicitly checked both ways of performing the non-relativistic limit as schematically indicated in Fig.~\ref{fig:qed_mapping}~(b). After some calculation we find (omitting the spatial and temporal dependences)
\begin{widetext}
\begin{align}
\label{NonRelEOM}
 \imagi \partial_0 \J_{k} &= -\imagi \left\{  \partial^{l}\hat{T}_{kl} - \hat{W}_{k} - \frac{e}{mc^2} \partial_l \A^{l}_{\mathrm{tot}} \J^{\mathrm{p}}_{k}  - \frac{e}{mc^2} \left(\partial_k \A^{l}_{\mathrm{tot}}  \right) \J^{\mathrm{p}}_k + \frac{e}{mc^2} \left(\partial_k \partial_l \A^{\mathrm{tot}}_m   \right) \epsilon^{lmn} \hat{M}_{n} - \frac{e}{mc^2}\left[\partial_k \left(\frac{1}{2mc^2} \A_{\mathrm{tot}}^2 + A_{0}^{\mathrm{tot}}  \right)  \right]  \J_{0}\right\} \nonumber
\\
& - \epsilon_{klj} \partial^{l} \left\{ -\frac{e \hbar^3}{4 m^2}\p^{\dagger} \left( \overset{\smash{\raisebox{-1.5pt}{\tiny$\leftarrow$}}}{\partial}^{n} \overset{\smash{\raisebox{-1.5pt}{\tiny$\leftarrow$}}}{\partial}_{n} \sigma^{j} - \sigma^{j}\partial^{n} \partial_n \right)\p +  \frac{i e}{mc^2} \partial_n \A^n_{\mathrm{tot}} \hat{M}^{j} - \frac{i e}{2 mc^2} \left[\left(\partial^{j} \A_n^{\mathrm{tot}} \right) - \left(\partial_n{j} \A^j_{\mathrm{tot}} \right)   \right] \hat{M}^{n} \right\} \nonumber
\\
&- \frac{1}{mc^2} \left\{\left(\imagi \partial_0 \A_{k}^{\mathrm{tot}}   \right)\J_{0} + \A_k^{\mathrm{tot}}\left(\frac{\imagi e}{mc^2} \partial_l \A^{l}_{\mathrm{tot}} \J_{0} - \imagi \partial^{l} \J^{\mathrm{p}}_l   \right)   \right\},
\end{align}
\end{widetext}
where 
\begin{align*}
\hat{T}_{kl}= \frac{e \hbar^2}{2 m^2 c} \left[ \left(\partial_k \p^{\dagger}\right) \partial_l \p + \left(\partial_l \p^{\dagger}\right) \partial_k \p - \frac{1}{2} \partial_k \partial_l \p^{\dagger} \p \right]
\end{align*}
is the usual momentum-stress tensor and
\begin{align*}
 \hat{W}_k(\vec{r})= \frac{e^3}{mc} \int \diff ^3 r' \, \p^{\dagger}(\vec{r}) \left( \partial_k \frac{\p^{\dagger}(\vec{r}')\p(\vec{r}')}{4 \pi \epsilon_0 |\vec{r}- \vec{r}'|}   \right) \p(\vec{r})
\end{align*}
is the interaction-stress force (the divergence of the interaction-stress tensor) \cite{ tokatly-2005a, ullrich-TDDFT, stefanucci-MBT}. If we would have started with an uncoupled problem, we would find a similar equation with the replacement $\hat{\vec{A}}_{\mathrm{tot}} \rightarrow \vec{a}_{\ext}$ and $\hat{W}_k \rightarrow 0$. Further, the equation for the electromagnetic field does not change, except that we now have to employ the non-relativistic current (see appendix~\ref{app:NonRelEOM}). 

In a next step we perform the non-relativistic limit for the equation of motion of the polarization, i.e. Eq.~(\ref{GordonDecomposition}). We find to order $1/mc^2$
\begin{align}
\label{NonRelGordon}
& \imagi \partial_0 \hat{P}^{k} 
\\
&\approx \frac{2 e mc}{\hbar} \J^{k} - \frac{2 e mc}{\hbar } \left( \J^{k}_{\mathrm{p}} - \epsilon^{klj} \partial_l \hat{M}_{j} - \frac{e}{mc^2} \J_{0}\A^{k}_{\mathrm{tot}} \right) \nonumber
\\
& = 0 \nonumber.
\end{align}
Thus, at this level of approximation the polarization does not change in time.


\subsection{QEDFT for the Pauli-Fierz Hamiltonian}

\label{subsec:PauliFierz}

In this subsection we derive the basic formulation of non-relativistic QEDFT for the full Pauli-Fierz Hamiltonian. We show how the Gordon-decomposition, i.e. the equation of motion for the polarization $P_k$, makes the current $J_k$ a unique functional of $(a^k_{\ext},j^{k}_{\ext})$ and thus becomes the basic variable for the matter part in this limit. Further we demonstrate how the non-relativistic limit of the above Kohn-Sham construction produces the Kohn-Sham construction for the Pauli-Fierz Hamiltonian. A comparison of this level of approximation with relativistic QEDFT and with other approximations is presented schematically in appendix~\ref{app:overview}.

We start by performing the non-relativistic limit of Eq.~(\ref{PolRelQEDFT}). Irrespective of the difference between $\vec{a}_{\ext}$ and $\vec{\tilde{a}}_{\ext}$ (note, that we again employ the $a_{\ext}^{0}=0$ gauge for the external potentials as explained in Sec.~\ref{subsec:FoundRelQEDFT}) the equation in this limit is always zero. However, by employing Eq.~(\ref{NonRelGordon}) we can rearrange the non-relativistic limit to
\begin{align}
\label{NonRelContra}
 \vec{J}^{(\alpha)}(\vec{r}) - \vec{\tilde{J}}^{(\alpha)}(\vec{r}) = -\frac{J_{0}^{(0)}(\vec{r})}{mc^2} \left(\vec{a}_{\ext}^{(\alpha)}(\vec{r}) - \vec{\tilde{a}}_{\ext}^{(\alpha)}(\vec{r})   \right) \neq 0,
\end{align}
which is non-zero provided the density obeys $J^{(0)}_0 \neq 0$. The form of Eq.~(\ref{PotRelQEDFT}) does not change and thus we have in the non-relativistic limit that
\begin{align}
\label{NonRelPairMapping}
(a^{k}_{\ext},j^{k}_{\ext})\; {\buildrel\rm 1:1 \over \leftrightarrow }  \; (J_{k}, A_{k}). 
\end{align} 
Accordingly we can label all physical wave functions by the non-relativistic internal pair $(J_k,A_{k})$. Since $J_{k}$ does no longer have an internal structure (no positronic degrees of freedom), our approach of Sec.~\ref{subsec:EOMQED} to determine the conjugate pairs based on a Legendre-transformation argument now works just fine. The Pauli-Fierz Lagrangian (determined from the Pauli-Fierz Hamiltonian of Eq.~(\ref{NonRelHam})) has a similar structure as the full QED Lagrangian of Eq.~(\ref{QEDaction}) and thus allows to identify the conjugate pairs. Indeed, the Legendre-transformation argument holds for all further non-relativistic approximations, especially for our model system of Sec.~\ref{sec:Model}.

Now we can, instead of solving for the wave function, solve the coupled equations
\begin{align}
\label{NonRelCurrentFunBasic}
 \imagi \partial_0 \vec{J}(x)&  = \vec{q}_{\mathrm{p}}([J_k,A_k, a^{k}_{\ext}];x) + \vec{q}_{\mathrm{M}}([J_k,A_k, a^{k}_{\ext}];x)  \nonumber
\\
&+ \vec{q}_{0}([J_k,A_k, a^{k}_{\ext}];x), 
\\
\label{NonRelPotFunBasic}
\Box \vec{A}(x) &- \vec{\nabla} \left(\frac{1}{c} \int \diff^3 r' \frac{\vec{\nabla}'\cdot\vec{j}_{\ext}(x') + \vec{\nabla}'\cdot\vec{J}(x')}{4 \pi \epsilon_0 |\vec{r} - \vec{r}'|}  \right) \nonumber
\\
&=  \mu_0 c \left(\vec{j}_{\ext}(x) + \vec{J}(x) \right),
\end{align}
for a fixed initial state and external pair $(a^k_{\ext},j^{k}_{\ext})$, where $\hat{q}^{k}_{\mathrm{p}}$ is the first term on the right hand side of Eq.~(\ref{NonRelEOM}), $\hat{q}^{k}_{\mathrm{M}}$ corresponds to the second term and $\hat{q}^{k}_{0}$ corresponds to the third. The initial state and the fixed external pair $(a^k_{\ext},j^{k}_{\ext})$ determine the initial conditions for the above coupled equations. The explicit appearance of the external potential in several terms in the equation of motion and in the initial condition is due to the non-relativistic limit. The main advantage of this limit is, that we do no longer need an explicit approximation for functionals in the Maxwell equation, since we now consider the current directly. 


In a next step we can then perform the non-relativistic limit of the Kohn-Sham scheme of Eqs.~(\ref{IntPolFunKS}) and (\ref{IntPotFunKS}) which leads to
\begin{align*}
&J_{0}([\Phi_0,J_n,A_{n}];x) a_{\mathrm{KS}}^{k}(x) 
\\
&= J_{0}([\Psi_0,J_n,A_{n}];x) a_{\ext}^{k}(x) + \langle \A^{k} \J_0 \rangle ([\Psi_0,J_n,A_{n}];x) \nonumber
\\
&+ \frac{mc}{e}\left( J_{\mathrm{p}}^{k}([\Phi_0,J_n,A_{n}];x) -  J_{\mathrm{p}}^{k}([\Psi_0,J_n,A_{n}];x) \right) \nonumber
\\
&+ \frac{mc}{e}\epsilon^{klj}\partial_l\left( M_{j}([\Psi_0,J_n,A_{n}];x) -M_{j}([\Phi_0,J_n,A_{n}];x) \right) \nonumber
\\
&j_{\mathrm{KS}}^{k}(x) =  j^{k}_{\ext}(x) + J^{k}(x). \nonumber
\end{align*}
If we then further assume that the different initial states fulfill
\begin{align*}
 \braket{\Psi_0|\J_{0}(\vec{r})}{\Psi_0} = \braket{\Phi_0|\J_{0}(\vec{r})}{\Phi_0},
\end{align*}
(due to the continuity equations the zero components stay equivalent) we can define the so-called Hartree-exchange-correlation (Hxc) potential by
\begin{align*}
 \vec{a}_{\mathrm{KS}}[\Psi_0,\Phi_0, J_k, A_{k}, a^k_{\ext}] = \vec{a}_{\ext} + \vec{a}_{\mathrm{Hxc}}[\Psi_0,\Phi_0, J_k, A_{k}], 
\end{align*}
and we end up with
\begin{align*}
&J_{0}(x) a_{\mathrm{Hxc}}^{k}(x) = \langle \A^{k} \J_0 \rangle ([\Psi_0,J_n,A_{n}];x)
\\
&+ \frac{mc}{e}\left( J_{\mathrm{p}}^{k}([\Phi_0,J_n,A_{n}];x) -  J_{\mathrm{p}}^{k}([\Psi_0,J_n,A_{n}];x) \right) \nonumber
\\
&+ \frac{mc}{e}\epsilon^{klj}\partial_l\left( M_{j}([\Psi_0,J_n,A_{n}];x) -M_{j}([\Phi_0,J_n,A_{n}];x) \right).  \nonumber
\end{align*}
Thus, assuming that we have given an appropriate initial state of the form $\ket{\Phi_0}= \ket{\mathrm{M}_0}\otimes \ket{\mathrm{EM}_0}$ that has the same initial current, initial potential and electric field (corresponding to the first time-derivative of the potential) we can solve simultaneously
\begin{align}
\label{NonRelKSI}
& \imagi \hbar c \partial_0 \ket{\mathrm{M}(t)} = \left[\hat{H}_{\mathrm{M}} - \frac{1}{c}\int \diff^3 r \hat{\vec{J}}(x) \cdot \vec{a}_{\mathrm{KS}}(x) \right. 
\\
& \left. \hspace{2.5cm}- \frac{e}{2mc^3}\int \diff^3 r \hat{J}_0(\vec{r})\vec{a}_{\mathrm{KS}}^{2}(x) \right]\ket{\mathrm{M}(t)}, \nonumber
\\
\label{NonRelKSII}
&\Box A^{k}(x) + \partial^{k} \left(\frac{1}{c} \int \diff^3 r' \frac{\vec{\nabla}'\cdot\vec{j}_{\ext}(x') + \vec{\nabla}'\cdot\vec{J}(x')}{4 \pi \epsilon_0 |\vec{r} - \vec{r}'|}  \right) \nonumber
\\
& \qquad=  \mu_0 c \left(j^{k}_{\ext}(x) + J^{k}(x) \right).
\end{align}
If we further assume that the initial state $\ket{\mathrm{M}_0}$ is given as a Slater determinant of orbitals $\varphi(\vec{r})$, we can solve single-orbital Kohn-Sham equations. The simplest approximate Hxc potential is just the non-relativistic limit of the mean-field approximation of Eq.~(\ref{RelMFApprox}), i.e.
\begin{align*}
 \vec{a}_{\mathrm{Hxc}}(x) = \vec{A}(x).
\end{align*}
Note again, that without a further gauge transformation we now also have a scalar potential in the Kohn-Sham Hamiltonian due to $A_{0}$.

We point out, that we could alternatively use Eq.~(\ref{NonRelEOM}) directly to show the one-to-one correspondence between the external pair $(a^k_{\ext},j^{k}_{\ext})$ and the non-relativistic internal pair $(J_k,A_{k})$ \cite{vignale-2004}. However, besides being more involved, also the connection to relativistic QEDFT becomes less clear. Nevertheless, for the construction of approximations to the Kohn-Sham potential, Eq.~(\ref{NonRelEOM}) seems better suited, since it is a more explicit equation.


\subsection{QEDFT for approximate non-relativistic theories}

\label{subsec:approxQEDFT}

In this subsection we show how, by introducing further approximations, we can find a family of non-relativistic QEDFTs, which in the lowest-order approximation leads to the model QEDFT of Sec.~\ref{sec:Model}.

As pointed out before, in the non-relativistic situation the initial guess for the conjugate pairs, i.e. by identifying a Legendre-type transformation in the Lagrangian of the problem, holds true. Thus we can now derive all sorts of approximate QEDFTs by investigating different conserved currents and restrictions to the photonic degrees of freedom. In the currents this holds since approximating the conserved current $J_k$ implies approximating the Hamiltonian in Eq.~(\ref{NonRelHam}) accordingly. Thus, e.g., by assuming a negligible magnetic density $M_{l}(x) \approx 0$, i.e.
\begin{align*}
 \J_{k}(x) = \J_k^{\mathrm{p}}(\vec{r}) - \frac{1}{mc^2} \J_{0}(\vec{r}) \A_{k}^{\mathrm{tot}}(x),
\end{align*}
the according Hamiltonian as well as the defining Eqs.~(\ref{NonRelEOM}) and (\ref{NonRelGordon}) change. Actually, all terms $\hat{M}_l$ and $\hat{q}^{\mathrm{M}}_l$ vanish in these equations for this approximation. We again find due to the corresponding Eq.~(\ref{NonRelContra}) that we have
\begin{align}
\label{NonRelPairMappingII}
(a^{k}_{\ext},j^{k}_{\ext})\; {\buildrel\rm 1:1 \over \leftrightarrow }  \; (J_{k}, A_{k}), 
\end{align} 
and we can consider the corresponding coupled Eqs.~(\ref{NonRelCurrentFunBasic}) and (\ref{NonRelPotFunBasic}). The Kohn-Sham current becomes accordingly $j^k_{\mathrm{KS}} = j^k_{\ext} + J^k$ and the Hxc potential in this limit reduces to
\begin{align*}
J_{0}(x)& a_{\mathrm{Hxc}}^{k}(x) = \langle \A^{k} \J_0 \rangle ([\Psi_0,J_k,A_{k}];x)
\\
&+ \frac{mc}{e}\left( J_{\mathrm{p}}^{k}([\Phi_0,J_k,A_{k}];x) -  J_{\mathrm{p}}^{k}([\Psi_0,J_k,A_{k}];x) \right). \nonumber
\end{align*}

On the other hand we can also restrict the allowed photonic modes. For instance we can assume a perfect cubic cavity (zero-boundary conditions) of length $L$ \footnote{Actually also other boundaries are possible, but then the expansion in accordance to the Coulomb-gauge condition in the eigenfunctions of the Laplacian becomes more involved.}. Then, with the allowed wave vectors $\vec{k}_{\vec{n}} = \vec{n} (\pi / L)$ and the corresponding dimensionless creation and annihilation operators $\ad_{\vec{n},\lambda}$ and $\a_{\vec{n},\lambda}$ (see appendix (\ref{app:Hubbard}) for more details) we find
\begin{align*}
 \A_{k}(\vec{r})= \sqrt{\frac{\hbar c^2}{\epsilon_0}} \sum_{\vec{n},\lambda} \frac{\epsilon_{k}(\vec{n},\lambda)}{\sqrt{2 \omega_n}} \left[\a_{\vec{n},\lambda}  + \ad_{\vec{n},\lambda}  \right] \mathcal{S}(\vec{n}\cdot{\vec{r}}) ,
\end{align*}
where the mode function $\mathcal{S}$ is given in Eq.~(\ref{ModeFunction}). If we further restrict the modes by introducing a square-summable regularization function $f_{\mathrm{EM}}(\vec{n})$ \footnote{In the case of continuous frequencies one accordingly uses a square-integrable function.}, e.g. $f_{\mathrm{EM}}=1$ for $|\vec{n}| < mcL/(2 \pi \hbar)$ (energy smaller than rest-mass energy) and $0$ otherwise, the resulting regularized field
\begin{align}
\label{CavityPotential}
 \A_{k}(\vec{r})=  \sqrt{\frac{\hbar c^2}{\epsilon_0}} \sum_{\vec{n},\lambda} f_{\mathrm{EM}}(\vec{n})\frac{\epsilon_{k}(\vec{n},\lambda)}{\sqrt{2 \omega_n}} \left[\a_{\vec{n},\lambda}  + \ad_{\vec{n},\lambda}  \right] \mathcal{S}(\vec{n}\cdot{\vec{r}})
\end{align}
makes the coupled Pauli-Fierz Hamiltonian self-adjoint without any further renormalization procedure \cite{hiroshima-2002}. Such a restriction is assumed in the following. These approximations are then directly reflected in the Hamiltonian and the derived equations of motion. While the basic Eq.~(\ref{NonRelGordon}) does not change, and thus $J_k$ is the basic matter-variable, the basic equation of motion for the potential $A_k$ has to reflect the restriction to specific modes. By multiplying Eq.~(\ref{NonRelPotFunBasic}) from the left by 
\begin{align*}
 \epsilon_{k}(\vec{n}, \lambda) \mathcal{S}(\vec{n} \cdot \vec{r} )
\end{align*}
and integrating we find the mode expansions
\begin{align}
\label{ModeExpansion}
\sqrt{\frac{\hbar c^2}{\epsilon_0}}&\frac{f_{\mathrm{EM}}(\vec{n})}{\sqrt{2 \omega_n}} \left(\partial_0^2 + \vec{k}_{\vec{n}}^2  \right) q_{\vec{n}, \lambda}(t) 
\\
&=  \mu_0 c\left(j_{\vec{n}, \lambda}^{\ext}(t) + J_{\vec{n}, \lambda}(t) \right), 
\end{align}
where $\hat{q}_{\vec{n}, \lambda} = \a_{\vec{n},\lambda}  + \ad_{\vec{n},\lambda} $ and we use the definition
\begin{align*}
 j_{\vec{n}, \lambda}^{\ext}(t) = \int \diff^3 r \; \vec{\epsilon}(\vec{n}, \lambda)\cdot \vec{j}_{\ext}(x) \, \mathcal{S}( \vec{n} \cdot \vec{r}).
\end{align*}
The Coulomb part vanishes since we employ a partial integration and the fact that $\vec{\epsilon}(\vec{n}, \lambda) \cdot \vec{n} = 0$. Of course, one finds the same equations by a straightforward calculation of the Heisenberg equation of motion for the Maxwell-field (\ref{CavityPotential}) with the according Pauli-Fierz Hamiltonian (\ref{NonRelHam}). From the restriction to specific modes the field $A_k$ is restricted in its spatial form and therefore the photonic variable changes from $A_k$ to the set of mode expectation values 
\begin{align*}
 A_{k}(x) \rightarrow \left\{ A_{\vec{n}, \lambda}(t) \right\}.
\end{align*}
This change in basic variable is also reflected in the conjugate external variable which is given via Eq.~(\ref{ModeExpansion}) as
\begin{align*}
j_{\vec{n}, \lambda}^{\ext}(t) =  \frac{f_{\mathrm{EM}}(\vec{n}) \epsilon_0 }{\sqrt{2 \hbar \omega_n}} \left(\partial_0^2 + \vec{k}_{\vec{n}}^2  \right) q_{\vec{n}, \lambda}(t) - J_{\vec{n}, \lambda}(t).\nonumber
\end{align*}
Thus we accordingly find
\begin{align*}
 j^{k}_{\ext}(x) \rightarrow \left\{ j_{\vec{n}, \lambda}^{\ext}(t) \right\},
\end{align*}
and the conjugate pairs become
\begin{align*}
(a^{k}_{\ext},\left\{ j_{\vec{n}, \lambda}^{\ext} \right\})\; {\buildrel\rm 1:1 \over \leftrightarrow }  \; (J_{k}, \left\{ A_{\vec{n}, \lambda} \right\}). 
\end{align*} 
Thus we have to solve the mode Eqs.~(\ref{ModeExpansion}) together with the according equation of motion for the current. Correspondingly also the Kohn-Sham scheme and the mean-field approximation for $\vec{a}_{\mathrm{Hxc}}$ change to its mode-equivalents.

If we then also employ the dipole-approximation, i.e. we assume that the extension of our matter system is small compared to the wavelengths of the allowed photonic modes, we find
\begin{align}
\label{DipolePotential}
 \A_{k} =  \sqrt{\frac{\hbar c^2}{L^3 \epsilon_0 }} \sum_{\vec{n},\lambda} f_{\mathrm{EM}}(\vec{n})\frac{\epsilon_{k}(\vec{n},\lambda)}{\sqrt{2 \omega_n}} \left[\a_{\vec{n},\lambda}  + \ad_{\vec{n},\lambda}  \right].
\end{align}
This only changes the definition of effective currents that couple to the modes, i.e.  
\begin{align*}
 j_{\vec{n}, \lambda}^{\ext}(t) = \int \frac{\diff^3 r}{L^{3/2}} \; \vec{\epsilon}(\vec{n}, \lambda)\cdot \vec{j}_{\ext}(x),
\end{align*}  
but leaves the structure of the QEDFT reformulation otherwise unchanged. If we assume the magnetization density $M_l$ to be negligible, we have from first principles rederived the QEDFT formulation presented in \cite{tokatly-2013}. In this work the situation of only scalar external potentials, i.e. $\vec{a}_{\ext}=0$ and $a^{0}_{\ext} \neq 0$, has been considered as a second case. In this situation, the gauge freedom is only up to a spatial constant, which is usually fixed by choosing $a^0_{\ext} \rightarrow 0$ for $|\vec{r}| \rightarrow \infty$. Since $a^{0}_{\ext}$ couples to the zero component of the current, i.e. the density $\J_0$, the conjugate pair becomes 
\begin{align*}
(a_{\ext}^{0},\left\{ j_{\vec{n}, \lambda}^{\ext} \right\})\; {\buildrel\rm 1:1 \over \leftrightarrow }  \; (J_{0}, \left\{ A_{\vec{n}, \lambda} \right\}). 
\end{align*} 
To demonstrate this mapping, the first time derivative of $\J_0$ is obviously not enough, since this amounts to the continuity equation and no direct connection between the two conjugate variables of the matter part of the quantum system is found. Therefore, one has to go to the second time derivative of $\J_0$ \cite{tokatly-2013}. If we then further simplify this physical situation (see appendix (\ref{app:Hubbard}) for a detailed derivation) we find the model Hamiltonian of Sec.~\ref{sec:Model}. In a similar manner, by imposing the restrictions on the corresponding equations of motion, we can rederive the model QEDFT of Sec.~\ref{subsec:FoundModel} \footnote{We note, that one could have also derived the model QEDFT by employing the gauge of Eq.~(\ref{GaugeCondition}) for the external potentials. By applying the dipole approximation also to the external vector potential the conjugate variable becomes the density (dipole moment). For clarity of presentation, though, we have chosen to start from the scalar-
potential case.}. 

Finally, for a simple overview, we have collected the different QEDFTs that we have explicitly considered in this work in appendix~\ref{app:overview}.     



\section{Conclusion and outlook}
\label{sec:Conclusion}

In this work we have shown how one can extend the ideas of TDDFT to quantized coupled matter-photon systems. We have first explained the basic ideas of QEDFT for a model system of a two-site Hubbard model coupled to a single photonic mode. By rewriting the problem in terms of an effective theory for a pair of internal functional variables and proving the uniqueness of solutions for the resulting non-linear coupled equations, we have demonstrated how an explicit solution for the coupled photon-matter wave function can be avoided. Further we have discussed how an auxiliary quantum system, the so-called Kohn-Sham system, can be used to construct approximations for the implicit functionals appearing in the effective equations. The Kohn-Sham construction gives rise to effective fields and effective currents, which are termed Kohn-Sham potential and Kohn-Sham current, respectively. By numerically constructing the exact Kohn-Sham potential and Kohn-Sham currents, we have 
illustrated the capability 
of this new approach to exactly 
describe the dynamics of coupled matter-photon systems and contrasted these exact fields with the mean-field approximation.

In the following, instead of reformulating every possible approximate treatment of coupled matter-photon systems seperately, we have shown how these QEDFTs for approximate Hamiltonians are merely approximations to relativistic QEDFT, which itself is based on QED. To avoid problems with the Kohn-Sham construction, we have based relativistic QEDFT on the expectation value of the polarization and the vector potential of the quantum system. By then performing the non-relativistic limit of QEDFT we have demonstrated that the resulting theory is the QEDFT reformulation of the Pauli-Fierz Hamiltonian. The non-relativistic limit automatically makes the (non-relativistic) current the basic variable for the matter system. Accordingly, the non-relativistic limit of the Kohn-Sham potentials and currents leads to the corresponding Kohn-Sham fields for the Pauli-Fierz Hamiltonian. By performing further approximations for non-relativistic QEDFT, e.g. assuming the magnetic density negligible, we have shown how other QEDFTs (
that reformulate the corresponding approximate Hamiltonians) can be derived. Depending on the level of approximation, the basic internal functional variables change, e.g. if we confine the electromagntic field with a cavity, the (allowed) mode expectation values become the new internal variable of the photons. In a final step we restricted to a two-site model coupled to only one mode, recovering the model QEDFT of the beginning.

We point out, that at every level of QEDFT we recover the corresponding (standard) time-dependent density-functional reformulations \cite{runge-1984, vignale-2004} if we assume the quantized nature of the photons negligible, i.e. the charged particles interact via the classical Coulomb interaction only. This will be the case in most standard situation of condensed matter theory, e.g. when investigating dynamics of atoms or molecules in free space. However, we expect that interesting effects happen when the boundary conditions for the Maxwell field are changed, e.g. for atoms in a cavity. Thus we have a potential tool that can treat complex electronic systems in the setting of quantum optics. Also, we can investigate the explicit interplay of photons with molecules or nanostructures, e.g. in nano-plasmonics. However, for this theory to be practical we are in need of reliable approximations to the basic functionals. In \cite{pellegrini-2014} functionals based on an optimized effective potential approach \cite{
ullrich-TDDFT, marques-TDDFT} are constructed, which provide good results even in the situation of strongly-coupled systems. Although the currently available approximations have only been tested for simple model systems, the hierarchy of QEDFT approximations allows to simply scale up these functionals to more complex situations. Thus we can develop approximations for simple systems, e.g. only one mode couples to the matter system, and then extend these approximations to more involved problems, e.g. considering more modes. In this way we can easily control the validity of our approximations. In this respect, we are also working on a fixed-point approach in the spirit of \cite{ruggenthaler-2011, nielsen-2013}, which allows us to construct the exact Kohn-Sham potentials and compare the approximate potentials to the (numerically) exact expressions. Details of this approach will be part of a forthcoming publication \cite{flick-2014}. On the other hand, the fixed-point approach is also a way to 
extend the validity of QEDFT beyond Taylor-expandable fields. A different way, especially for discretized matter systems, is the non-linear Schr\"odinger equation approach introduced in \cite{tokatly-2011, farzanehpour-2012}. 
Certain theoretical and mathematical details of the model QEDFT of Sec.~\ref{sec:Model}, that are beyond the scope of the current manuscript, will be discussed in \cite{farzanehpour-2014}. 

Finally, since we are aiming at investigating quantum optical settings, we also need to discuss the cavity and the problem of open quantum systems. In this work we focused on closed systems and on a perfect (cubic) cavity. It is straightforward (but tedious) to extend the current work to an arbitrary shape of the perfect cavity. We have to use an expansion of the photon field in the according eigenfunctions of the cavity, such that these modes obey the Coulomb-gauge condition. However, in actual quantum optical experiments, the cavities are not perfect but rather an open quantum system, which allows for an exchange with the environment. To take care of this channel of decoherence there are several possible ways. One can employ the current formulation of QEDFT and derive a master equation, as has also been done for standard TDDFT \cite{yuen-2009,yuen-2010}. Also extensions to stochastic equations \cite{dagosta-2007,appel-2009,appel-2011} are possible. On the other hand, one 
can couple further bosonic degrees of freedom to the system and prescribe a bath spectral density, making these degrees of freedom a bath for the system \cite{tokatly-2013}.  Since the present framework allows for a consistent treatment of interacting fermionic and bosonic particles, the inclusion of a bath and coupling to other fields, e.g. phonons, will be the subject of future work.


\section{Acknowledgment}

MR acknowledges useful discussions with C.\ Genes and H.\ Ritsch and financial support by the Austrian Science Fonds (FWF projects J 3016-N16 and P 25739-N27).
We further acknowledge financial support from the European Research Council Advanced
Grant DYNamo (ERC-2010-AdG-267374), Spanish Grant (FIS2010-21282-C02-01),
Grupos Consolidados UPV/EHU del Gobierno Vasco (IT578-13), Ikerbasque and the
European Commission projects CRONOS (Grant number 280879-2 CRONOS CP-FP7).


\appendix


\section{Conventions}
\label{app:con}

In this work we employ the standard covariant notation $x^{\mu} = (ct, \vec{r})$ with greek letters indicating four vectors, e.g. $\mu \in \{0,1,2,3 \}$, and roman letters indicating spatial vectors, i.e. $k \in \{1,2,3\}$. To lower (or raise) the indices, i.e. going from contravariant vectors to covariant vectors (or vice versa), we adopt the convention
\begin{align*}
 g_{\mu \nu} =
 \begin{pmatrix}
  1 & 0 & 0 & 0 \\
  0 & -1 & 0 & 0 \\
  0  & 0  & -1 & 0  \\
  0 & 0 & 0 & -1
 \end{pmatrix}
\end{align*}
for the Minkwoski metric. We denote spatial (contravariant) vectors with the vector-symbol, i.e. $A^{k} \equiv \vec{A}$, and the derivatives with respect to the space-time vectors $x^{\mu}$ by $\partial_{\mu} = \partial/ \partial x^{\mu}$. With these definitions the divergence can be written as $\partial_k A^{k} = \vec{\nabla} \cdot \vec{A}$, where we also adopt the Einstein summation convention. Further we note that $J_k A^{k} = - \vec{J} \cdot \vec{A}$. With the help of the Levi-Civita symbol $\epsilon^{ijk}$ we can write the curl as $\epsilon^{ijk} \partial_j A_k \equiv - \vec{\nabla} \times \vec{A}$ and the multiplication of Pauli matrices becomes $\sigma^{k}\sigma^l = (1/2)( \{\sigma^{k}, \sigma^{l} \} + [\sigma^{k}, \sigma^{l} ]) = -g^{kl}-\imagi \epsilon^{klm} \sigma_{m}$.

Further, for notational simplicity we only point out in the text (when necessary), whether we are in the Schr\"odinger or Heisenberg picture, and do not explicitly indicate the picture used in the operators. In the Schr\"odinger picture, operators which are not explicitly time-dependent only carry a purely spatial dependence, e.g. $\A^{k}(\vec{r})$. We indicate explicit time-dependence in the Schr\"odinger picture by either carring the full space-time dependence, e.g. $\J^{k}(x)$ (for the Pauli-Fierz current density) of Eq.~(\ref{NonRelCurrent}), or by a dependence on $t$, e.g. $\hat{H}(t)$ in Eq.~(\ref{NonRelHam}). In the Heisenberg picture, every operator also depends on time, e.g. $\spin(x)$ in Eq.~(\ref{QuantizedDirac}).


\section{Quantum Electrodynamics in Coulomb gauge}
\label{app:Coulomb}

In this appendix we give a detailed derivation of QED in Coulomb gauge. We start from the (classical) coupled QED Lagrangian with external fields  $a_{\mu}^{\mathrm{ext}}(x)$ and $j_{\mu}^{\mathrm{ext}}(x)$ given by \cite{greiner-FQ} 
\begin{align}
\label{QEDLagrangian}
 \mathcal{L}_{\mathrm{QED}}(x) = & \; \mathcal{L}_{\mathrm{M}}(x) - \frac{1}{c} J^{\mu}(x) a_{\mu}^{\mathrm{ext}}(x)
\\
&+ \; \mathcal{L}_{\mathrm{E}}(x) - \frac{1}{c} \left(J_{\mu}(x) + j_{\mu}^{\mathrm{ext}}(x) \right)A^{\mu}(x)   \nonumber
\end{align}
Here we use the standard definitions for the (classical) Dirac fields, i.e.
\begin{align*}
 \mathcal{L}_{\mathrm{M}}(x) = \bar{\psi}(x)\left( \imagi \hbar c \gamma^\mu \partial_u - m c^2 \right) \psi(x),
\end{align*}
where 
\begin{align*}
 \psi(x) = \left(\phi(x) \atop \chi(x)  \right)
\end{align*}
is a Dirac four-spinor with the two-component (spin) functions $\phi(x)$ and $\chi(x)$, the Gamma matrices are given by
\begin{align*}
\gamma^i =
 \begin{pmatrix}
  0 & \sigma^i \\
  -\sigma^i & 0
 \end{pmatrix},
\;\gamma^0 =
 \begin{pmatrix}
  \mathds{1} & 0 \\
  0 & - \mathds{1}
 \end{pmatrix},
\end{align*}
with $\sigma^i$ the usual Pauli matrices, $\bar{\psi} = \bar{\psi} \gamma^0$ and
\begin{align*}
 J_{\mu}(x) = e c \bar{\psi}(x)\gamma_{\mu} \psi(x),
\end{align*}
is the conserved (Noether) current. Further we use the Minkowski metric $g_{\mu \nu}=(+,-,-,-)$ to raise and lower the indices. For the (classical) Maxwell field we have
\begin{align}
\label{photonLagrangian}
 \mathcal{L}_{\mathrm{E}}(x) = -\frac{\epsilon_0}{4} F^{\mu \nu}(x)F_{\mu \nu}(x),
\end{align}
where $F_{\mu \nu}(x) = \partial_{\mu}A_{\nu}(x) - \partial_{\nu}A_{\mu}(x)$ is the electric field tensor and $A_{\mu}(x)$ is the vector potential. 

Now we employ the Coulomb gauge condition for the Maxwell field, i.e. $\vec{\nabla}\cdot \vec{A}(x) = 0$. Then it holds that
\begin{align}
\label{CoulombPotential}
 -\Delta A^{0}(x) = \frac{1}{\epsilon_0 c} \left(J^{0}(x) + j^{0}_{\mathrm{ext}}(x)\right),
\end{align}
where $\Delta$ is the Laplacian. If we impose square-integrability on all of $\mathbb{R}^3$ \footnote{If we consider the situation of a finite volume, e.g. due to a perfect cavity, the boundary conditions change. These different boundary conditions, in principle, change the Green's function of the Laplacian and thus the instantaneous interaction. We ignore these deviations from the Coulomb interaction in this work for simplicity.} the Green's function of the Laplacian becomes $\Delta^{-1}= 1/(4 \pi |\vec{r}-\vec{x'}|)$ and therefore
\begin{align}
\label{A0}
 A^0(x) = \frac{1}{c} \int \diff ^3 x' \,  \frac{J^{0}(x') + j^{0}_{\mathrm{ext}}(x')}{ 4 \pi \epsilon_0 |\vec{r}-\vec{x'}|} .
\end{align}
Since the zero component of the four potential $A_{\mu}(x)$ is given in terms of the full current, it is not subject to quantization. The conjugate momenta of the photon field (that need to be quantized) are the same as in the current-free theory and thus the usual canonical quantization-procedure applies \cite{greiner-FQ}, i.e.
\begin{align}
\label{PerpDelta}
 \left[ \A_k(\vec{r}), \epsilon_0 \E_{l}(\vec{r}') \right] = -\imagi \hbar c \delta_{kl}^{\perp}(\vec{r} -\vec{r}'),
\end{align}
where $\E_{k}$ is the electric field operator, $\delta_{kl}^{\perp}(\vec{r} -\vec{r}') = (\delta_{kl}- \partial_k \Delta^{-1} \partial_l)\delta^{3}(\vec{r} -\vec{r}')$ is the transverse delta-function and $k,l$ are spatial coordinates only. Equivalently we can define these operators by their respective plane-wave expansions 
\begin{align*}
 &\hat{\vec{A}}(\vec{r}) \!\!= \!\! \sqrt{\frac{\hbar c^2}{\epsilon_0}} \!\! \int \!\! \frac{\diff^3 k}{\sqrt{2 \omega_k (2 \pi)^3}} \!\!\sum_{\lambda=1}^2 \! \vec{\epsilon}(\vec{k}, \lambda) \!\! \left[\a_{\vec{k},\lambda} e^{\imagi \vec{k}\cdot\vec{r}}\! +\! \ad_{\vec{k},\lambda} e^{- \imagi \vec{k}\cdot \vec{r}}  \right],
\\
 &\hat{\vec{E}}(\vec{r}) \!\!=\!\! \sqrt{\frac{\hbar}{\epsilon_0}}\!\! \int \!\! \frac{\diff^3 k \, \imagi \omega_{k}}{\sqrt{2 \omega_k (2 \pi)^3}} \!\!\sum_{\lambda=1}^2 \!\! \vec{\epsilon}(\vec{k}, \lambda) \!\! \left[\a_{\vec{k},\lambda} e^{\imagi \vec{k}\cdot \vec{r}} \!-\! \ad_{\vec{k},\lambda} e^{- \imagi \vec{k}\cdot \vec{r}}  \right],
\end{align*}
where $\omega_k = c k$, $\vec{\epsilon}(\vec{k}, \lambda)$ is the transverse-polarization vector \cite{greiner-FQ}, and the annihilation and creation operators obey
\begin{align*}
 \left[\a_{\vec{k}',\lambda'}, \ad_{\vec{k},\lambda}\right] = \delta^3 (\vec{k}-\vec{k}') \delta_{\lambda \lambda'}.
\end{align*}
If we further define the magnetic field operator by $c \hat{\vec{B}} = \vec{\nabla}\times \hat{\vec{A}}$, the Hamiltonian corresponding to $\mathcal{L}_{\mathrm{E}}$ is given in Eq.~(\ref{MaxwellHam}). We used normal ordering to get rid of the infinite zero-point energy in this expression. Also for the Dirac field, the coupling does not change the conjugate momenta. Therefore we can perform the usual canonical quantization procedure for Fermions which leads to the (equal-time) anti-commutation relations \cite{greiner-FQ}
\begin{align*}
 \{ \spin_{\alpha}(\vec{r}), \dspin_{\beta}(\vec{r}') \} = \gamma^{0}_{\alpha \beta} \delta^{3}(\vec{r} -\vec{r}').
\end{align*}
The Hamiltonian corresponding to $\mathcal{L}_{\mathrm{M}}$ thus becomes the one of Eq.~(\ref{DiracHam}), where we used $\vec{r} \cdot \vec{y} = - x_{k} y^{k}$.

It is straightforward to give the missing terms of the QED Hamiltonian due to the coupling to the external fields as well as due to the coupling between the quantized fields. If we apply the definition of the quantized current $\hat{J}_{\mu}$ of Eq.~(\ref{4current}) to Eq.~(\ref{A0}) we find (using normal ordering, i.e. rearranging the annihilation parts of the operators to the right, to discard the corresponding zero-point energy)
\begin{align*}
 :\!\! \left(J_{0}(\vec{r}) \! + \! j_{0}^{\mathrm{ext}}(x) \right)& \! A^{0}(x) \!:=\! \frac{1}{2 c} \!\! \int \!\! \frac{\diff^{3}r \, \diff^{3} r'}{4 \pi \epsilon_0 |\vec{r} -\vec{r}'|} \left( j_0^{\mathrm{ext}}(x) j^0_{\mathrm{ext}}(x') \right. \nonumber
\\
& \left.+ 2\J_0(\vec{r}) j^0_{\mathrm{ext}}(x')  + :\J^0(\vec{r}) \J_0(\vec{r}'):\right).
\end{align*}
Here, if we disregard the purely multiplicative first term on the right-hand side, we arrive at Eq.~(\ref{CoulombHam}). The rest is given in Eqs.~(\ref{ExternalHam}) and (\ref{InteractionHam}). Alternatively, with the definition of $A^{0}$ in Eq.~(\ref{A0}) and
\begin{align*}
\hat{H}'_{\mathrm{int}}(t) = \hat{H}_{\mathrm{int}} + \frac{1}{c} \int \diff^3 r : \hat{J}_0 (\vec{r}) A^{0}(x): 
\end{align*}
the full QED Hamiltonian can also be written as
\begin{align}
\label{QEDHam'}
 \hat{H}(t)&=  \hat{H}_{\mathrm{M}} + \hat{H}_{\mathrm{E}} + \hat{H}'_{\mathrm{int}}(t) 
\\
&+ \frac{1}{c}\int \diff^3 x \left(\hat{J}_{\mu}(\vec{r}) a^{\mu}_{\ext}(x) + \hat{A}_{\mu}(\vec{r}) j^{\mu}_{\ext}(x) \right). \nonumber
\end{align}


\section{Non-relativistic equations of motion}
\label{app:NonRelEOM}

To find the non-relativistic limit of Eq.~(\ref{HeisenbergCurrent}), we cannot straightaway apply the decoupling to Eq.~(\ref{Decoupling}). Since we have to apply the decoupling consistently to the Hamiltonian as well as the current we need to rewrite the equation of motion. We start (in the Heisenberg picture) by
\begin{align*}
&\imagi \partial_0 \left[ e c \spin^{\dagger}\gamma^0 \gamma^k \spin \right] = \frac{2 e mc ^2}{\hbar} \left[ \c^{\dagger} \sigma^k \p - \p^{\dagger} \sigma^k \c \right]
\\
& - \imagi e c \left[ \p^{\dagger} \left( \sigma^{k}\sigma^{l}\partial_l + \overset{\smash{\raisebox{-1.5pt}{\tiny$\leftarrow$}}}{\partial}_{l} \sigma^{l} \sigma^{k}  \right)\p + \c^{\dagger} \left( \sigma^{k}\sigma^{l}\partial_l + \overset{\smash{\raisebox{-1.5pt}{\tiny$\leftarrow$}}}{\partial}_{l} \sigma^{l} \sigma^{k}  \right)\c  \right] \nonumber
\\
&- \frac{2 \imagi e^2}{\hbar} \epsilon^{klj} \A_{l}^{\mathrm{tot}} \left[\p^{\dagger} \sigma_j \p + \c^{\dagger}\sigma_{j} \c  \right]. \nonumber
\end{align*}
This leads with $\sigma^l \sigma^k = -g^{lk} - \imagi \epsilon^{lkj} \sigma_j$ and $\Im\{\p^{\dagger}\A^{k}_{\mathrm{tot}}  \p\} \equiv 0$ to
\begin{align*}
 &\imagi \hbar \partial_0 \J^{k} = 2 \Im \left\{ -2 e m c^2 \c^{\dagger} \sigma^k \p  +e^2  \A_{l}^{\mathrm{tot}} \left[\p^{\dagger} \sigma^k \sigma ^l \p - \c^{\dagger}\sigma^{l} \sigma^{k} \c  \right] \right.
\\
&\left. - \imagi e \hbar c \c^{\dagger} \overset{\smash{\raisebox{-1.5pt}{\tiny$\leftarrow$}}}{\partial}_{l} \sigma^{l} \sigma^{k} \c  - \imagi e \hbar c \p^{\dagger} \sigma^{k} \sigma^{l} \partial_{l}  \p \right\}. \nonumber
\end{align*}
Adding and subtracting on the right hand side the term $e \p^{\dagger} \sigma^{k}\left( \imagi \hbar  c \partial_0 - \hat{D}\right) \c$ and employing Eq.~(\ref{ChiComponent}) we find
\begin{align*}
&\imagi \hbar \partial_0 \J^{k} = 2 e \Im \left\{ \left[ \c^{\dagger} \left(- \imagi \hbar c \overset{\smash{\raisebox{-1.5pt}{\tiny$\leftarrow$}}}{\nabla} + e \hat{\vec{A}}^{\mathrm{tot}} \right) \cdot \vec{\sigma} - \p^{\dagger} e A_0^{\mathrm{tot}} \right. \right. \nonumber
\\
& \left. - \p^{\dagger}  e^2 \! \!\! \int \! \! \diff^3 r' \frac{:\!\p^{\dagger}(\vec{r}')\p(x') \! + \! \c^{\dagger}(x')\c(x')\!:}{4 \pi \epsilon_0 |\vec{r}-\vec{r}'|} - mc^2 \p^{\dagger}\right] \sigma^{k} \c \nonumber
\\
&\left. + c \p^{\dagger} \sigma^{k} \imagi \hbar c \partial_0 \c \right\}. 
\end{align*}
With the help of the definition $[...] = [\hat{D} + mc^2]^{-1}$ this can be rewritten as
\begin{widetext}
\begin{align*}
& \imagi \hbar \partial_0 \J^{k} = 2 e \im \left\{ \left[- \p^{\dagger}\left(\imagi \hbar c \overset{\smash{\raisebox{-1.5pt}{\tiny$\leftarrow$}}}{\nabla} -e \hat{\vec{A}}^{\mathrm{tot}}\right) \cdot \vec{\sigma}[...]^{\dagger} \left(\imagi \hbar c \overset{\smash{\raisebox{-1.5pt}{\tiny$\leftarrow$}}}{\nabla} -e \hat{\vec{A}}^{\mathrm{tot}}\right)\cdot \vec{\sigma} - \p^{\dagger} e A_0^{\mathrm{tot}} - \p^{\dagger}  e^2 \! \!\! \int \! \! \diff^3 r' \frac{:\!\p^{\dagger}(x')\p(x') \! + \! \c^{\dagger}(x')\c(x')\!:}{4 \pi \epsilon_0 |\vec{r}-\vec{r}'|} - mc^2 \p^{\dagger}\right] \right. \nonumber
\\
& \sigma^{k} [...] \vec{\sigma} \cdot \left(- \imagi \hbar c \vec{\nabla} - e \hat{\vec{A}}^{\mathrm{tot}}  \right) \p + \p^{\dagger} \sigma^{k}\left[\imagi \hbar c \partial_0 [...] \vec{\sigma} \cdot \left(- \imagi \hbar c \vec{\nabla} - e \hat{\vec{A}}^{\mathrm{tot}}  \right) \right] \p + \p^{\dagger} \sigma^{k}[...] \vec{\sigma} \cdot \left(- \imagi \hbar c \vec{\nabla} - e \hat{\vec{A}}^{\mathrm{tot}}  \right) \nonumber
\\
&\left. \left[  \vec{\sigma} \cdot \left(- \imagi \hbar c \vec{\nabla} - e \hat{\vec{A}}^{\mathrm{tot}}  \right) [...] \vec{\sigma} \cdot \left(- \imagi \hbar c \vec{\nabla} - e \hat{\vec{A}}^{\mathrm{tot}}  \right) + e A_0^{\mathrm{tot}} -  e^2 \! \!\! \int \! \! \diff^3 r' \frac{:\!\p^{\dagger}(x')\p(x') \! + \! \c^{\dagger}(x')\c(x')\!:}{4 \pi \epsilon_0 |\vec{r}-\vec{r}'|} - mc^2 \right] \p \right\} \nonumber
\end{align*}
\end{widetext}
Now, if we employ the approximation $[...]\approx 1/2 mc^2$ (also in the Coulomb terms) we end up with
\begin{align*}
\imagi \hbar \partial_0 \J^{k} \approx \imagi \hbar \partial_0 2 ec  \Re \left\{ \p^{\dagger} \sigma^k \frac{\vec{\sigma}}{2 m c^2} \cdot \left(-\imagi \hbar c \vec{\nabla} - e \hat{\vec{A}}^{\mathrm{tot}}  \right) \p \right\},
\end{align*}
which is just the equation of motion for the non-relativistic current (\ref{NonRelCurrent}) with the Pauli-Fierz Hamiltonian.

For the Maxwell field the non-relativistic limit of Eq.~(\ref{IntPotential}) is with the help of Eq.~(\ref{NonRelCurrent}) straightforward. It is only important to see, that this does agree with the equation of motion for $\A_{k}$ due to the Pauli-Fierz Hamiltonian (\ref{NonRelHam}). The main difference to the fully relativistic derivation is, that now we have a term of the form
\begin{align*}
 \frac{e}{2mc^2} \int\!\! \diff^3 r \, \J_{0}(x)\left(\A^{k}(x) + a^{k}_{\ext}(x)\right)\left(\A_{k}(x) + a_{k}^{\ext}(x)\right).
\end{align*}
This term does not change anything in the first order equation, i.e. $\partial_0 \A_{k} = - \E_k$. In the second order we find due to Eq.~(\ref{PerpDelta}) that
\begin{align*}
 \int \diff^3 r' \, &\left[ \E^{k}(x); \A^l(x')\A_l(x') \right] \J_{0}(x')
\\
&=2 \frac{\imagi \hbar c}{\epsilon_0} \A^{l}(x) \J_{0}(x) - 2 \frac{\imagi \hbar c}{\epsilon_0} \partial^k \Delta^{-1}\partial^l \A_l(x) \J_{0}(x) \nonumber
\end{align*}
and 
\begin{align*}
 2 \int \diff^3 r' \, & \left[ \E^{k}(x); \A^l(x') \right] a_{l}^{\ext}(x')\J_{0}(x')
\\
&=2 \frac{\imagi \hbar c}{\epsilon_0} a^{l}_{\ext}(x) \J_{0}(x) - 2 \frac{\imagi \hbar c}{\epsilon_0}  \partial^k  \Delta^{-1} \partial^l a_l^{\ext}(x) \J_{0}(x) \nonumber
\end{align*}
Now, with the above definition for $\Delta^{-1}$ used in Eq.~(\ref{A0}) we find that these commutators lead to the terms
\begin{align*}
 - \partial^{k} &\left(\frac{1}{c} \int \diff^3 r' \frac{\vec{\nabla}'\cdot\hat{\vec{A}}^{\mathrm{tot}}(x')\frac{e}{mc^2}\J_{0}(x')}{4 \pi \epsilon_0 |\vec{r} - \vec{r}'|}  \right) \nonumber
\\
 &+  \mu_0 c \left(\A^{k}_{\mathrm{tot}}(x)\frac{e}{mc^2}\J_{0}(x) \right),
\end{align*}
of the equation of motion for the Maxwell field in the non-relativistic limit. The rest of the derivation is similar to the relativistic situation.


\section{Mode expansion}
\label{app:ModeExpansion}

If we restrict the allowed space for the photonic modes we also need to impose according boundary conditions. Let us first start with a cubic cavity of length $L$ with periodic boundary condition. We then find with the allowed wave vectors $\vec{k}_{n} = \vec{n} (2 \pi / L)$ and the corresponding dimensionless creation and annihilation operators $\ad_{\vec{n},\lambda}$ and $\a_{\vec{n},\lambda}$, which are connected to their continuous counterparts by
\begin{align*}
 \lim_{L \rightarrow 0} L^{3/2} \a_{\vec{n},\lambda} \rightarrow \a_{\vec{k},\lambda}, 
\end{align*}
that
\begin{align*}
 \A_{k}(\vec{r})= \sqrt{\frac{\hbar c^2}{\epsilon_0 L^3}  } \sum_{\vec{n},\lambda} \frac{\epsilon_{k}(\vec{n},\lambda)}{\sqrt{2 \omega_n}} \left[\a_{\vec{n},\lambda}  e^{\imagi \vec{k}_n \cdot \vec{r}}  + \ad_{\vec{n},\lambda} e^{- \imagi \vec{k}_n \cdot \vec{r}} \right].
\end{align*}
Here $\omega_n = c |\vec{n}| (2\pi / L)$. If we change the conditions at the boundaries to zero-boundary conditions, then the allowed wave-vectors change to $\vec{k}_{n} = \vec{n} (\pi / L)$ and 
the discrete operators obey
\begin{align*}
 \lim_{L \rightarrow 0} (2 L)^{3/2} \imagi \a^{\dagger}_{\vec{n},\lambda} \rightarrow \a^{\dagger}_{\vec{k},\lambda}. 
\end{align*}
With the normalized mode functions
\begin{align}
\label{ModeFunction}
 \mathcal{S}(\vec{n}\cdot \vec{r}) = \left(\frac{2}{L}\right)^{3/2}\prod_{i=1}^{3} \sin\left( \frac{\pi n_i }{L} r_i \right),
\end{align}
the field operator therefore reads as
\begin{align*}
 \A_{k}(\vec{r})= \sqrt{\frac{\hbar c^2}{\epsilon_0}  } \sum_{\vec{n},\lambda} \frac{\epsilon_{k}(\vec{n},\lambda)}{\sqrt{2 \omega_n}} \left[\a_{\vec{n},\lambda} + \ad_{\vec{n},\lambda} \right] \mathcal{S}(\vec{n}\cdot \vec{r}).
\end{align*}
Here $\omega_n = c |\vec{n}| (\pi / L)$.


\section{Derivation of the model Hamiltonian}
\label{app:Hubbard}

We start with the non-relativistic Hamiltonian of Eq.~(\ref{NonRelHam}) where we assume that the magnetization density $M_k$ is negligible. We further assume a perfect cubic cavity of length $L$ and employ the dipole approximation, i.e. $e^{\pm \imagi \vec{k}_n \cdot \vec{r} } \approx 1$. Thus we find a Maxwell field defined by Eq.~(\ref{DipolePotential}). At this level of approximation our starting point coincides with that adopted in \cite{tokatly-2013}.

In a next step we allow only scalar external potentials. In the following we present a detailed derivation of the length-gauge Hamiltonian employed in \cite{tokatly-2013} for the formulation of the electron-photon TDDFT.  For simplicity we restrict our derivations to the case of one mode and one particle. The case of several modes and particles works analogously and leads to the Hamiltonian (13) of Ref.~\cite{tokatly-2013}. 

With the definition of the dimensionless photon coordinate $q$ and the conjugate momentum $\imagi \diff/ \diff q$, the single-mode vector potential is given by
\begin{align}
\label{SimplePotential}
\hat{\vec{A}} = \mathcal{C} \frac{q \vec{\epsilon} }{\sqrt{\omega}},
\end{align}
where we use the definition
\begin{align*}
 \mathcal{C} = \left(\frac{\hbar c^2}{\epsilon_0 L^3}  \right)^{1/2},
\end{align*}
and assume $f_{\mathrm{EM}} = 1$. The resulting Hamiltonian in first quantized notation reads
\begin{align}
\label{StartingHam}
\hat{H}(t) &= \frac{1}{2 m} \left(\imagi\hbar\vec{\nabla}+\frac{e}{c}\hat{\vec{A}}\right)^2 - \frac{\hbar \omega}{2} \frac{\diff^2}{\diff q^2} + \frac{\hbar \omega}{2} q^2  
\\
& + e a^0_{\ext}(x) - \frac{1}{c}\vec{j}_{\ext}(t) \cdot \hat{\vec{A}} , \nonumber
\end{align}
since at this level of approximation $\vec{\nabla}\cdot \vec{j}_{\ext} = 0$ due to the expansion in Coulomb-gauged eigenmodes. In Eq.~(\ref{StartingHam}) we introduced the notation
$$
\vec{j}_{\ext}(t) = \int \frac{\diff^3r}{L^{3/2}} \vec{j}_{\ext}(x)
$$
In a next step we transform the Hamiltonian into its length-gauge form \cite{faisal-MPT} by the unitary transformation
\begin{align*}
\hat{U} = \exp\left[\frac{\imagi}{\hbar}\left(\frac{\mathcal{C} e }{c}\frac{\vec{\epsilon} \cdot \vec{r}}{\sqrt{\omega}} q  \right)  \right].
\end{align*}
If we then perform a canonical variable transformation of the photon-coordinate $\imagi \diff/\diff q \rightarrow p$ and $q \rightarrow - \imagi \diff/\diff p$ (leaving the commutation relations unchanged) we find
\begin{align}
\label{first_stepHam}
\hat{H}(t) &= -\frac{\hbar^2}{2 m} \vec{\nabla}^2 - \frac{\hbar \omega}{2} \frac{\diff^2}{\diff p^2} +
\frac{\hbar \omega}{2} \left(p - \frac{\mathcal{C} e }{\hbar c}\frac{\vec{\epsilon} \cdot \vec{r}}{\sqrt{\omega}} \right)^2   \nonumber
\\
& + e a^0_{\ext}(x)  + \frac{\imagi \mathcal{C}}{c\sqrt{\omega}} \vec{\epsilon}\cdot \vec{j}_{\ext}(t) \frac{\diff}{\diff p}.
\end{align}

Then we perform yet another time-dependent gauge transformation 
\begin{align*}
\hat{U}(t) = \exp\left[ \frac{\imagi\mathcal{C}}{\hbar c\omega^{\frac{3}{2}}}
\left( j_{\ext}(t) p - \frac{\mathcal{C}}{2c\sqrt{\omega}}\int_0^t j_{\ext}^{2}(t')dt' \right) \right]
\end{align*}
where ${j}_{\ext}(t) = \vec{\epsilon}\cdot \vec{j}_{\ext}(t)$ is the projection of the external current on the direction of the photon polarization. The above transformation is aimed at eliminating the linear in $p$-derivative term in Eq.~(\ref{first_stepHam}). Using the general transformation rule $H\mapsto -\imagi\hbar\hat{U}^{\dagger}\partial_t \hat{U} + \hat{U}^{\dagger}\hat{H}\hat{U}$ we obtain
\begin{align}
\label{2nd_stepHam}
\hat{H}(t) &= -\frac{\hbar^2}{2 m} \vec{\nabla}^2 - \frac{\hbar \omega}{2} \frac{\diff^2}{\diff p^2} + \frac{\hbar \omega}{2}\left(p - \frac{\mathcal{C} e }{\hbar c}\frac{\vec{\epsilon} \cdot \vec{r}}{\sqrt{\omega}} \right)^2   \nonumber
\\
& + e a^0_{\ext}(x) - \frac{1}{\omega c} \frac{\mathcal{C}}{\sqrt{\omega}} p\, \partial_{t}j_{\ext}(t).
\end{align}
Here we see, that the photonic variable $p$ is coupled to the dipole moment $e \vec{r}$, which indicates that $p$ is actually proportional to the electric field. 

In a last step we then discretize the matter-part of the problem and employ a two-site approximation such that
\begin{align*}
-\frac{\hbar^2}{2 m} \vec{\nabla}^2 &\rightarrow -t_{\mathrm{kin}} \s_{x},
\\
e \omega \vec{\epsilon}\cdot\vec{r}& \rightarrow e \omega \vec{\epsilon}\cdot\vec{l}\s_z
\equiv e \J,
\\
e a^0_{\ext}(x) &\rightarrow e a^0_{\ext}(t)\s_z,
\end{align*}
where $t_{\mathrm{kin}}$ is the kinetic (hopping) energy, $\vec{l}$ is the vector connecting two sites, and $a^0_{\ext}(t)$ corresponds to the potential difference between the sites. To highlight the general structure of the photon-matter Hamiltonian (and bring it to the form used in Sec.~\ref{sec:Model}) we also redefine the external current, the external potential, and the photon field as follows
\begin{align*}
\partial_t j_{\ext}(t) &\rightarrow \omega\tilde{j}_{\ext}(t),
\\
e a^0_{\ext}(t)\s_z& \rightarrow - \frac{1}{c}a_{\ext}(t)\J,
\\
\frac{\mathcal{C}}{\sqrt{\omega}} p &\rightarrow \hat{A} 
= \frac{\mathcal{C}}{\sqrt{2\omega}}(\ad + \a).
\end{align*}
After implementing the above redefinitions in Eq.~(\ref{2nd_stepHam}) and neglecting irrelevant constant terms we arrive at the following Hamiltonian 
\begin{align}
\label{Final2site-Ham}
\hat{H}(t) &= -t_{\mathrm{kin}} \s_{x} + \hbar \omega \ad \a -\frac{1}{c} \J \hat{A} 
- \frac{1}{c}a_{\ext}(t)\J  \nonumber
\\
&- \frac{1}{c} \tilde{j}_{\ext}(t) \hat{A}
\end{align}
With the choice of an appropriate dimensionless coupling constant $\lambda$ Eq.~(\ref{Final2site-Ham}) reduces to the simple model Hamiltonian of Eq.~(\ref{Hubbard}). 

We note, that the same model Hamiltonian could have been derived by assuming an external vector potential in a gauge such that $a_{\ext}^{0} = 0$ and $\vec{a}_{\ext} \neq 0$. Then by the dipole approximation the corresponding Hamiltonian to Eq.~(\ref{StartingHam}) we would have terms of the form $\vec{a}_{\ext} \cdot \vec{\nabla}$, $\vec{a}_{\ext}^2$ and mixed terms of internal and external vector potential. By going into length gauge also for the external potential and performing the same steps as above, one ends up with the same two-site one-mode Hamiltonian.


\section{Overview of QEDFTs}
\label{app:overview}

Here we give an overview of the different QEDFTs that we have discussed explicitly. We employ for the Kohn-Sham scheme an uncoupled auxiliary quantum system with an initial state $\ket{\Phi_0} = \ket{\mathrm{M}_0} \otimes \ket{\mathrm{EM}_0}$. For the different levels of approximation the prerequisites for this initial state change, i.e. we might have different initial conditions that have to be fulfilled. Further we use the notational convention that the super-index s refers to the (uncoupled) Kohn-Sham quantity, e.g. $P_{0}[\Phi_0, P_k, A_k] = P_{0}^{\mathrm{s}}$.

\begin{widetext}
\begin{tabular}{|c| c| c| c| c|}
\hline 
Level of & External and & Kohn-Sham & Kohn-Sham & Initial \\
Approximation & Internal variables &  fields & Equations & Conditions \\[0.5ex] 
\hline 
QED & $(a_{\ext}^{k}, j_{\ext}^{k})$ & $\begin{aligned} & \!\!\!\!\!  P_{0}^{\mathrm{s}} \vec{a}_{\mathrm{KS}} \! \! = \! \! \frac{\imagi \hbar c} {2e} \!\! \left( \! \vec{Q}_{\mathrm{kin}}  \!\! -  \! \vec{Q}_{\mathrm{kin}}^{\mathrm{s}}  \!\! +  \! \vec{Q}_{\mathrm{int}} \! \right) \\ &+ mc^2 \left( \vec{J} - \vec{J}^{\mathrm{s}} \right) + P_{0} \vec{a}_{\mathrm{ext}}   \end{aligned}$ & $\begin{aligned} \!\imagi \hbar \partial_t \! \ket{\!\mathrm{M}\!}\!\! =\!\! \left[\!\hat{H}_{\mathrm{M}} \!\!-\!\! \frac{1}{c}\!\int \!\! \hat{\vec{J}} \!\cdot \!\vec{a}_{\mathrm{KS}} \!\right]\!\!\ket{\!\mathrm{M}\!}   \end{aligned}$ & $\begin{aligned}  \hspace{0.0cm} P_{k}^{(0)}    \end{aligned}$
\\
(Sec.~\ref{sec:RelQEDFT}) & $\begin{aligned}  \vec{P} & = e c\,  \langle :\!\hat{\psi}^{\dagger} \vec{\gamma} \hat{\psi} \! : \rangle \\ \vec{A} & =  \langle \hat{\vec{A}} \rangle  \end{aligned}$ & $\begin{aligned}\!\!\!\!\!\!\! \vec{j}_{\mathrm{KS}} \!\! =  \!\! \vec{j}_{\ext} \!\! + \! \underbrace{e c \langle :\! \dspin \vec{\gamma} \spin \!: \rangle}_{:= \vec{J}^{\mathrm{s}}} \!\!+\! (\!\vec{J} \!\!-\! \!\vec{J}_{\mathrm{s}} \!) \end{aligned}$ & $\begin{aligned}&\Box \vec{A}\!-\! \!\vec{\nabla} \!\left(\!\frac{1}{c} \!\int\! \frac{\vec{\nabla}'\!\!\cdot\!\vec{j}_{\mathrm{KS}}}{4 \pi \epsilon_0 |\vec{r}\! - \! \vec{r}'|}  \right) \nonumber
\\
& \qquad=  \mu_0 \, c  \,\vec{j}_{\mathrm{KS}}  \end{aligned}$ & $\begin{aligned}\!\!\! \left( \!A_{k}^{(0)} \!, A_{k}^{(1)} \!\right)  \end{aligned}$\\
\hline 
NR limit& $(a_{\ext}^{k}, j_{\ext}^{k})$ & $\begin{aligned}  \!\!\!\!\!J_{0}^{\mathrm{s}} & \vec{a}_{\mathrm{Hxc}} \!\!\!=\!\! \langle \hat{\vec{A}} \J_0 \rangle \!\!+\!\! \frac{mc}{e}\!\left(\! \vec{J}_{\mathrm{p}}\! - \! \vec{J}_{\mathrm{p}}^{\mathrm{s}} \right) 
\\
&- \frac{mc}{e}\vec{\nabla} \! \times  \! \left( \vec{M}-\vec{M}^{\mathrm{s}}\right)  \end{aligned}$ & $\begin{aligned} \imagi \hbar \partial_t \! \ket{\!\mathrm{M}\!}\!\! &=\!\! \left[\!\hat{H}_{\mathrm{M}} \!-\! \frac{1}{c}\!\int \!\! \hat{\vec{J}} \!\cdot \!\vec{a}_{\mathrm{KS}} \right. \\ & \left.  \!- \frac{e}{2mc^3}\!\int \!\! \hat{J}_0 \vec{a}_{\mathrm{KS}}^{2}\!\right]\!\ket{\!\mathrm{M}\!}   \end{aligned}$ & $\begin{aligned} \!\!\! \left(\!J_{0}^{(0)}\!, J_{k}^{(0)}\! \right)  \end{aligned}$ \\    
(Sec.~\ref{subsec:PauliFierz})&  $\begin{aligned} \vec{J} & \!\!=  \!\vec{J}_{p}\!\!+ \!\vec{\nabla} \! \times  \! \vec{M} \\ & -\!\! \frac{e}{mc^2} \langle \! \J_0 \hat{\vec{A}}_{\mathrm{tot}}  \rangle  \\ \vec{A} & =  \langle \hat{\vec{A}} \rangle  \end{aligned}$ & $\begin{aligned} \!\!\!\!\!\!\!\!\! \vec{j}_{\mathrm{KS}} & \!=  \! \vec{j}_{\ext}  \! +  \!\vec{J}_{\mathrm{p}}^{\mathrm{s}}\!\!+  \! \vec{\nabla} \! \times  \!\vec{M}_{\mathrm{s}} \\ & -\!\! \frac{e}{mc^2} J_0^{\mathrm{s}}  \underbrace{(\vec{a}_{\ext} + \vec{a}_{\mathrm{Hxc}})}_{\vec{a}_{\mathrm{KS}}}     \end{aligned}$ & $\begin{aligned}&\Box \vec{A}\!-\!\! \vec{\nabla} \!\left(\!\frac{1}{c} \!\int\! \frac{\vec{\nabla}'\!\!\cdot\!\vec{j}_{\mathrm{KS}}}{4 \pi \epsilon_0 |\vec{r}\! - \! \vec{r}'|}  \right) \nonumber
\\
& \qquad=  \mu_0 \, c  \,\vec{j}_{\mathrm{KS}}  \end{aligned}$& $\begin{aligned} \!\!\!  \left( \!A_{k}^{(0)}\!, A_{k}^{(1)}\! \right)   \end{aligned}$ \\ 
\hline 
No Mag & $(a_{\ext}^{k}, j_{\ext}^{k})$ &$\begin{aligned} & J_{0}^{\mathrm{s}} \vec{a}_{\mathrm{Hxc}} \!\!\!=\!\! \langle \hat{\vec{A}} \J_0 \rangle \!\!+\!\! \frac{mc}{e}\!\left(\! \vec{J}_{\mathrm{p}}\! - \! \vec{J}_{\mathrm{p}}^{\mathrm{s}} \right)  \end{aligned}$ & $\begin{aligned} \imagi \hbar \partial_t \! \ket{\!\mathrm{M}\!}\!\! &=\!\! \left[\!\hat{H}_{\mathrm{M}} \!-\! \frac{1}{c}\!\int \!\! \hat{\vec{J}} \!\cdot \!\vec{a}_{\mathrm{KS}} \right. \\ & \left.  \!- \frac{e}{2mc^3}\!\int \!\! \hat{J}_0 \vec{a}_{\mathrm{KS}}^{2}\!\right]\!\ket{\!\mathrm{M}\!}   \end{aligned}$ & $\begin{aligned}\!\!\! \left(\!J_{0}^{(0)}\!, J_{k}^{(0)} \! \right)  \end{aligned}$  \\    
(Sec.~\ref{subsec:approxQEDFT}) & $\begin{aligned} \!\!\!\!\! \vec{J}  = \vec{J}_{p} \!-\! &\frac{e}{mc^2} \langle \J_0 \hat{\vec{A}}_{\mathrm{tot}}  \rangle  \\ \vec{A} & =  \langle \hat{\vec{A}} \rangle  \end{aligned}$ & $\begin{aligned} \!\!\!\!\!\! \vec{j}_{\mathrm{KS}} & \!=  \! \vec{j}_{\ext}  \! +  \!\vec{J}_{\mathrm{p}}^{\mathrm{s}} \\ & -\!\! \frac{e}{mc^2} J_0^{\mathrm{s}}  \underbrace{(\vec{a}_{\ext} + \vec{a}_{\mathrm{Hxc}})}_{\vec{a}_{\mathrm{KS}}}     \end{aligned}$ & $\begin{aligned}&\Box \vec{A}\!-\!\! \vec{\nabla} \!\left(\!\frac{1}{c} \!\int\! \frac{\vec{\nabla}'\!\!\cdot\!\vec{j}_{\mathrm{KS}}}{4 \pi \epsilon_0 |\vec{r}\! - \! \vec{r}'|}  \right) \nonumber
\\
& \qquad=  \mu_0 \, c  \,\vec{j}_{\mathrm{KS}}  \end{aligned}$ & $\begin{aligned}   \!\!\!  \left(\! A_{k}^{(0)}\!, A_{k}^{(1)} \! \right)  \end{aligned}$ \\
\hline 
Cavity & $(a_{\ext}^{k}, \{ j^{\ext}_{\vec{n},\lambda} \})$ & $\begin{aligned} & J_{0}^{\mathrm{s}} \vec{a}_{\mathrm{Hxc}} \!\!\!=\!\! \langle \hat{\vec{A}} \J_0 \rangle \!\!+\!\! \frac{mc}{e}\!\left(\! \vec{J}_{\mathrm{p}}\! - \! \vec{J}_{\mathrm{p}}^{\mathrm{s}} \right)  \end{aligned}$ & $\begin{aligned} \imagi \hbar \partial_t \! \ket{\!\mathrm{M}\!}\!\! &=\!\! \left[\!\hat{H}_{\mathrm{M}} \!-\! \frac{1}{c}\!\int \!\! \hat{\vec{J}} \!\cdot \!\vec{a}_{\mathrm{KS}} \right. \\ & \left.  \!- \frac{e}{2mc^3}\!\int \!\! \hat{J}_0 \vec{a}_{\mathrm{KS}}^{2}\!\right]\!\ket{\!\mathrm{M}\!}   \end{aligned}$ &  $\begin{aligned}\!\!\!    \left(\!J_{0}^{(0)}\!, J_{k}^{(0)} \! \right)\end{aligned}$ \\ 
(Sec.~\ref{subsec:approxQEDFT}) & $\begin{aligned}  & \hspace{-0.6cm} \vec{J}  = \vec{J}_{p} \!\!-\!\! \frac{e}{mc^2} \langle \J_0 \hat{\vec{A}}_{\mathrm{tot}}  \rangle  \\ \! A_{\vec{n}, \lambda}  \!\!&= \!\! \sqrt{\!\!\frac{\hbar c^2}{\epsilon_0}}\!  \frac{f_{\mathrm{EM}}}{\sqrt{2 \omega_n}} q_{\vec{n}, \lambda}  \end{aligned}$ & $\begin{aligned} & \!\!\!\!\! j^{\mathrm{KS}}_{\vec{n},\lambda}  = j^{\ext}_{\vec{n},\lambda} +(J_{\mathrm{p}}^{\mathrm{s}})_{\vec{n},\lambda}  \\ & -\!\! \frac{e}{mc^2} (J_0^{\mathrm{s}}  \underbrace{(\vec{a}_{\ext} + \vec{a}_{\mathrm{Hxc}})}_{\vec{a}_{\mathrm{KS}}})_{\vec{n},\lambda}  \end{aligned}$ & $\begin{aligned}\frac{f_{\mathrm{EM}}}{\sqrt{2 \omega_n}}& \left(\partial_0^2 + \vec{k}_{\vec{n}}^2  \right) q_{\vec{n}, \lambda}(t) 
\\
&=  \sqrt{\frac{\mu _0}{\hbar c^2}} j_{\vec{n}, \lambda}^{\mathrm{KS}}  \end{aligned}$ & $\begin{aligned} \!\! \left\{\! A_{\vec{n}, \lambda}^{(0)}, A_{\vec{n}, \lambda}^{(1)} \! \right\}  \end{aligned}$ \\ 
\hline 
Dipole  & $(a_{\ext}^{k}, \{ j^{\ext}_{\vec{n},\lambda} \})$ & $\begin{aligned} & J_{0}^{\mathrm{s}} \vec{a}_{\mathrm{Hxc}} \!\!\!=\!\! \langle \hat{\vec{A}} \J_0 \rangle \!\!+\!\! \frac{mc}{e}\!\left(\! \vec{J}_{\mathrm{p}}\! - \! \vec{J}_{\mathrm{p}}^{\mathrm{s}} \right)  \end{aligned}$ &  $\begin{aligned} \imagi \hbar \partial_t \! \ket{\!\mathrm{M}\!}\!\! &=\!\! \left[\!\hat{H}_{\mathrm{M}} \!-\! \frac{1}{c}\!\int \!\! \hat{\vec{J}} \!\cdot \!\vec{a}_{\mathrm{KS}} \right. \\ & \left.  \!- \frac{e}{2mc^3}\!\int \!\! \hat{J}_0 \vec{a}_{\mathrm{KS}}^{2}\!\right]\!\ket{\!\mathrm{M}\!}   \end{aligned}$  &  $\begin{aligned}  \!\!\!\! \left(\!J_{0}^{(0)}\!, J_{k}^{(0)} \!\right)\end{aligned}$  \\    
(Sec.~\ref{subsec:approxQEDFT}) & $\begin{aligned} & \hspace{-0.6cm} \vec{J}  = \vec{J}_k\!\! -\!\! \frac{e}{mc^2} \langle \J_0 \hat{\vec{A}}_{\mathrm{tot}}  \rangle  \\   A_{\vec{n}, \lambda} \!\! &= \!\! \sqrt{\!\!\frac{\hbar c^2}{L^3 \epsilon_0}}\!  \frac{ f_{\mathrm{EM}}}{\sqrt{2 \omega_n}} q_{\vec{n}, \lambda}  \end{aligned}$ &$\begin{aligned} & \!\!\!\!\!  j^{\mathrm{KS}}_{\vec{n},\lambda}  = j^{\ext}_{\vec{n},\lambda} +(J_{\mathrm{p}}^{\mathrm{s}})_{\vec{n},\lambda}  \\ & -\!\! \frac{e}{mc^2} (J_0^{\mathrm{s}}  \underbrace{(\vec{a}_{\ext} + \vec{a}_{\mathrm{Hxc}})}_{\vec{a}_{\mathrm{KS}}})_{\vec{n},\lambda}  \end{aligned}$ & $\begin{aligned}\frac{f_{\mathrm{EM}}}{\sqrt{2 \omega_n}}& \left(\partial_0^2 + \vec{k}_{\vec{n}}^2  \right) q_{\vec{n}, \lambda}(t) 
\\
&=  \sqrt{\frac{L^3 \mu _0 }{\hbar c^2}} j_{\vec{n}, \lambda}^{\mathrm{KS}}  \end{aligned}$ & $\begin{aligned} \!\!  \left\{\! A_{\vec{n}, \lambda}^{(0)}, A_{\vec{n}, \lambda}^{(1)} \! \right\}  \end{aligned}$ \\ 
\hline 
Model & $(a_{\ext},  j_{\ext})$ & $\begin{aligned} n_{s} a_{\mathrm{KS}} = \lambda \langle \hat{n} \A \rangle + n a_{\ext}(t) \end{aligned}$ &  $\begin{aligned} \imagi \hbar  \partial_t \ket{\!\mathrm{M}\!}\! &= \!\left[-t_{\mathrm{kin}} \s_x \right. \\ & \left.  \!- \frac{1}{c} \J  a_{\mathrm{KS}}\!  \right]\!\!\ket{\!\mathrm{M}\!}
\\
&   \end{aligned}$ & $\begin{aligned} \!\!\! \left(\!J^{(0)}\!, J^{(1)} \!\right)    \end{aligned}$ \\    
(Sec.~\ref{sec:Model}) & $\begin{aligned} J & = e \omega l \sigma_z  \\  A = & \sqrt{\frac{\hbar c^2}{\epsilon_0 L^3}} \frac{q}{\sqrt{2 \omega}}  \end{aligned}$ &$\begin{aligned} &j_{\mathrm{KS}} =  j_{\ext} + J \end{aligned}$ & $\begin{aligned} \left( \partial_0^2 + k^2\right) A  =   \frac{\mu_0 c}{L^3} j_{\mathrm{KS}}
 \end{aligned}$ & $\begin{aligned} \!\! \left(\! A^{(0)}\!, A^{(1)} \!\right) \end{aligned}$ \\ 
\hline 
\end{tabular}
\end{widetext}

We point out, that due to the change of the physical current $\vec{J}$ through out the hierachy of QEDFTs also the inhomogeneity in the according Maxwell equations change. This inhomogeneity describes how the photons are coupled to the charged quantum particles, which effectively also leads to a coupling between the photons. This can be most easily seen in the non-relativistic limit, where the inhomogeneity contains terms like $\langle \J_0 \A_k \rangle$. Since the current of the auxiliary Kohn-Sham system is by construction equal to the exact current (at least for the non-relativistic limit), this coupling between the photons is also present in the Kohn-Sham Maxwell equation. The term $J_0^{\mathrm{s}}\vec{a}_{\mathrm{Hxc}}$ of the Kohn-Sham current contains these non-trivial couplings as functionals of the initial states and internal pair. When restricting the photons to a cavity, the Kohn-Sham current is then responsible to couple the different photon modes. The coupling terms in the Kohn-Sham current are 
specifically relevant in the context of, e.g. nano-plasmonics, where the electromagnetic fields are enhanced due to the presence of the plasmons, or in the optical control of currents in solids \cite{schultze-2013}.



\begin{thebibliography}{10}

\bibitem{ryder-QFT}
L.~H. Ryder,
\newblock {\em Quantum field theory},
\newblock Cambridge Univ.\ Press, Cambridge, 2006.

\bibitem{greiner-FQ}
W.~Greiner and J.~Reinhard,
\newblock {\em Field Quantization},
\newblock Springer-Verlag, Berlin, 1996.

\bibitem{greiner-QED}
B.~M. W.~Greiner and J.~Rafelski,
\newblock {\em Quantum Electrodynamics of Strong Fields},
\newblock Springer-Verlag, Berlin, 1985.

\bibitem{fetter-MBT}
A.~Fetter and J.~Walecka,
\newblock {\em Quantum Theory of Many-Particle Systems},
\newblock Dover Publications, Mineola, New York, 2003.

\bibitem{stefanucci-MBT}
G.~Stefanucci and R.~van Leeuwen,
\newblock {\em Nonequilibrium Many-Body Theory of Quantum Systems},
\newblock Cambridge University Press, Cambridge, 2013.

\bibitem{bonitz-QKT}
M.~Bonitz,
\newblock {\em Quantum Kinetic Theory},
\newblock Teubner-Verlag, Stuttgart/Leipzig, 1998.

\bibitem{dreizler-DFT}
R.~Dreizler and E.~Gross,
\newblock {\em Density Functional Theory - An Approach to the Quantum Many-Body
  Problem},
\newblock Springer-Verlag, Berlin, 1990.

\bibitem{engel-DFT}
E.~Engel and R.~Dreizler,
\newblock {\em Density Functional Theory - An Advanced Course},
\newblock Springer-Verlag, Berlin, 2011.

\bibitem{ullrich-TDDFT}
C.~A. Ullrich,
\newblock {\em Time-Dependent Density-Functional Theory},
\newblock Oxford University Press, Oxford, 2012.

\bibitem{marques-TDDFT}
M.~A. Marques, N.~T. Maitra, F.~M. Nogueira, E.~K. Gross, and A.~Rubio,
\newblock {\em Fundamentals of time-dependent density functional theory},
  volume 837,
\newblock Springer, 2012.

\bibitem{burke-2012}
K.~Burke,
\newblock J. Chem. Phys. {\bf 136},  (2012).

\bibitem{bleiziffer-2013}
P.~Bleiziffer, A.~He{\ss}elmann, and A.~G{\"o}rling,
\newblock J. Chem. Phys. {\bf 139},  (2013).

\bibitem{andrade-2012}
X.~Andrade et~al.,
\newblock J. Phys. Cond. Matt. {\bf 24}, 233202 (2012).

\bibitem{gupta-2001}
A.~K. Gupta and D.~Neuhauser,
\newblock Int. J. Quant. Chem. {\bf 81}, 260 (2001).

\bibitem{iwasa-2009}
T.~Iwasa and K.~Nobusada,
\newblock Phys. Rev. A {\bf 80}, 043409 (2009).

\bibitem{chen-2010}
H.~Chen, J.~M. McMahon, M.~A. Ratner, and G.~C. Schatz,
\newblock J. Phys. Chem. C {\bf 114}, 14384 (2010).

\bibitem{fratalocchi-2011}
A.~Fratalocchi and G.~Ruocco,
\newblock Phys. Rev. Lett. {\bf 106}, 105504 (2011).

\bibitem{yabana-2012}
K.~Yabana, T.~Sugiyama, Y.~Shinohara, T.~Otobe, and G.~F. Bertsch,
\newblock Phys. Rev. B {\bf 85}, 045134 (2012).

\bibitem{scully-QO}
M.~Scully and M.~Zubairyh,
\newblock {\em Quantum Optics},
\newblock Cambridge University Press, Cambridge, 1997.

\bibitem{gardiner-QO}
C.~Gardiner and P.~Zoller,
\newblock {\em Quantum Noise},
\newblock Springer-Verlag, Berlin, 2004.

\bibitem{dicke-1954}
R.~Dicke,
\newblock Phys. Rev. {\bf 93}, 99 (1954).

\bibitem{chen-2008}
Q.-H. Chen, Y.-Y. Zhang, T.~Liu, and K.-L. Wang,
\newblock Phys. Rev. A {\bf 78}, 051801 (2008).

\bibitem{braak-2013}
D.~Braak,
\newblock arXiv preprint arXiv:1304.2529  (2013).

\bibitem{rzazewski-1975}
K.~Rza\ifmmode~\dot{z}\else \.{z}\fi{}ewski, K.~W\'odkiewicz, and
  W.~\ifmmode~\dot{Z}\else \.{Z}\fi{}akowicz,
\newblock Phys. Rev. Lett. {\bf 35}, 432 (1975).

\bibitem{rzazewski-1991}
K.~Rza\ifmmode~\dot{z}\else \.{z}\fi{}ewski and K.~W\'odkiewicz,
\newblock Phys. Rev. A {\bf 43}, 593 (1991).

\bibitem{vukics-2012}
A.~Vukics and P.~Domokos,
\newblock Phys. Rev. A {\bf 86}, 053807 (2012).

\bibitem{raimond-2001}
J.~M. Raimond, M.~Brune, and S.~Haroche,
\newblock Rev. Mod. Phys. {\bf 73}, 565 (2001).

\bibitem{walther-2006}
H.~Walther, B.~T. Varcoe, B.-G. Englert, and T.~Becker,
\newblock Rep. Prog. Phys. {\bf 69}, 1325 (2006).

\bibitem{mekhov-2012}
I.~B. Mekhov and H.~Ritsch,
\newblock J. Phys. B {\bf 45}, 102001 (2012).

\bibitem{ritsch-2013}
H.~Ritsch, P.~Domokos, F.~Brennecke, and T.~Esslinger,
\newblock Rev. Mod. Phys. {\bf 85}, 553 (2013).

\bibitem{blais-2004}
A.~Blais, R.-S. Huang, A.~Wallraff, S.~M. Girvin, and R.~J. Schoelkopf,
\newblock Phys. Rev. A {\bf 69}, 062320 (2004).

\bibitem{wallraff-2004}
A.~Wallraff et~al.,
\newblock Nature {\bf 431}, 162 (2004).

\bibitem{todorov-2010}
Y.~Todorov et~al.,
\newblock Phys. Rev. Lett. {\bf 105}, 196402 (2010).

\bibitem{you-2011}
J.~You and F.~Nori,
\newblock Nature {\bf 474}, 589 (2011).

\bibitem{schwartz-2011}
T.~Schwartz, J.~A. Hutchison, C.~Genet, and T.~W. Ebbesen,
\newblock Phys. Rev. Lett. {\bf 106}, 196405 (2011).

\bibitem{imorral-2012}
A.~F. i~Morral and F.~Stellacci,
\newblock Nature Materials {\bf 11}, 272 (2012).

\bibitem{ciuti-2006}
C.~Ciuti and I.~Carusotto,
\newblock Phys. Rev. A {\bf 74}, 033811 (2006).

\bibitem{hutchison-2012}
J.~A. Hutchison, T.~Schwartz, C.~Genet, E.~Devaux, and T.~W. Ebbesen,
\newblock Ang. Chem. Int. Ed. {\bf 51}, 1592 (2012).

\bibitem{fausti-2011}
D.~Fausti et~al.,
\newblock Science {\bf 331}, 189 (2011).

\bibitem{tokatly-2013}
I.~V. Tokatly,
\newblock Phys. Rev. Lett. {\bf 110}, 233001 (2013).

\bibitem{rajagopal-1994}
A.~K. Rajagopal,
\newblock Phys. Rev. A {\bf 50}, 3759 (1994).

\bibitem{ruggenthaler-2011b}
M.~Ruggenthaler, F.~Mackenroth, and D.~Bauer,
\newblock Phys. Rev. A {\bf 84}, 042107 (2011).

\bibitem{vanleeuwen-2001}
R.~Van~Leeuwen,
\newblock Int. J. Mod. Phys. B {\bf 15}, 1969 (2001).

\bibitem{leeuwen-1999}
R.~van Leeuwen,
\newblock Phys. Rev. Lett. {\bf 82}, 3863 (1999).

\bibitem{farzanehpour-2012}
M.~Farzanehpour and I.~V. Tokatly,
\newblock Phys. Rev. B {\bf 86}, 125130 (2012).

\bibitem{ruggenthaler-2011}
M.~Ruggenthaler and R.~van Leeuwen,
\newblock Europhys. Lett. {\bf 95}, 13001 (2011).

\bibitem{runge-1984}
E.~Runge and E.~K.~U. Gross,
\newblock Phys. Rev. Lett. {\bf 52}, 997 (1984).

\bibitem{Tokatly2011ChemPhys}
I.~V. Tokatly,
\newblock Chem. Phys. {\bf 391}, 78  (2011).

\bibitem{tokatly-2011}
I.~V. Tokatly,
\newblock Phys. Rev. B {\bf 83}, 035127 (2011).

\bibitem{li-2008}
Y.~Li and C.~A. Ullrich,
\newblock J. Chem. Phys. {\bf 129},  (2008).

\bibitem{baer-2008}
R.~Baer,
\newblock J. Chem. Phys. {\bf 128},  (2008).

\bibitem{kurth-2011}
S.~Kurth and G.~Stefanucci,
\newblock Chem. Phys. {\bf 391}, 164  (2011),
\newblock Open problems and new solutions in time dependent density functional
  theory.

\bibitem{vignale-2008}
G.~Vignale,
\newblock Phys. Rev. A {\bf 77}, 062511 (2008).

\bibitem{shore1990theory}
B.~Shore,
\newblock {\em The Theory of Coherent Atomic Excitation: Multilevel atoms and
  incoherence},
\newblock The Theory of Coherent Atomic Excitation, Wiley, 1990.

\bibitem{Shore1993}
B.~W. Shore and P.~L. Knight,
\newblock J. Mod. Opt. {\bf 40}, 1195 (1993).

\bibitem{gerry2005introductory}
C.~Gerry and P.~Knight,
\newblock {\em Introductory Quantum Optics},
\newblock Cambridge University Press, 2005.

\bibitem{Braak2011}
D.~Braak,
\newblock Phys. Rev. Lett. {\bf 107}, 100401 (2011).

\bibitem{Fuks2013}
J.~I. Fuks et~al.,
\newblock Phys. Rev. A {\bf 88}, 062512 (2013).

\bibitem{nielsen-2013}
S.~E.~B. Nielsen, M.~Ruggenthaler, and R.~van Leeuwen,
\newblock Europhys. Lett. {\bf 101}, 33001 (2013).

\bibitem{flick-2014}
J.~Flick, M.~Ruggenthaler, H.~Appel, and A.~Rubio,
\newblock arXiv:  (2014).

\bibitem{Glauber1963}
R.~J. Glauber,
\newblock Phys. Rev. {\bf 130}, 2529 (1963).

\bibitem{Glauber1963a}
R.~J. Glauber,
\newblock Phys. Rev. {\bf 131}, 2766 (1963).

\bibitem{Narozhny1981}
N.~B. Narozhny, J.~J. Sanchez-Mondragon, and J.~H. Eberly,
\newblock Phys. Rev. A {\bf 23}, 236 (1981).

\bibitem{maitra-2001}
N.~T. Maitra and K.~Burke,
\newblock Phys. Rev. A {\bf 63}, 042501 (2001).

\bibitem{maitra-2002b}
N.~T. Maitra, K.~Burke, and C.~Woodward,
\newblock Phys. Rev. Lett. {\bf 89}, 023002 (2002).

\bibitem{kuemmelRMP}
S.~K\"ummel and L.~Kronik,
\newblock Rev. Mod. Phys. {\bf 80}, 3 (2008).

\bibitem{pellegrini-2014}
C.~Pellegrini, J.~Flick, I.~Tokatly, H.~Appel, and A.~Rubio,
\newblock arXiv:  (2014).

\bibitem{takaesu-2009}
T.~Takaesu,
\newblock J. Math. Phys. {\bf 50},  (2009).

\bibitem{hainzl-2003}
C.~Hainzl and H.~Siedentop,
\newblock Commun. Math. Phys. {\bf 243}, 241 (2003).

\bibitem{nelson-1964}
E.~Nelson,
\newblock J. Math. Phys. {\bf 5}, 1190 (1964).

\bibitem{bachmann-2012}
S.~Bachmann, D.-A. Deckert, and A.~Pizzo,
\newblock Journal of Functional Analysis {\bf 263}, 1224  (2012).

\bibitem{vignale-2004}
G.~Vignale,
\newblock Phys. Rev. B {\bf 70}, 201102 (2004).

\bibitem{hainzl-2002}
C.~Hainzl and R.~Seiringer,
\newblock Adv. Theor. Math. Phys {\bf 6}, 847 (2002).

\bibitem{hiroshima-2002}
F.~Hiroshima,
\newblock Annales Henri Poincar{\'e} {\bf 3}, 171 (2002).

\bibitem{strange-RQM}
P.~Strange,
\newblock {\em Relativistic Quantum Mechanics: with applications in condensed
  matter and atomic physics},
\newblock Cambridge University Press, Cambridge, 1998.

\bibitem{tokatly-2005a}
I.~V. Tokatly,
\newblock Phys. Rev. B {\bf 71}, 165104 (2005).

\bibitem{farzanehpour-2014}
M.~Farzanehpour and I.~Tokatly,
\newblock arXiv:  (2014).

\bibitem{yuen-2009}
J.~Yuen-Zhou, C.~Rodr{\'\i}guez-Rosario, and A.~Aspuru-Guzik,
\newblock Phys. Chem. Chem. Phys. {\bf 11}, 4509{\textendash}4522 (2009),
\newblock n/a.

\bibitem{yuen-2010}
J.~Yuen-Zhou, D.~G. Tempel, C.~A. Rodr{\'\i}guez-Rosario, and A.~Aspuru-Guzik,
\newblock Phys. Rev. Lett. {\bf 104}, 043001 (2010).

\bibitem{dagosta-2007}
M.~Di~Ventra and R.~D'Agosta,
\newblock Phys. Rev. Lett. {\bf 98}, 226403 (2007).

\bibitem{appel-2009}
H.~Appel and M.~Di~Ventra,
\newblock Phys. Rev. B {\bf 80}, 212303 (2009).

\bibitem{appel-2011}
H.~Appel and M.~D. Ventra,
\newblock Chem. Phys. {\bf 391}, 27  (2011).

\bibitem{faisal-MPT}
F.~H. Faisal,
\newblock {\em Theory of multiphoton processes},
\newblock Springer-Verlag, Berlin, 1987.

\bibitem{schultze-2013}
M.~Schultze et~al.,
\newblock Nature {\bf 493}, 75 (2013).

\end{thebibliography}

\end{document}